\documentclass[usenatbib]{emulateapj}
\usepackage{graphicx}
\usepackage[normalem]{ulem}
\usepackage[flushleft]{threeparttable}
\usepackage[usenames,dvipsnames,svgnames,table]{xcolor}
\usepackage{hyperref}
\definecolor{darkblue}{rgb}{0.0,0.0,0.9}
\hypersetup{colorlinks,breaklinks,
            linkcolor=darkblue,urlcolor=darkblue,
            anchorcolor=darkblue,citecolor=darkblue}
\usepackage{amsmath,amssymb}
\usepackage{amsmath}

\usepackage{natbib}
\begin{document}

\def\etal{et al.\ \rm}
\def\ba{\begin{eqnarray}}
\def\ea{\end{eqnarray}}
\def\etal{et al.\ \rm}
\def\Fdw{F_{\rm dw}}
\def\Tex{T_{\rm ex}}
\def\Fdis{F_{\rm dw,dis}}
\def\Fnu{F_\nu}
\def\WD{\rm WD}

\newcommand{\ra}{r_\mathrm{a}}
\newcommand{\rb}{r_\mathrm{b}}
\newcommand{\rc}{r_\mathrm{c}}
\newcommand{\md}{\mathrm{d}}
\newcommand{\me}{\mathrm{e}}
\newcommand{\mi}{\mathrm{i}}
\newcommand{\rp}{r_\mathrm{p}}
\newcommand{\fmer}{f_\mathrm{m}}
\newcommand{\rt}{\mathrm{t}}

\newcommand{\fnow}{f_\mathrm{m}(12\,\mathrm{Gyr})}
\def\p{\partial}
\newcommand{\epsGR}{\epsilon_\mathrm{GR}}
\newcommand{\nn}{\nonumber}
\newcommand{\jmin}{j_\mathrm{min}}
\newcommand{\Rg}{\mathbf{R}_\mathrm{g}}
\newcommand{\Rb}{\mathbf{R}_\mathrm{b}}

\newcommand{\epsstr}{\epsilon_\mathrm{strong}}
\newcommand{\epsweak}{\epsilon_\mathrm{weak}}
\newcommand{\epspibytwo}{\epsilon_{\pi/2}}

\newcommand{\tmin}{t_\mathrm{min}}
\newcommand{\imin}{i_\mathrm{min}}
\newcommand{\jmax}{j_\mathrm{max}}
\newcommand{\pmin}{p_\mathrm{min}}
\newcommand{\emax}{e_\mathrm{max}}
\newcommand{\emin}{e_\mathrm{min}}
\newcommand{\elim}{e_\mathrm{lim}}
\newcommand{\tsec}{t_\mathrm{sec}}
\newcommand{\calA}{$\mathcal{A}$}
\newcommand{\calB}{$\mathcal{B}$}
\newcommand{\calC}{$\mathcal{C}$}
\newcommand{\calD}{$\mathcal{D}$}

\newcommand\cmtrr[1]{{\color{red}[RR: #1]}}
\newcommand\cmtch[1]{{\color{red}[CH: #1]}}

\newcommand\newrr[1]{{\color{red} #1}}
\newcommand\newch[1]{{\color{magenta} #1}}

%%%%%%%%%%%%%%%%%%%%%%%%%%%%%%%%%%%
%%%%%%%%%%%%%%%%%%%%%%%%%%%%%%%%%%%
%%%%%%%%%%%%%%%%%%%%%%%%%%%%%%%%%%%
%%%%%%%%%%%%%%%%%%%%%%%%%%%%%%%%%%%

\title{Anatomy of a slow merger: dissecting secularly-driven inspirals of LIGO/Virgo gravitational wave sources}

%\title{Analytical study of compact object binary mergers in hierarchical triples and stellar clusters}

\author{Chris Hamilton\altaffilmark{1,2,$\dagger$} \& Roman R. Rafikov\altaffilmark{1,2}}
\altaffiltext{1}{Institute for Advanced Study, Einstein Drive, Princeton, NJ 08540}
\altaffiltext{2}{Department of Applied Mathematics and Theoretical Physics, University of Cambridge, Wilberforce Road, Cambridge CB3 0WA, UK}
\altaffiltext{$\dagger$}{chamilton@ias.edu}

%%%%%%%%%%%%%%%%%%%%%%%%%%%%%%%%%%%
%%%%%%%%%%%%%%%%%%%%%%%%%%%%%%%%%%%
%%%%%%%%%%%%%%%%%%%%%%%%%%%%%%%%%%%
%%%%%%%%%%%%%%%%%%%%%%%%%%%%%%%%%%%

\begin{abstract}

The dozens of compact object mergers detected by LIGO/Virgo raise a key theoretical question: how do initially wide binaries shrink sufficiently quickly that they are able to merge via gravitational wave (GW) radiation within a Hubble time? 
One promising class of answers involves secular driving of binary eccentricity by some external tidal perturbation. This perturbation can arise due to the presence of a tertiary point mass, in which case the system exhibits Lidov-Kozai (LK) dynamics, or it can stem from the tidal field of the stellar cluster in which the binary orbits. 
While these secular tide-driven mechanisms have been studied exhaustively in the case of no GW emission, 
when GWs are included the dynamical behavior is still incompletely understood.
In this paper we consider compact object binaries driven to merger via high eccentricity excitation by (doubly-averaged, test-particle quadrupole level) cluster tides --- which includes LK-driven mergers as a special case --- and include the effects of both general relativistic precession and GW emission.  We provide for the first time an analytical understanding of the different evolutionary stages of the binary's semimajor axis, secular oscillation timescale, and phase space structure all the way to merger. 
%%
%We also derive a new and improved analytical formula for the merger timescale given the binary's initial conditions.
%%
Our results will inform future population synthesis calculations of compact object binary mergers from hierarchical triples and stellar clusters.
\end{abstract}
%%%%%%%%%%%%%%%%%%%%%%%%%%%%%%%%%%%%%%%%%%%%%%%%%%

%%%%%%%%%%%%%%%%%%%%%%%%%%%%%%%%%%%%%%%%%%%%%%%%%%

%%%%%%%%%%%%%%%%% BODY OF PAPER %%%%%%%%%%%%%%%%%%

%%%%%%%%%%%%%%%%%%%%%%%%%%%%%%%%%%%%%%%%%%%%%%%%%%
%%%%%%%%%%%%%%%%%%%%%%%%%%%%%%%%%%%%%%%%%%%%%%%%%%
\section{Introduction}
\label{sec:Introduction}
%%%%%%%%%%%%%%%%%%%%%%%%%%%%%%%%%%%%%%%%
%%%%%%%%%%%%%%%%%%%%%%%%%%%%%%%%%%%%%%%%
%\subsection{Background}

The LIGO/Virgo
Collaboration has now detected around 90 compact object binary mergers
\citep{Abbott2021gwtc}. However, there is still ambiguity on the theoretical side about which mechanisms drive these mergers. 

One much-studied candidate is the Lidov-Kozai (LK) mechanism which operates in hierarchical triple systems \citep{Lidov1962-tu,Kozai1962-ck,Naoz2016-qj}, in which a compact object binary is both orbited and tidally torqued by a bound tertiary perturber. Provided the inclination angle between the two orbital planes of the hierarchical triple system is sufficiently large, the
binary's eccentricity $e$ can be driven periodically towards values approaching unity on secular timescales (i.e. much longer than any of the orbital periods in the system).  This
greatly reduces the pericentre distance $p\equiv a(1-e)$, where $a$ is the binary's semimajor axis. 
Gravitational wave (GW) radiation is strongly enhanced at small $p$, i.e. is
strongly concentrated around the maximum $e \to
e_\mathrm{max} \approx 1$, so there is a `burst' of orbital energy loss at each peak. By energy conservation, this corresponds to a decay in binary semimajor axis by some amount $\Delta a$.
Such losses accumulate over multiple secular cycles until the binary is so compact that it
effectively decouples from the tidal perturbations and undergoes a GW-dominated inspiral.
This basic understanding has inspired scores of papers that aim to understand binary black hole mergers and predict their rates (e.g.
\citealt{Miller2002-co,Blaes2002-rx,Wen2003-jf,Thompson2011-wv,Antonini2012-fv,Antognini2014-jj,Silsbee2017-bo,Liu2017-ax}).

More generally, every binary that resides in a stellar cluster feels
the cluster's gravitational potential.  As we showed in \cite{Hamilton2019-zl,Hamilton2019-jn}
--- hereafter referred to as Papers I and II respectively ---
the cluster potential
provides a tidal torque on the binary just as a tertiary perturber would.
This cluster-tidal torque can drive LK-like eccentricity oscillations in binaries on astrophysically relevant timescales.
%the resulting secular equations of motion can always be recast as a Hamiltonian dynamical system with one degree of freedom. The effective Hamiltonian contains a dimensionless parameter $\Gamma$, which is related to the curvature of the cluster potential; the LK theory is recovered precisely in the limit $\Gamma=1$.  
One can therefore consider the cluster itself to be a ubiquitous `third body' which has the capacity to induce `cluster tide-driven' mergers \citep{Hamilton2019-mq,Bub2019-vn,Arca_Sedda2020-vg}.
%Of course, to do
%this quantitatively means we must include GW emission in our equations of
%motion, which we did not do in Papers I-III. In this paper we will introduce
%GW emission into our calculations and understand the physics of
%cluster tide-driven compact object mergers.
%At this stage the skeptical reader might ask: what is to be gained from a
%Paper on cluster tide-driven mergers that is not already known from LK merger
%theory?  Aren't we going to just end up with the usual formulae, but with a few
%new factors of $\Gamma$ floating around?  Isn't the physics essentially
%understood already? And indeed, 
%:
%\begin{enumerate}
%\item The binary is torqued periodically to very high eccentricity.
%\end{enumerate}
The primary caveat to both the LK-driven and cluster-tide driven merger mechanisms is that 1pN general relativistic (GR) apsidal precession acts to quench the oscillations and hence delay mergers. In \cite{Hamilton2021-nk} --- hereafter Paper III --- we provided a systematic account of how the LK-like dynamics of binaries are modified as the strength of GR precession is varied.

Guided the basic understanding outlined above,
most of the aforementioned studies either opt for direct numerical integration
of the binary equations of motion (e.g. \citealt{Silsbee2017-bo}), or they aim at a parameterization of the
total merger timescale as a function of $e_\mathrm{max}$ (e.g. \citealt{Thompson2011-wv,Liu2018-kg}), and very little
theoretical understanding has been developed beyond this (but see \citealt{Randall2018-uq}).
%\footnote{Or a version of $\emax$ that is
%modified by non-ideal effects, e.g. the breakdown of the test-particle
%approximation, 
%(i.e. allowing the binary's inner orbit to have comparable angular
%momentum to the tertiary's outer orbit), 
%the DA approximation,
%(allowing for
%short-timescale fluctuations in the tidal torque)
%and the quadrupole
%approximation %(including octupole or higher terms in the perturbing potential)
%--- see \citealt{Antonini2014-oa,Anderson2017-cb,Liu2018-kg,Grishin2018-da} and references
%therein.} 
On the other hand, a detailed look at numerical integrations suggests that LK-driven and cluster tide-driven mergers are in fact very dynamically rich, even in the doubly-averaged, test-particle quadrupole limit\footnote{In this limit we average the dynamics over both the binary's inner Keplerian orbit and its barycentric orbit around the cluster/perturber; we assume the outer orbit is unchanging; and we perform a tidal expansion of the perturbation upon the binary only to quadrupole order. For more details see \cite{Antonini2014-oa,Naoz2016-qj,Grishin2018-da}.} that we consider exclusively in this paper.
They exhibit non-trivial time evolution of the binary's semimajor
axis, maximum and minimum eccentricity, phase space location, secular period, etc.  
To demonstrate this, we now provide a numerical example of a LK-driven merger.

\subsection{Example of a LK-driven merger}

%%%%%%%%%%%%%%%%%%%%%%%%%%%%%%%%%%%%%%%%%%%%%%%%
\begin{figure*}
\centering
   \includegraphics[width=0.99\linewidth]{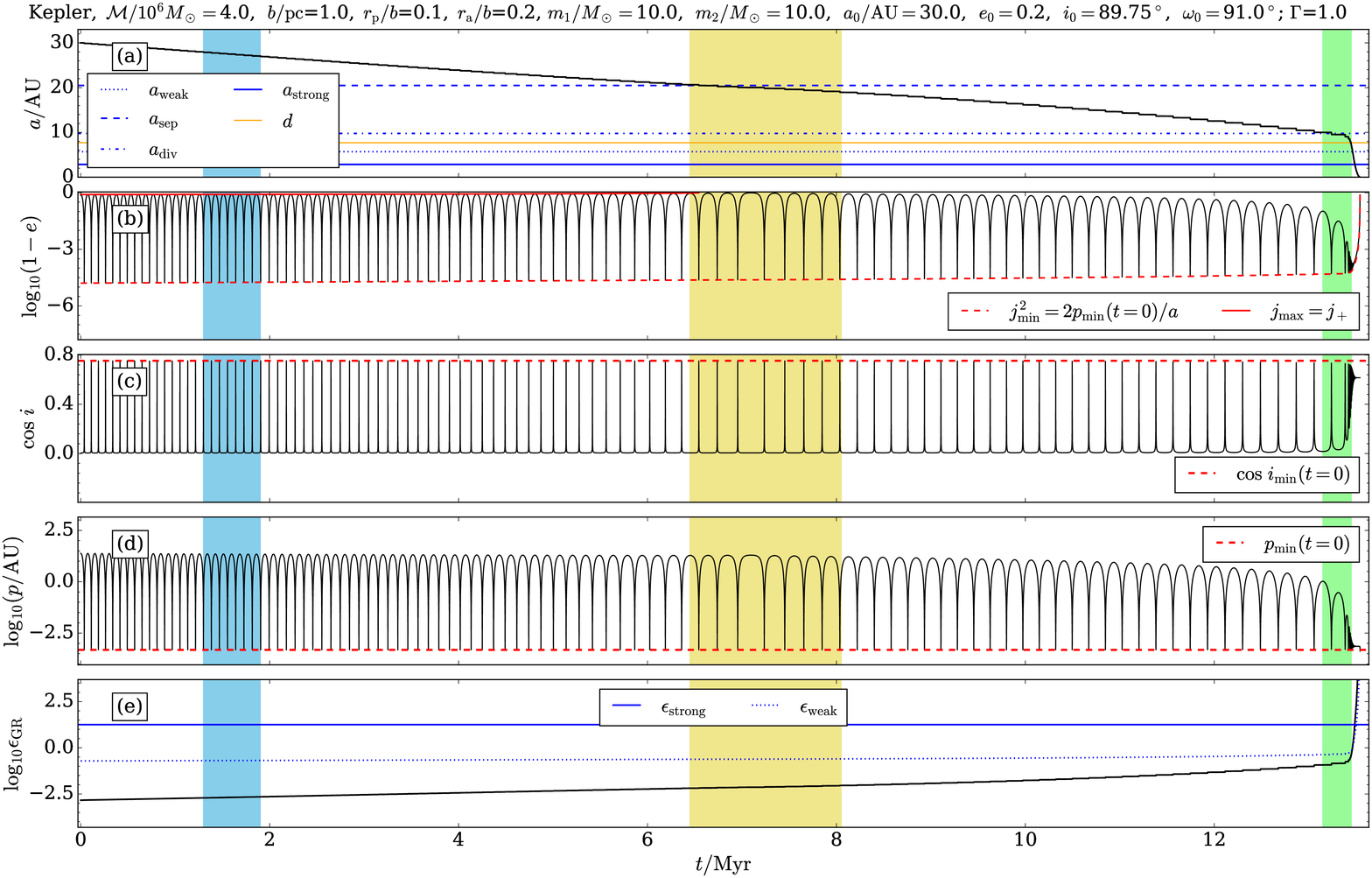}
    \includegraphics[width=0.99\linewidth]{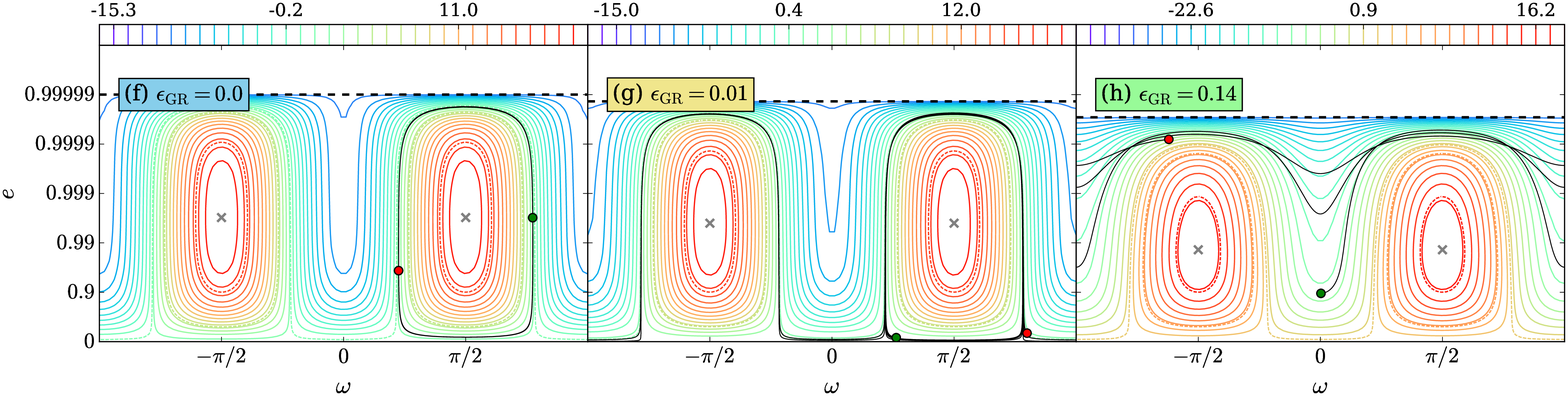}
    \includegraphics[width=0.99\linewidth]{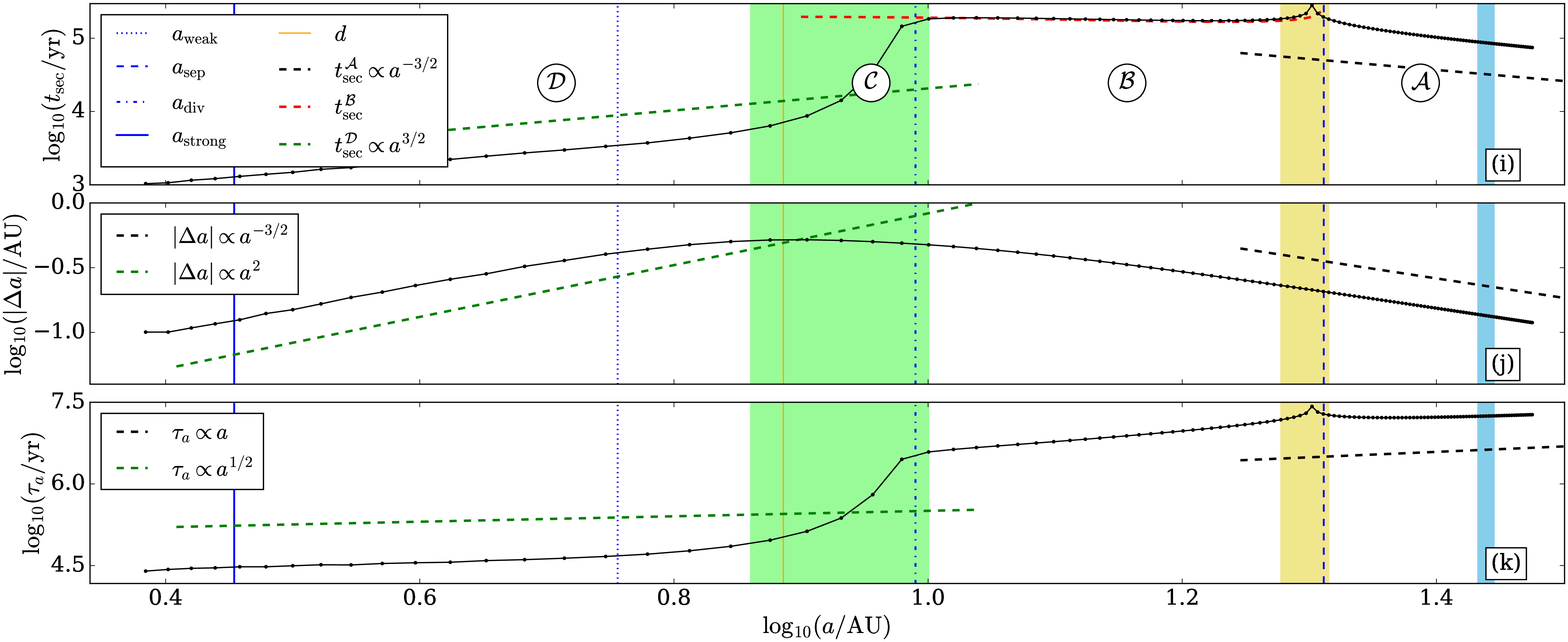}
\caption{\textit{Example 1}. A binary ($m_1=m_2=10M_\odot$) orbits a SMBH with mass $\mathcal{M}=4\times 10^6M_\odot$.
Panels (a)-(e) show the time evolution of $a, e, i, p$ and $\epsGR$
respectively. Panels (f), (g) and (h) show the phase space evolution during the
time intervals shaded in blue, yellow and green respectively in panels (a)-(e).
Finally, panels (i), (j) and (k) show with black dots the values of secular period $\tsec$,
semimajor axis jump $\vert \Delta a\vert$ during the eccentricity peak, and characteristic orbital decay time $\tau_a$ as functions of $a$.}
    \label{fig:Numerical_LK}
\end{figure*}
%%%%%%%%%%%%%%%%%%%%%%%%%%%%%%%%%%%%%%%%%%%%%%%%

Figure \ref{fig:Numerical_LK}
shows the result of
integrating the doubly-averaged (hereafter `DA') test-particle quadrupole
Lidov-Kozai equations of motion for a binary with constituent masses $m_1
= m_2 = 10M_\odot$ orbiting a supermassive black hole (SMBH) of mass $\mathcal{M}=4\times 10^6M_\odot$.
The peri/apocentre of the binary's `outer' orbit around the SMBH, $(\rp,\ra)$, as well as its initial `inner' 
semimajor axis $a$, eccentricity $e$, inclination $i$
and argument of pericentre $\omega$ are given in the text at the top of the
figure\footnote{For the precise definitions of these quantities see \S\ref{sec:dynamical_framework}.}.
%Note that these initial conditions were chosen for pedagogical reasons, as the
%resulting dynamical evolution exhibits a clean separation of asymptotic
%behaviors, allowing us to illustrate clearly several different analytic results
%that we will derive later, all in a single representative
%example.  
%However, in reality they may prove an unrealistic choice, in particular because the
%initial semimajor axis is very large ($a_0 = 250\mathrm{AU}$). }.
%In a real stellar cluster
% one would expect
%such a wide binary to experience at least one hard encounter within the runtime of
%the integration ($\sim 7$ Gyr) --- see \S\ref{PaperII_scattering}. We will ignore
%such caveats here, focusing only on illustrating the physics of the `idealised'
%secular problem considered in this paper.
%We will begin by describing how Figure
%\ref{fig:Numerical_Initially_WeakLib} is organised, before
%moving on to a discussion of its physical content. 
In panels (a)-(d) we plot the time
evolution of $a$,  $e$, $i$ and
pericentre distance $p= a(1-e)$ with black lines.

As expected, the binary undergoes large-amplitude eccentricity oscillations, which drive a decay in semimajor axis and ultimately lead to merger $(a\to 0)$.
%, alongside various
%critical values shown with colored lines which will become important later.
%Panel (e) shows the
%corresponding run of $\epsGR(t)$, which is a dimensionless measure of the strength of general relativistic (GR) apsidal precession (Paper III).
%as well as the two critical GR strengths
%$\epsweak$ and $\epsstr$, defined in equations \eqref{eqn:epsweak} and
%\eqref{eqn:epsstrong} respectively. 
Moreover, the reader will notice three color-shaded vertical stripes in each of these panels blue, yellow and green) defining three representative time intervals of the dynamical evolution. In panels (f)-(h) we plot in black the trajectory of the binary in the $(\omega, e)$ phase space (familiar from Papers II and III) during those
respective segments, starting at the green dot and ending at the red dot. Panel (e) shows $\epsGR$, the dimensionless measure of the strength of GR precession (Paper III).
In panels (i) and (j) we show with black
dots the secular oscillation timescale $\tsec$ and the decay in semimajor axis $\vert
\Delta a\vert$, respectively, as functions of $a$, evaluated at the end of each secular oscillation. 
Panel (k) shows the corresponding orbital decay timescale\footnote{We note that the secular timescale was computed by finding the time elapsed between adjacent eccentricity maxima in panel (b), while $\Delta a$ was computed by calculating the semimajor
axis before ($a_\mathrm{bef}$) and after ($a_\mathrm{aft}$) each peak, and defining $\Delta a(a_\mathrm{bef}) \equiv a_\mathrm{aft}-a_\mathrm{bef}$.
%Note
%that both axes in panels (i) and (j) are logarithmic. 
To draw the colored contours (which are of constant DA Hamiltonian $H^*$, see equation (\ref{eqn:DA_Hamiltonian})) in panels (f)-(h) we set $\Gamma = 1$ and took the values of $\Theta$ and $\epsGR$ at the midpoint of the corresponding colored stripe (see \S\ref{sec:dynamical_framework} for the meaning of these constants). Note also that to not overload the plots we refrain from showing explicit separatrices.} $ \tau_a \equiv \vert \md \ln a/\md t  \vert^{-1}$. The binary evolves from right to left in these last three panels, and the blue, yellow and green shaded segments are again indicated. There are also various critical values and analytical scalings shown with colored lines throughout the Figure, which will be explained in \S\S\ref{sec:GW_emission}-\ref{sec:phase_space_evolution}.

This Figure exhibits several striking features, some of which are rarely mentioned in analyses of LK-driven mergers.  For instance:
\begin{itemize}
    \item Despite the fact that $a$ is changing with time (panel (a)), for most of the evolution there exist two approximately conserved quantities, namely the minimum inclination $\imin$ and the minimum pericentre distance $\pmin$ reached during each secular cycle --- see the red dashed lines in panels (c) and (d).
    \item The timescale for secular oscillations $\tsec$ changes by several orders of magnitude throughout the evolution, and exhibits a highly non-trivial dependence on semimajor axis (panel (i)).
    Viewed as a function of time (right to left in that panel), it first increases with time (up until the yellow shaded stripe), then is almost constant (up to the green stripe), and thereafter \textit{decreases} with time, ultimately becoming much smaller than its initial value. This decrease is counter-intuitive, since naively one might expect a more tightly bound binary to exhibit slower tide-driven secular evolution.
    \item The binary's phase space trajectory evolves as its semimajor axis shrinks. At early times it follows a librating trajectory around the fixed point at $\omega = \pi/2$ (panel (f)),
    then it transitions to a circulating trajectory (panel (g)), which is ultimately pushed to very high minimum eccentricity (panel (h)).
\end{itemize}
Of these three observations, the first has been discussed by e.g. \cite{Wen2003-jf}.  The second was mentioned by \cite{Randall2018-uq}, although their explanation of this phenomenon was incomplete as we show in Appendix \ref{sec:RX18}. The third was briefly signposted by \citet{Blaes2002-rx} and \citet{Antonini2016-mv} --- see \S\ref{sec:phase_space_evolution}.  There is very little literature that goes into quantitative detail about these features and nobody has considered their interplay (for instance, how the changing phase space structure affects the evolution of the secular timescale).

A central purpose of the present paper is to explain such dynamical characteristics. It will turn out that the qualitatively different
behaviors exhibited by $\Delta a(a)$, $\tsec(a)$, $\tau_a(a)$ etc. map onto different GR regimes and phase space features explored in Paper III.

%As in our previous work we highlight the importance of the dynamical bifurcation that separates $\Gamma > 1/5$ from $0 < \Gamma \leq 1/5$, of the distinction between librating from circulating phase space trajectorys, and so on. %We also provide a new formula for the merger time of a binary given its initial conditions, which is more accurate than previous results (at least in the DA approximation --- short-timescale effects will be considered in future work).

%%%%%%%%%%%%%%%%%%%%%%%%%%%%%%%%%%%%%%%%%%%%%%%%

\subsection{Plan for the rest of this paper}

This paper is organised as follows. In
\S\ref{sec:dynamical_framework} we describe briefly the setup of our system without GW emission, and 
establish notation. We also gather
in Appendix \ref{sec:high_ecc_no_GWs} some results from Papers I-III which we will refer to throughout this work.
In \S\ref{sec:GW_emission} we introduce the effect of GW emission, and
derive two approximate conservation laws, namely conservation of the minimum
pericentre distance and the minimum inclination reached by the binary.
In \S\ref{sec:phase_space_evolution} we outline how a binary evolves through phase space and different GR regimes as its semimajor axis shrinks (much more detail is given in Appendix \ref{sec:phase_space_appendix}).
This allows us to establish several asymptotic regimes in which we can make analytic progress.  Using these regimes, in \S\ref{sec:secular_timescale} we write down approximate expressions for the secular timescale $\tsec(a)$, while in \S\ref{sec:SMA_Evolution} we derive expressions for $\Delta a(a)$ and $\tau_a(a)$.  Again, details of the derivations are relegated to Appendices \ref{sec:tsec_type2_appendix}-\ref{sec:Delta_a_appendix}.
%To make these formulae
%simple and explicit we must specify the value of $\Gamma$, work out whether an
%orbit librates or circulates, etc. Thus, 
In \S\ref{sec:Numerical_Examples} we verify our results via several more
numerical examples akin to Figure \ref{fig:Numerical_LK}, this time for binaries in (non-Keplerian) cluster potentials. In Appendix \ref{sec:RX18} we compare our work with the LK calculations of \citet{Randall2018-uq}.
%The more complicated (and, in practice,
%less relevant) regime of $0 < \Gamma \leq 1/5$ is treated in Appendix
%\ref{sec:Gamma_Regime_II}. 
In \S\ref{sec:Discussion} we discuss our results, including their
implications for LK-driven mergers, and summarize in
\S\ref{sec:Summary}.

%We will find that essentially all important quantities can be expressed in terms
%of four dimensionless constants, namely $(\Gamma, \Theta, \epsGR, \jmin)$.  When
%we introduce GW emission we break the constancy of $\Theta, \epsGR, \jmin$, but
%find that they can be expressed very simply in terms of $a$ and two new
%conserved quantities, the minimum pericentre distance $\pmin$ and the minimum
%inclination $\imin$. Finally we gwt $a(t)$.

\section{Dynamical framework} 
\label{sec:dynamical_framework}

Here we briefly describe our setup in order to establish notation.  For more detail see Papers I-III.

Consider a binary with component masses $m_1$, $m_2$, orbiting in a fixed, smooth, axisymmetric background potential $\Phi$ whose symmetry axis is $Z$. Let $(X,Y)$ describe the plane perpendicular to $Z$.
Then on long timescales the orbit of the binary's barycenter (hereafter the `outer' orbit) usually fills an axisymmetric torus. However, if $\Phi$ is spherically symmetric then the outer orbit is actually confined to a plane, which we can choose to be the $(X,Y)$ plane.
Apart from phase information the orbit in this plane can be described by its peri/apocentre  $(r_\mathrm{p}, r_\mathrm{a})$.
In this case, on timescales much longer than an outer orbital period the binary will fill an annulus with inner and outer radii $(r_\mathrm{p}, r_\mathrm{a})$.  In the special case of Keplerian $\Phi$ (the LK limit), the outer orbit describes not an annulus but a fixed ellipse with semimajor axis $a_\mathrm{g} = (\ra+\rp)/2 $ and eccentricity $e_\mathrm{g} = (\ra-\rp)/(\ra+\rp)$.

The binary's internal (`inner') Keplerian orbital motion is described by the usual orbital elements: semi-major axis $a$, eccentricity $e$, inclination $i$ (relative to the $(X,Y)$ plane), longitude of the ascending node $\Omega$ (relative to the $X$ axis, which is arbitrary but fixed in the cluster frame), argument of pericentre $\omega$ and mean anomaly $\eta$.
It is also useful to introduce Delaunay actions $L=\sqrt{G(m_1+m_2) a}, J=L\sqrt{1-e^2}$, and $J_z = J\cos i$, and their conjugate angles $\eta$, $\omega$ and $\Omega$, as well as the dimensionless variables
%%%%%%%%%%%%
\begin{align}  
\label{eqn:def_Theta}
\Theta &\equiv J_z^2/L^2 = (1-e^2)\cos^2 i,
\\
j &\equiv J/L = \sqrt{1-e^2}.
\label{eq:AMdefs}
\end{align} 
%%%%%%%%%%%%
Clearly, $j$ must obey $\Theta^{1/2}\leq j\leq 1$
to be physically meaningful for a given $\Theta$. The minimum/maximum $j$ achieved in a given secular cycle is called $j_\mathrm{min/max}$.

Ignoring GW emission, the evolution of the inner orbit is dictated by the mutual Newtonian gravitational attraction of the binary components, 1pN GR apsidal precession, and the perturbing tidal influence of the potential $\Phi$. Expanding the tidal force due to the cluster to quadrupole order, and averaging over the inner and outer orbital motion (i.e. performing a weighted integral over the torus, annulus or ellipse mentioned above) we find the test-particle quadrupole doubly-averaged (`DA') equations of motion, given explicitly in equations (12)-(14) of Paper III.  More succinctly, these can be derived from the DA Hamiltonian
%%%%%%%%%%%%
\begin{align} 
H = \frac{Aa^2}{8}H^*, \,\,\,\,\,\,\,\,\,\,\, H^* \equiv H_1^* + H_\mathrm{GR}^*,
\label{eqn:DA_Hamiltonian} 
\end{align} 
%%%%%%%%%%%%
where $A$ is a constant (with units of s$^{-2}$) that depends on the potential and outer orbit. In the LK case, $A=G\mathcal{M}/[2
a_\mathrm{g}^3(1-e_\mathrm{g}^2)^{3/2}]$, where $a_\mathrm{g}$ and $e_\mathrm{g}$ are respectively the semimajor axis and eccentricity of the outer orbit and $\mathcal{M}$ is the perturber mass. Next,
%%%%%%%%%%%%
\begin{align}
    &H_1^* = (2+3e^2)(1-3\Gamma\cos^2i)-15\Gamma e^2 \sin^2 i \cos 2\omega, 
    \label{H1star} \\ 
    &H_\mathrm{GR}^* = -\epsilon_\mathrm{GR}(1-e^2)^{-1/2}, 
    \label{HGR}
\end{align}
%%%%%%%%%%%%
are dimensionless Hamiltonians encoding the effects of cluster tides and GR precession respectively. The quantity $\Gamma$ is a scalar parameter which, like $A$, depends on the cluster potential $\Phi$ and the choice of outer orbit. It takes values  $\in (0,1)$ for binaries in spherically symmetric potentials $\Phi$, and LK theory is recovered in the limit $\Gamma=1$.
Due to a dynamical bifurcation, it turns out that very high eccentricities are much more readily achieved by binaries with $\Gamma>1/5$ than those with $\Gamma < 1/5$ (Paper II). Finally, the parameter $\epsGR$ measures the strength of GR precession:
\begin{align} 
\epsilon_\mathrm{GR} &\equiv \frac{24G^2(m_1+m_2)^2}{c^2Aa^4} 
\label{eq:epsGRformula} \\
&=  0.258 \times \left( \frac{A^*}{0.5}\right)^{-1}\left( \frac{\mathcal{M}}{10^5M_\odot}\right)^{-1}\left( \frac{b}{\mathrm{pc}}\right)^{3} \nn \\ & \times \left( \frac{m_1+m_2}{M_\odot}\right)^{2}  \left( \frac{a}{20 \, \mathrm{AU}}\right)^{-4}. 
\label{eqn:epsGRnumerical}
\end{align}
In the numerical estimate
\eqref{eqn:epsGRnumerical} we have assumed a spherical cluster
of mass $\mathcal{M}$ and scale radius $b$, and $A^* \equiv A/(G\mathcal{M}/b^3)$ --- see Paper I\footnote{In the LK limit one should set $A^* = 0.5$ and $b =
a_\mathrm{g}(1-e_\mathrm{g}^2)^{1/2}$.}. The physical effect of GR precession is typically to quench the cluster tide-driven eccentricity oscillations, as we explored in detail in Paper III, and as has been long established in LK theory \citep{Miller2002-co,Fabrycky2007-tu,Bode2014-xh}. As we have shown in Paper III, there are typically no large $e$ oscillations in the `strong GR' regime $\epsGR \gtrsim \epsstr \equiv 3(1+5\Gamma)$. On the other hand one can ignore GR precession in the `weak GR' regime $\epsGR \ll \epsweak$ (see equation \eqref{eqn:epsweak}). In the intermediate regime of `moderate GR', $\epsweak \lesssim \epsGR \ll \epsstr$, high eccentricity excitation is modified but is not prohibited.  

In Papers I-III we have gone into great detail about the phase space dynamics that follows from the Hamiltonian $H$, shown how phase space trajectories are split into librating and circulating families, and so on. We will not repeat the arguments here. However, it will be important throughout that we have expressions for various key quantities characterising secular evolution in terms of $\Gamma$ and the three dimensionless numbers $\Theta,\epsGR,\jmin$, which are constants of motion when GW emission is ignored. It will also be important to split circulating phase space trajectories into two asymptotic regimes, characterised by $\jmax \sim 1$ (`high-$\jmax$') and $\jmax \ll 1$ (`low-$\jmax$'), which is a distinction we did not make in previous papers. We gather all the relevant results in Appendix \ref{sec:high_ecc_no_GWs}, and draw upon them freely hereafter.

%%%%%%%%%%%%%%%%%%%%%%%%%%%%%%%%%%%%%%%%%%%%%%%%
\section{Secular dynamics including gravitational wave emission}
\label{sec:GW_emission}
%%%%%%%%%%%%%%%%%%%%%%%%%%%%%%%%%%%%%%%%%%%%%%%%

Gravitational wave emission modifies the conservative dynamical picture described in \S\ref{sec:dynamical_framework}, allowing compact object binaries to merge. Throughout this paper we concentrate on those binaries whose merger timescale is significantly shortened by secular eccentricity excitation --- the so-called `cluster tide-driven mergers' (which includes LK-driven mergers as a special case). Following \citet{Randall2018-uq} we can separate these binaries further into `fast mergers' and `slow mergers'.  Fast mergers are those that occur after only one (or at most a few) secular eccentricity cycles.
%, and so are essentially fully characterised by their initial $e_\mathrm{max}$. 
%In
%this case the binary is either in the strong GR regime at $t=0$, or it reaches
%the strong GR regime after $\sim 1$ secular cycle.  
Slow mergers occur after many secular eccentricity cycles, and
inevitably involve a gradual transition of the binary from the
weak-to-moderate GR regime to the strong GR regime (with $\emax$ and other important quantities changing over time) --- see Figure \ref{fig:Numerical_LK}. 
%In fact, one can consider
%slow mergers to be a somewhat more general case, since one can imagine any fast merger
%merely as the final few secular cycles of a slow merger.
In this paper
we will focus on understanding the physics of slow mergers.

We will assume throughout this paper that $\Gamma>0$, and that the binary's
maximum eccentricity $e_\mathrm{max}$ is achieved at $\omega=\pm\pi/2$.
As we have seen in Papers I-III these conditions cover almost all cases of practical interest (at least in spherical clusters), and are always satisfied in the important special case of LK dynamics ($\Gamma=1$).

\subsection{Slow mergers}

Consider a binary with initial eccentricity not close to unity,
and suppose that, unless excited to high $e_\mathrm{max} \approx 1$, it will not merge within a Hubble time.  
For the required cluster tide-driven eccentricity excitation to be possible, we know from Paper III 
that the binary must begin its life in the
weak-to-moderate GR regime. Supposing this is the case, and that the binary does
indeed achieve high values of $e$ periodically, then during each high-eccentricity episode its semimajor axis $a$ will
be decreased by some amount $\vert \Delta a \vert$ 
because of GW emission. For a slow merger, by assumption, each individual decrease is small, $\vert \Delta a \vert \ll a$, though of course $\Delta a$ itself depends on the value of $a$ (see Figure \ref{fig:Numerical_LK}j). Away from $e\approx 1$, GW emission will be completely negligible, so
we can treat $a$ as constant there.  Next, the time between these
high-eccentricity episodes is $\tsec(a)$. Therefore on timescales longer than a few
secular periods we can approximate the slow decay of $a(t)$ as
%we can  
%To calculate the long-term evolution of $a$ we need to work out the amount by which 
%the semimajor axis of a binary is decreased during one secular eccentricity cycle, $\Delta a$. Then, denoting by $\langle .. \rangle$ an average over a few secular timescales, we can write
%%%%%%%%%%%%%
\begin{align}
\frac{\md a}{\md t}  \approx  \frac{\Delta a(a)}{t_\mathrm{sec}(a)}.
%\,\,\,\,\, \implies \,\,\,\, \,
%t-t_\mathrm{i} \approx \int_{a_\mathrm{i}}^{a(t)} \md a'\,  \frac{t_\mathrm{sec}(a')}{\Delta a(a')}.
\label{eqn:t_of_a_WTM}
\end{align}
%%%%%%%%%%%%%
%where $a_\mathrm{i} \equiv a(t_\mathrm{i})$ and $t_\mathrm{i} < t$ is some earlier
%reference time. 
Equation \eqref{eqn:t_of_a_WTM} is an implicit equation for $a(t)$ for slow-merging binaries. 
We can use it to define a characteristic orbital decay timescale at a given $a$:
\begin{align}
\tau_a(a) \equiv \bigg\vert \frac{\md \ln a}{\md t} \bigg\vert^{-1} \approx \Big\vert\frac{a \times \tsec(a)}{\Delta a} \Big\vert.  
\label{eqn:decay_timescale}
\end{align}
This is the quantity we plotted in Figure \ref{fig:Numerical_LK}k.

Eventually, $a$ will become small enough that the binary
reaches the strong GR regime and gets `trapped' at high eccentricity. When this happens, equation
\eqref{eqn:t_of_a_WTM} breaks down and we must use a different prescription to follow $a(t)$ accurately all the
way to merger. 

%%%%%%%%%%%%%%%%%%%%%%%%%%%%%%%%%%%%%%%%%%%%%%%%

%To take just
%one example, in LK literature one is frequently confronted with the
%counter-intuitive result that as a binary's semimajor axis $a$ shrinks, the
%timescale $t_\mathrm{sec}$ for the next secular eccentricity oscillation
%\textit{decreases} (\citealt{Randall2018-uq}). This trend is rarely mentioned and
%until now has not been fully explained. %Nor has anyone written down an
%explicit formula for the long-term decay in semimajor axis $a(t)$ that results
%from repeated bursts of GW energy at the peak of each secular cycle. 

\subsection{Equations of motion}

The main goals of \S\ref{sec:phase_space_evolution}  will be to understand the behavior of $\Delta a(a)$, $\tsec(a)$ and $\tau_a(a)$ during a slow merger, and to appreciate how this behavior is intimately linked with the binary's phase space trajectory (librating or circulating) and the strength of GR precession (value of $\epsGR$).
To achieve these goals we must first consider how GW emission affects our equations of motion. 

Our DA theory without GW emission consists of equations (12)-(14) of Paper III, which govern the evolution of $\omega$, $j$ and $\Omega$ respectively under the combined effect of secular cluster tides and GR precession. In addition to this, to 2.5th post-Newtonian order GW emission causes the binary's semimajor axis and eccentricity to evolve according to the \citet{Peters1964-fb} equations\footnote{Of course, in principle one can carry the
post-Newtonian expansion to higher order, and higher-order terms are important for e.g. LIGO/Virgo templates of inspiralling
binary waveforms.  However, since this is only important at very late times when
compared to the long secular evolution that we are considering here,
 we always truncate at 2.5pN.}:
%%%%%%%%%%
\begin{align}
\left(\frac{\md a}{\md t} \right)_\mathrm{GW} =& -\frac{64G^3m_1m_2(m_1+m_2)}{5c^5} \nn \\ &\times \frac{1}{a^3(1-e^2)^{7/2}}   \left( 1+\frac{73}{24}e^2 + \frac{37}{96}e^4 \right),
\label{eqn:dadt_GW}
\\
\left(\frac{\md e}{\md t} \right)_\mathrm{GW} =& -\frac{304G^3m_1m_2(m_1+m_2)}{15c^5} \nn \\ &\times \frac{1}{a^4(1-e^2)^{5/2}}   \left( 1+\frac{121}{304}e^2\right).
\label{eqn:dedt_GW}
\end{align}
%%%%%%%%%%
To include GW emission in our theory we therefore add the following terms to our equations of motion for $L$, $j$ and $j_z \equiv j\cos i$ respectively:
%%%%%%%%%%
\begin{align}
\label{eqn:dLdt_GW}
&\left(\frac{\md L}{\md t} \right)_\mathrm{GW} = \frac{1}{2}\sqrt{\frac{G(m_1+m_2)}{a}} \left(\frac{\md a}{\md t} \right)_\mathrm{GW},
\\
\label{eqn:dJdt_GW}
 &\left(\frac{\md j}{\md t} \right)_\mathrm{GW} = -\frac{e}{j} \left(\frac{\md e}{\md t} \right)_\mathrm{GW},
    \\
    \label{eqn:dJzdt_GW}
  &\left(\frac{\md j_z}{\md t} \right)_\mathrm{GW} = \frac{j_z}{j}
  \left(\frac{\md j}{\md t} \right)_\mathrm{GW}.
\end{align}
%%%%%%%%%%
We see that GW emission affects both $L$ and $j_z$ (and hence also $\Theta \equiv j_z^2$), which were constants when GW emission was ignored.
Note also that at this order, GW emission does not directly affect the argument of pericentre $\omega$ or
longitude of ascending node $\Omega$ (so equations (12) and (14) of Paper III are unchanged), nor does it affect inclination $i$. 

%%%%%%%%%%%%%%%%%%%%%%%%%%%%%%%%
\subsection{Conserved quantities}
\label{sec:conservation_laws}
%%%%%%%%%%%%%%%%%%%%%%%%%%%%%%%%

When GW emission is switched off, the DA dynamics respects three exact
conservation laws. The first is the conservation of $a$, which results from the `adiabatic' assumption that the binary's inner orbital period
is much shorter than timescale of variation of the weak cluster
perturbation (i.e. the outer orbital timescale), allowing us to average over the inner orbit (`single averaging').  The second is the conservation of the dimensionless Hamiltonian $H^*$ (equation \eqref{eqn:DA_Hamiltonian}), which follows from the fact that cluster tides are sufficiently weak that we can also average over the outer orbital period and thus treat the perturbation as time-independent (`double averaging'). The third is conservation of the $z$ component of the binary's angular momentum, or equivalently $\Theta$, which follows from the axisymmetry of the DA time-averaged perturbation as viewed from the binary frame.

Now that we are including GW emission, the binary's binding energy and inner
orbital angular momentum will be dissipated according to equations
\eqref{eqn:dadt_GW}-\eqref{eqn:dJzdt_GW}, and so none of
$a, H^*$ or $\Theta$ will be strictly conserved. On the other hand,
it is clear from equations \eqref{eqn:dadt_GW} and
\eqref{eqn:dedt_GW} that for a fixed semimajor axis $a$, changes in orbital elements due to GW emission
are very strongly concentrated around peak eccentricity $e \to 1$, as we already
anticipated. Thus, in the weak-to-moderate GR regime
we expect these GW contributions to the equations of motion to be completely
negligible except in the vicinity of $e\approx 1$.
This means that binding energy and angular momentum
\textit{can} be treated as roughly conserved away from eccentricity peaks, since in the weak-to-moderate GR regime GW emission is negligible except for short
bursts around $e\approx e_\mathrm{max}$. 

Additionally,  as we saw in Figure \ref{fig:Numerical_LK}, two new
(approximate) conservation laws emerge which are valid on much longer timescales ($\gg \tsec$) --- these are the conservation of the minimum pericentre distance $\pmin \equiv a(1-\emax)$, and the conservation of the minimum inclination
$i_\mathrm{min}$.  The fact that these conservation laws hold almost all the way to merger will facilitate our analytical understanding. We now derive each of them in turn.

\subsubsection{Conservation of $\pmin$}
%%%%%%%%%%%%%%%%%%%%%%%%%%%%%%%%%%%%%%%

Approximate conservation of $\pmin$ follows from the fact that, as $e\to 1$, GW emission dissipates orbital energy of the binary much more efficiently than its angular momentum\footnote{The following argument is not unique to the GW emission and can be generalized for any short-range dissipative force, e.g. due to the fluid tides acting inside the binary components.}. As mentioned above, because of the steep dependence on $1-e^2$ in \eqref{eqn:dadt_GW}-\eqref{eqn:dedt_GW}, GW emission is most effective at changing $a$, $e$ (and $p$) only at the eccentricity `peak',
when $e\to 1$ (which is the case only during a small fraction $\sim \jmin \ll 1$ of each secular period), and can be neglected during the rest of the secular cycle. Thus, during each eccentricity peak GW emission causes changes of the binary orbital elements predominantly over the short time intervals $\delta t \ll T_\mathrm{b}$ during the periastron passages (where $T_\mathrm{b} = 2\pi \sqrt{a^3/[G(m_1+m_2)]}$ is the inner orbital period of the binary). 

Let the characteristic relative velocity of the binary components at periastron be $v_p\sim \sqrt{G(m_1+m_2)/p}$, which follows from energy conservation and the fact that $p=a(1-e)\ll a$. Representing the effect of the GW radiation reaction as an impulsive force $F$ acting over time $\delta t$, we can estimate the change in the binary orbital energy $E$ over each periastron passage to be $\delta E \sim Fv_p \,\delta t$. Thus, the characteristic timescale $t_E$ on which $E$ evolves is
%%%%%%%%%%
\begin{align}
    t_E \sim \frac{E}{\delta E} T_\mathrm{b} \sim \frac{G(m_1+m_2)}{aFv_p \, \delta t} T_\mathrm{b}.
\end{align}
%%%%%%%%%%
Similarly, during each periastron passage GW emission changes the binary angular momentum $J\sim pv_p$ by $\delta J \sim F p \, \delta t$, so that the characteristic time on which $J$ evolves is
%%%%%%%%%%
\begin{align}
    t_J \sim \frac{J}{\delta J} T_\mathrm{b} \sim \frac{v_p}{F\, \delta t}T_\mathrm{b}.
\end{align}
%%%%%%%%%%
The ratio of these two timescales is 
%%%%%%%%%%
\begin{align}
    \frac{t_E}{t_J} \sim \frac{G(m_1+m_2)}{a v_p^2} \sim \frac{p}{a} = 1-e.
    \label{eq:t_rat}
\end{align}
%%%%%%%%%%
Thus for $e\approx e_\mathrm{max}\to 1$, we have $t_E/t_J \sim 1-\emax \ll 1$.
In other words, when the binary is near peak eccentricity, its energy is dissipated much more rapidly than its angular momentum, so that one can assume that $J \approx J_\mathrm{min}$ is almost constant even though $E$ (and $a$) evolves substantially. Since  $J_\mathrm{min} \propto [a(1 -
e_\mathrm{max}^2)]^{1/2} \approx (2p_\mathrm{min})^{1/2}$ for high $e_\mathrm{max}$, this implies that the minimum periastron distance $\pmin$ does not change appreciably as a result of GW emission over a single eccentricity peak (\citealt{Wen2003-jf}, \S 3.1). And since the system undergoes quasi-periodic secular oscillations, the binary returns to the same value of
$J_\mathrm{min}$ (and hence the same $\pmin$) at the $e$-peak of the following
secular cycle. 

%%%  STOPPED HERE

To be more precise, let us consider the rate of change of the pericentre distance with respect to semimajor axis:
\begin{align}
p \equiv a(1-e) \,\,\,\,\,\, \implies \,\,\,\,\,\,  \frac{\md p}{\md a} = 1- e -a \frac{\md e}{\md a}.
\label{eqn:dpda_intermadiate}
\end{align}
Dividing \eqref{eqn:dedt_GW} by \eqref{eqn:dadt_GW} to get $\md e/\md a$, plugging this in to the right hand side of \eqref{eqn:dpda_intermadiate} and
expanding near $e = 1$ gives
%%%%%%%%%%
\begin{align}
\frac{\md p}{\md a}
= (1-e)^2 + \mathcal{O}\left( (1-e)^3\right).
\label{eqn:dpda_expansion}
\end{align}
%%%%%%%%%%
In other words, the rate of change of pericentre distance $p$ vanishes in the limit $e\to 1$, so that $p=\pmin$ is constant at the eccentricity peak. Since each individual secular cycle is symmetric (in time, relative to its eccentricity minimum), $\pmin$ would then take the same value at the next eccentricity peak, be preserved there, and so on, just as we observed in Figure \ref{fig:Numerical_LK}d. Thus we arrive at the conservation of the
\textit{minimum pericentre distance}: 
%%%%%%%%%%
\begin{align}
p_\mathrm{min} \equiv a(1-e_\mathrm{max}) \approx \frac{1}{2}a\jmin^2 = \mathrm{const},
\label{eqn:pmin}
\end{align}
%%%%%%%%%%
which holds true over multiple secular cycles as long as $e_\mathrm{max}\to 1$.

Equation \eqref{eqn:pmin} implies
a simple scaling for $\jmin$ in the weak-to-moderate GR regime:
%%%%%%%%%%
\begin{align}
j_\mathrm{min}(a) \approx \left(\frac{2\pmin}{a}\right)^{1/2}, 
\label{eqn:jmin_of_a}
\end{align}
%%%%%%%%%%
where $\pmin = a(t_\mathrm{i}) \times (1-e(t_\mathrm{i}))$ at some reference time $t_\mathrm{i}$. This is an important result of this paper and will subsequently allow significant analytical simplification.

We note that an argument similar to the one above also applies when the binary gets
trapped at high eccentricity in the strong GR regime. In that case there are no
more secular oscillations (the binary having decoupled from cluster tides), but
energy is still being dissipated efficiently by GW emission while angular momentum is not (for more details see \cite{Wen2003-jf} and \cite{Antognini2014-jj}). As a result $p$ (rather than just $\pmin$) stays approximately constant, so that $j \propto a^{-1/2}$. We will use
this scaling when considering the strong GR regime in
\S\ref{sec:SMA_Evolution}.

\subsubsection{Conservation of $\imin$}
\label{eqn:inclination_conservation}
%%%%%%%%%%%%%%%%%%%%%%%%%%%%%%%%%%%%%%%

The conservation of $\imin$ --- the minimum value of the binary inclination which is reached at the eccentricity peak --- follows from the fact that the GW emission does not affect the orientation of the orbital plane of the binary and thus does not affect its inclination. Again, because of the time-symmetry of each secular cycle, at the next eccentricity peak the binary will arrive at the same value of $\imin$ (which, again, will not be changed by the GW emission, regardless of the decay in $a$), and so on. As a result, it follows that over time intervals much longer than each secular cycle
%%%%%%%%%%%%
\begin{align}
\imin = \mathrm{const}.
\label{eqn:imin_const}
\end{align}
%%%%%%%%%%%%
This is precisely what we saw in Figure \ref{fig:Numerical_LK}c.

For the remainder of this paper we will take $\pmin$ and $\imin$ as our two primary, $a-$independent constants of motion which persist over multiple secular cycles in the weak-to-moderate GR regime. We can then re-write the key secular evolution parameter $\Theta$ in terms only of $a$ and these conserved quantities as
\begin{align}
\label{eqn:Theta_of_a}
\Theta(a) \approx \frac{2\pmin}{a} \cos^2\imin,
\end{align}
which will greatly simplify our analytical understanding. We will discuss the circumstances in which the conservation of $\pmin$ and $\imin$ breaks down, invalidating equation (\ref{eqn:Theta_of_a}), in our detailed discussion of a numerical example in \S\ref{sec:Numerical_Example_3}.

%%%%%%%%%%%%%%%%%%%%%%%%%%%%%%%%%%%%%%%
\section{Evolution of a shrinking binary through time and phase space}
\label{sec:phase_space_evolution}
%%%%%%%%%%%%%%%%%%%%%%%%%%%%%%%%%%%%%%%

The binary in Figure \ref{fig:Numerical_LK}
started its life in the weak GR regime (see panel (e)).  Then as its semimajor axis shrank it entered the moderate GR regime and finally ended up in the strong GR regime before merging.
On a related note, we also saw that the binary's phase space trajectory evolved from librating (panel (f)), to circulating with $\emin \ll 1$ (panel (g); we call this a `high-$\jmax$' circulating trajectory), to circulating with $\emin \approx 1$ (panel (h); we call this a `low-$\jmax$' circulating trajectory).
This pattern of behavior is rather general, and in fact has been discussed briefly in the case of LK-driven mergers by \cite{Blaes2002-rx} and \cite{Antonini2016-mv}. %To quote directly from \cite{Blaes2002-rx}: 
%`gravitational radiation drives an initially librating inner
%binary over into a circulating inner binary, thereby causing the minimum eccentricity to drop until it crosses the separatrix. [...] Thereafter, the evolution is very similar to the case in which the inner binary starts off circulating: the minimum eccentricity rises until gravitational radiation becomes so strong that it starts to circularize the orbit'.

Physically, this evolution of phase space trajectory follows from
the way GR precession modifies the phase space structure, by encouraging rapid pericentre precession at high $e$ and hence expanding the region of circulating trajectories at the expense of librating trajectories. Hence, one can roughly think of increasing $\epsGR$ as pushing the separatrix `down' to lower eccentricity in the $(\omega, e)$ plane.
Since the increases in $\epsGR$ occur only when the binary is at $e \approx \emax$ (because this is where GW emission, and hence the decay of $a$, is concentrated), these downward shifts of the separatrix coincide with the binary's highest eccentricity.
In this way a binary on a librating trajectory inevitably moves `towards' the separatrix from below (or rather the separatrix moves closer to it from above) and ultimately ends up crossing the separatrix onto a circulating trajectory.  Eventually the binary gets trapped on a low-$\jmax$ circulating trajectory in the strong GR regime, where asymptotically there are no eccentricity oscillations at all.

In this section we wish to understand more quantitatively how the binary's phase space trajectory and GR regime evolves as a function of semimajor axis $a$.
The details of the various transitions between these regimes are quite technical, so we relegate most of our discussion to Appendix \ref{sec:phase_space_appendix} and retain here only the salient points.  

\subsection{Characteristic lengthscales}
\label{sec:lengthscales}

As shown in Appendix \ref{sec:phase_space_appendix}, the orbital evolution of a binary en route to merger admits a key lengthscale
%%%%%%%%%%%%%
\begin{align}
\label{eqn:d_def}
    d &\equiv  \left(\frac{4G^2(m_1+m_2)^2}{5c^2A\Gamma(2\pmin)^{1/2}}\right)^{2/7}
    \\
   &\approx 7.1\, \mathrm{AU} \times \Gamma^{-2/7}\left( \frac{A^*}{0.5}\right)^{-2/7}\left( \frac{\mathcal{M}}{10^6M_\odot}\right)^{-2/7} \nn \\ &\times \left( \frac{b}{\mathrm{pc}}\right)^{6/7}\left( \frac{m_1+m_2}{M_\odot}\right)^{4/7}
\left( \frac{\pmin }{10^{-2}\mathrm{AU}} \right)^{-1/7} ,
\label{eqn:lengthscale_numerical}
\end{align} 
%%%%%%%%%%%%%
which is independent of $a$.
There are then four critical semimajor axis values to contend with.

%All the key quantities in \S\ref{sec:Framework} were written in terms
%f $\Gamma$, $\epsGR$, $\Theta$ and $\jmin$. For a given outer orbit, $\Gamma$
%is constant. And from equations \eqref{eq:epsGRformula} and
%\eqref{eqn:jmin_of_a} we know how $\epsGR$ and $\jmin$ scale with semimiajor
%axis $a$.  All that is left is to work out how $\Theta$ depends on $a$.

A binary that starts its life in the weak GR regime will inevitably move into
the moderate GR regime at some point as its semimajor axis shrinks. Thus the
first critical value is $a_\mathrm{weak}$, which we define to be the semimajor
axis corresponding to $\epsGR = \epsweak$, i.e. to the transition between the
weak and moderate GR regimes. Using \eqref{eqn:gamma_of_a} and the fact that $\epsweak \ll \epsGR$, 
this is 
%which gives 
%$\zeta= (\sqrt{2}-1)/2$, i.e. it happens
%at a critical semimajor axis
%%%%%%%%%%%%%
\begin{align}
a_\mathrm{weak} \equiv
\left( \frac{\sqrt{2}-1}{2\cos^2 \imin} \right)^{2/7} d
\approx 0.63 (\cos\imin)^{-4/7}d.
\label{eqn:a_weak}
\end{align}
%%%%%%%%%%%%%
Note that $a_\mathrm{weak} \sim d$, unless $\cos \imin \ll 1$.

%, allowing us to simplify the preceding
%expressions for $j_\pm^2, \gamma$, etc. in the particular asymptotic regimes of
%weak ($(a/d)^{7/2} \gg 1$) or moderate ($(a/d)^{7/2} \ll 1$) GR. 

Next we define $a_\mathrm{strong}$, which demarcates the inevitable
transition between moderate and strong GR regimes, i.e. it corresponds to
$\epsGR = \epsstr \equiv 3(1+5\Gamma)$. Using \eqref{eq:epsGRformula}
we get
%%%%%%%%%%%%%
\begin{align}
a_\mathrm{strong} &\equiv \left( \frac{8G^2 (m_1+m_2)^2}{c^2A(1+5\Gamma)}\right)^{1/4}
\label{eqn:a_strong}
\\
&\approx 
6.1 \mathrm{AU} \times \left( \frac{1+5\Gamma}{6}\right)^{-1/4}
\left( \frac{A^*}{0.5} \right)^{-1/4} \nn
\\
&\left( \frac{\mathcal{M}}{10^6 M_\odot} \right)^{-1/4}
\left( \frac{b}{\mathrm{pc}} \right)^{3/4}
\left( \frac{m_1+m_2}{M_\odot} \right)^{1/2}.
\label{eqn:a_strong_numerical}
\end{align}
%%%%%%%%%%%%%
After entering the strong GR regime the binary
gets trapped at high eccentricity, and its semimajor axis decays while
$p=a(1-e)$ remains roughly constant. See \S\ref{sec:a_of_t_strong} for more details.

A binary initially on a librating phase space trajectory
will transition to a circulating trajectory once $a$ drops below some threshold
value, as we saw in Figure \ref{fig:Numerical_LK}. This threshold value is 
$a_\mathrm{sep}$, which is given by\footnote{To derive this we set $j_0^2=j_+^2=1$ in equations \eqref{eqn:jplus_of_a},
\eqref{eqn:j0_of_a} --- see Papers II-III.}
%%%%%%%%%%%%%
\begin{align}
\label{eqn:a_sep}
a_\mathrm{sep} \equiv  \left(\frac{1+5\Gamma}{10\Gamma} - \cos^2 \imin\right)^{-2/7}d.
\end{align}
%%%%%%%%%%%%%
Typically $a_\mathrm{sep} \gtrsim d$.
Of course, $a_\mathrm{sep}$ has physical meaning only if $\cos^2 \imin
< (1+5\Gamma)/10\Gamma$. This is because for $\cos^2\imin >
(1+5\Gamma)/10\Gamma$, the binary is already on a circulating trajectory even for
$a\gg d$, so it never crosses a separatrix on its way to $a \to 0$.
%This interpretation also holds when we include GR as long as we can ignore the
%explicit GR terms in \eqref{eq:djdtGR}, i.e. as long as the secular
%timescale is given by \eqref{eqn:t_sec_weak}).  
%Thus it works for all
%but low-$\jmax$ circulating trajectories (\S\ref{}). The latter caveat is not a problem
%since low-$\jmax$ circulation is the final stage that a binary goes through before
%moving into the strong GR regime. Other than this, $a> a_\mathrm{sep}$ is
%circulating while $a<a_\mathrm{sep}$ is for librating. We also note that
%although the right hand side of 
%\eqref{eqn:a_sep}
%is well defined for $0 < \Gamma \leq 1/5$, the physical interpretation is not
%the same --- in that case the separatrix for $\epsGR=0$ corresponds to $j_0^2 =
%\Theta$.

There is one further critical semimajor axis value, which we call
%%%%%%%%%%%%%
\begin{align}
a_\mathrm{div} \equiv (\sin\imin)^{-4/7}d.
\label{eqn:a_div}
\end{align}
%%%%%%%%%%%%%
At $a=a_\mathrm{div}$ the dimensionless numbers $\sigma$ and $\kappa$, which play a role in setting time spent at highest eccentricity (see Paper III), diverge.  This divergence will become important when we discuss the evolution of the secular timescale (\S\ref{sec:secular_timescale}). Moreover, if $a_\mathrm{sep}> a_\mathrm{div}$, then $a=a_\mathrm{div}$ corresponds approximately to the transition between high-$\jmax$ ($a_\mathrm{div} < a < a_\mathrm{sep}$) and low-$\jmax$ ($a<a_\mathrm{div}$) circulating trajectories --- see Appendix \ref{sec:phase_space_appendix} for details. For future reference we write down the ratio:
\begin{align}
    \frac{a_\mathrm{sep}}{a_\mathrm{div}} = \left(1+ \frac{5\Gamma-1}{1+5\Gamma-10\Gamma\cos^2 \imin} \right)^{2/7}.
    \label{eqn:asep_over_adiv}
\end{align}

To summarize, we have defined four critical semimajor axis values
$a_\mathrm{weak}, a_\mathrm{strong}, a_\mathrm{sep}, a_\mathrm{div}$. The weak
GR regime corresponds to $a > a_\mathrm{weak}$, while the moderate GR regime
corresponds to $a_\mathrm{strong} < a < a_\mathrm{weak}$. We emphasize that we
have purposely written e.g. $a > a_\mathrm{weak}$ rather than $a\gg
a_\mathrm{weak}$ here: it turns out that different dynamical regimes are not very
well separated in semimajor axes (in fact we usually have $a_\mathrm{weak} \sim a_\mathrm{sep} \sim
a_\mathrm{div} \sim d$), despite being well
separated in $\epsGR$ --- see e.g. Figure \ref{fig:Numerical_LK}. 
%In other words, for instance, a binary can be safely in
%%the moderate GR regime $\epsweak \ll \epsGR \ll \epsstr$, and yet in realistic
%cases there is almost never a sufficient semimajor axis separation such that one
%can find $a_\mathrm{weak} \ll a \ll a_\mathrm{strong}$.
In Appendix \ref{sec:phase_space_appendix} we show in more detail how the binary passes 
through these different regimes as $a$ shrinks for several different values of $\Gamma > 0$.
The results are quite complex:
transitions between different regimes do not
always happen in the same order, and the $\Gamma > 1/5$ case has to be considered separately from $0<\Gamma\leq 1/5$. 
%%%%%%%%%%%%%
%\begin{enumerate}
%item The asymptotic regime of weak GR, $a\gg a_\mathrm{weak}$, in which
%the orbit can be librating or (high-$\jmax$) circulating but we always %have $\jmin \approx
%j_- \sim \cos \imin$ and $\vert j_+^2 \vert, \vert j_0^2\vert \sim 1$.
%Moreover, in this weak GR regime both $\vert j_+^2 \vert$ and $\vert j_0^2\vert$
%scale weakly with $a$, as does $\kappa$. 
%\item The moderate GR regime in which crucially \textit{the binary is assumed
%to be on a low-$\jmax$ circulating trajectory} with $\jmin \approx \gamma j_- %\gg j_-$, $\jmax =
%-\sigma j_-$, and we assume that $a$ is sufficiently small that
%$\vert j_+^2 \vert, \vert j_0^2\vert \sim (a/d)^{7/2} \gg 1$, and
%$\vert \kappa \vert$ is roughly constant and not large compared to unity. 
%\item The strong GR regime, $a < a_\mathrm{strong}$, in which the binary
%gets trapped at high eccentricity, and its semimajor axis decays while
%$p=a(1-e)$ is kept roughly constant.
%\end{enumerate}
%%%%%%%%%%%%%
%Again, we emphasize that in general the separation between these three specific
%regimes is far from clean, and 
Nevertheless, the asymptotic regimes defined here will allow us to make analytical progress, and will give us a qualitative
understanding of the behavior of $\tsec$, $\Delta a$ and $\tau_a$ throughout slow
mergers, which is what we turn to next.

%%%%%%%%%%%%%%%%%%%%%%%%%%%%%%%%%%%%%%%%%%%%%%%%%%%%%%%%%%%%%%%%%%%%%%%%%%%%%%%%%%%%%%%%%%%%%%%%%%%%%%%%%%%%%%%%%%%%%%%%%%%%%%%%%%%%%%%%%%%%%%%%
%%%%%%%%%%%%%%%%%%%%%%%%%%%%%%%%%%%%%%%%%%%%%%%%%%%%%%%%%%%%%%%%%%%%%%%%%%%%%%%%%%%%%%%%%%%%%%%%%%%%%%%%%%%%%%%%%%%%%%%%%%%%%%%%%%%%%%%%%%%%%%%%
\subsection{The secular timescale}
\label{sec:secular_timescale}
%%%%%%%%%%%%%%%%%%%%%%%%%%%%%%%%%%%%%%%%%%%%%%%%%%%%%%%%%%%%%%%%%%%%%%%%%%%%%%%%%%%%%%%%%%%%%%%%%%%%%%%%%%%%%%%%%%%%%%%%%%%%%%%%%%%%%%%%%%%%%%%%
%%%%%%%%%%%%%%%%%%%%%%%%%%%%%%%%%%%%%%%%%%%%%%%%%%%%%%%%%%%%%%%%%%%%%%%%%%%%%%%%%%%%%%%%%%%%%%%%%%%%%%%%%%%%%%%%%%%%%%%%%%%%%%%%%%%%%%%%%%%%%%%%

One crucial quantity in any study of LK-driven or cluster tide-driven mergers is the period of
secular eccentricity oscillations, $\tsec$, since this gives the time elapsed
between each episode of GW emission. 
In Figure \ref{fig:Numerical_LK}i we plotted $\tsec$ as a function of semimajor axis $a$ for a binary undergoing LK oscillations as it shrank and ultimately merged (time runs from right to left in that panel).
We labelled four regimes of the $\tsec(a)$ curve, \calA, \calB, \calC, \calD, and each regime exhibited a different $a$-dependence.
In this section we will explain the behavior in each of these regimes in turn.

In DA theory without GW emission, the definition of $\tsec$ is
%%%%%%%%%%%%%%%%%%%%%%%%
\begin{align}
\label{eqn:tsec_exact}
\tsec  = 2\int_{\jmin}^{\jmax} \left( \frac{\md j}{\md t}\right)^{-1} \md j,
\end{align}
%%%%%%%%%%%%%%%%%%%%%%%%
with $\md j/\md t$ given in \eqref{eq:djdtGR}. The right hand side of
\eqref{eqn:tsec_exact} of course depends on $a$, which decays throughout a slow merger.% Here
%we want to derive a tractable approximation to $\tsec(a)$ valid in the regimes \calA-\calD.

\subsubsection{Regime $\mathcal{A}$: Librating trajectories, $a> a_\mathrm{sep}$}
%%%%%%%%%%%%%%%%%%%%%%%%%%%%%%%%%%%%%%%%%%%%%
%%%%%%%%%%%%%%%%%%%%%%%%%%%%%%%%%%%%%%%%%%%%%

Regime $\mathcal{A}$ in Figure \ref{fig:Numerical_LK}i corresponds approximately to $a> a_\mathrm{sep}$, i.e. to librating phase space trajectories (see Figure \ref{fig:Numerical_LK}f and \S\ref{sec:lengthscales}).
Binaries on librating phase space trajectories spend most of their time far away from $e\approx 1$ (see Figure \ref{fig:Numerical_LK}b).
In the weak-to-moderate GR regime this means that the explicitly $\epsGR$-dependent terms in \eqref{eq:djdtGR} 
are unimportant for most of the evolution, and so in this regime a good approximation to the secular period
is found by evaluating \eqref{eqn:tsec_exact} ignoring the explicitly $\epsGR$-dependent terms.
Technically speaking, such an approximation becomes exact in the asymptotic limit $\epsGR \ll \epsweak$ and $\sigma \to 0$ (equation \eqref{eqn:sigma_of_a}) --- see Paper III.

\begin{figure}
\centering
\includegraphics[width=0.85\linewidth,clip]{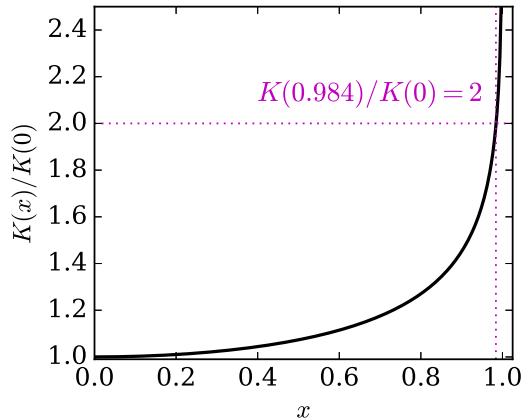}
\caption{Value of the elliptic integral $K(x)$ normalised by $K(0) \equiv \pi/2$ as a function of $x$.  The scaling is weak except for $x \gtrsim 0.99$.}
\label{fig:K_of_x}
\end{figure}

%In fact, we know from Paper III that in the case of weak GR we may ignore the explicit $\epsGR$ terms in
%\eqref{eq:djdtGR} thorughout the entire evolution
%simply act to reduce the time spent in the high-eccentricity state (Paper III),
%which is already a fraction $\sim \jmin \ll 1$ of
%the total secular period. Technically speaking we can ignore GR in this regime if 
%as long as $\sigma \ll 1$ (equation \eqref{eqn:sigma_of_a}).
%As a
%result, a good approximation to the secular period for most librating trajectories in the
%weak-to-moderate GR regime is found by evaluating \eqref{eqn:tsec_exact} and ignoring the explicitly $\epsGR$-dependent terms.
%in \eqref{eq:djdtGR}. 
Moreover, for librating trajectories we know that $\jmax \approx j_+ \sim 1$ (equation \eqref{eqn:jmax_librating}).
Defining $ \Delta \equiv
\mathrm{max}[j_+^2,j_-^2,j_0^2] - \mathrm{min}[j_+^2,j_-^2,j_0^2], $ and
assuming $\jmin^2 \ll \jmax^2$, we find (see eqations (32)-(34) of Paper II):
%%%%%%%%%%%%%
\begin{align} 
\label{eqn:t_sec_weak} 
t_\mathrm{sec}^\mathcal{A} \approx \frac{8}{3A}\sqrt{\frac{G(m_1+m_2)}{\vert 25\Gamma^2-1\vert}} \times \frac{1}{a^{3/2}\sqrt{\Delta}} K\left( \frac{j_+}{\sqrt{\Delta}}\right),
\end{align}
%%%%%%%%%%%%%%
where $K(x)\equiv \int_0^{\pi/2} \md \alpha /\sqrt{1-x^2\sin^2\alpha}$ is plotted in Figure \ref{fig:K_of_x}. We emphasize
that GR is still implicitly present in equation
\eqref{eqn:t_sec_weak} because it affects the values of $j_+$ and $\Delta$ that must be plugged into the right hand side. 
%%%%%%%%%%%%%
%%%%%%%%%%%%%

We cannot simplify the expression \eqref{eqn:t_sec_weak} further without specifying the ordering of $j_0^2$, $j_\pm^2$, which itself depends on $\Gamma$.  However, far from the separatrix between librating and circulating trajectories (i.e. $a$ sufficiently greater than $a_\mathrm{sep}$) we expect $j_+$, $\Delta$ to depend only
weakly on $a$ (Figures
\ref{fig:jPlus_of_a_Gamma_Regime_I}-\ref{fig:jPlus_of_a_Gamma_Regime_II}), so that
%%%%%%%%%%%%%%%
\begin{align}
t_\mathrm{sec}^\mathcal{A} \propto a^{-3/2}.
\label{eqn:tsec_scaling_A}
\end{align}
%%%%%%%%%%%%%%%
This scaling is confirmed in regime $\mathcal{A}$  of Figure \ref{fig:Numerical_LK}i with a black dashed line.
Physically it arises because the binary's inner orbital period is
proportional to $a^{3/2}$. It has been noted by many authors when estimating a
LK-driven merger timescale (e.g. \citealt{Wen2003-jf,Thompson2011-wv,Randall2018-uq}), although they did not tie it to the librating nature of the phase space trajectory.

When the binary's phase space trajectory gets close to the separatrix $a \to a_\mathrm{sep}$ we know (Appendix \ref{sec:high_ecc_no_GWs})
that $j_+^2, j_0^2 \to 1$ and so $\jmax/\sqrt{\Delta} \to 1$.
Figure \ref{fig:K_of_x} shows that $K(x)$ diverges for $x
\to 1$, so $\tsec$ should peak around this point. This again matches what we see\footnote{We should note that the peak in $\tsec$ is not centred
precisely on $a_\mathrm{sep}$.  This is because the expression
\eqref{eqn:a_sep} is only approximate,
derived in a particular high eccentricity limit and assuming exact conservation
of $\pmin$, $\imin$.}
in regime $\mathcal{A}$ of Figure \ref{fig:Numerical_LK}i.

%To evaluate \eqref{eqn:t_sec_weak} explicitly we need to know
%$\jmax$ and $\Delta$, which in turn requires knowledge of the $\Gamma$ regime,
%whether the orbit librates or circulates, etc. --- see Table
%\ref{PaperII_table_Ranges}. 

\subsubsection{Regime $\mathcal{B}$: Circulating trajectories, $a_\mathrm{div} < a < a_\mathrm{sep}$}

After crossing the separatrix to $a< a_\mathrm{sep}$, the binary ends up on a circulating phase space trajectory.  One can still use the results for $t_\mathrm{sec}$ obtained in Paper II (equations (32)-(34)), but for circulating trajectories the scaling \eqref{eqn:tsec_scaling_A} we found in regime $\mathcal{A}$ no longer holds. The behavior of $\tsec(a)$ becomes significantly more complicated, and $\tsec$ does not necessarily increase with decreasing $a$, as we will see.

The argument we used in deriving \eqref{eqn:t_sec_weak} relied on $e_\mathrm{min}$ being much smaller than unity, and the approximation $ \sigma \ll 1$.  If these approximations are good (which in particular now requires high-$\jmax$ circulation, i.e. $a_\mathrm{div}<a<a_\mathrm{sep}$, see Figure \ref{fig:jPlus_of_a_Gamma_Regime_I}) then we can say that we are in regime $\mathcal{B}$.  This time we have $\jmax = j_0$ (equation \eqref{eqn:jmax_type_I})
and $j_+^2 > 1$ (Figures \ref{fig:jPlus_of_a_Gamma_Regime_I}-\ref{fig:jPlus_of_a_Gamma_Regime_II}), so that $\Delta \approx j_+$ and instead of  \eqref{eqn:t_sec_weak}, equations (32)-(34) of Paper II give
%%%%%%%%%%%%%
\begin{align} 
\label{eqn:t_sec_B} 
t_\mathrm{sec}^\mathcal{B} \approx \frac{8}{3A}\sqrt{\frac{G(m_1+m_2)}{\vert 25\Gamma^2-1\vert}} \times \frac{1}{a^{3/2}j_+} K\left( \frac{j_0}{j_+}\right).
\end{align}
%%%%%%%%%%%%%%
Plugging equations \eqref{eqn:jplus_of_a}, \eqref{eqn:j0_of_a} into \eqref{eqn:t_sec_B} gives us an explicit expression for $\tsec^\mathcal{B}(a)$ which we plot with a red dashed line in Figure \ref{fig:Numerical_LK}i.  We see that $\tsec^\mathcal{B}(a)$
diverges as we take $a \to a_\mathrm{sep}$ from below, as expected.
But away from the divergence we expect the elliptic integral $K$ to scale weakly with $a$, and we get (using equation \eqref{eqn:jplus_of_a})
\begin{align}
\tsec^\mathcal{B} \propto \frac{(a/d)^{1/4}}{\sqrt{1+(a/d)^{7/2} \cos^2 \imin }}.
\label{eqn:tsec_scaling_TypeI}
\end{align}
%%%%%%%%%%%%%%%
Thus, $\tsec^\mathcal{B}$ can either increase or decrease with $a$, depending on the value of $(a/d)^{7/2}\cos^2 \imin$.

\subsubsection{Regime $\mathcal{C}$: Circulating trajectories, $a \lesssim a_\mathrm{div}$}
\label{sec:RegC}

We see from Figure \ref{fig:Numerical_LK}i that equation \eqref{eqn:t_sec_B} becomes inaccurate once $a \lesssim a_\mathrm{div}$. This is expected because $\sigma \to \infty$ as $a \to a_\mathrm{div}$ (Figure \ref{fig:jPlus_of_a_Gamma_Regime_I}), implying that the non-GR approximation for $t_\mathrm{sec}$ derived in Paper II (which we used for regimes $\mathcal{A}, \mathcal{B}$) becomes invalid. Instead, we now need to use equations (\ref{eqn:tsec_exact}), (\ref{eq:djdtGR}) to  evaluate $t_\mathrm{sec}$. 

Figure \ref{fig:Numerical_LK}i shows that $t_\mathrm{sec}$ very rapidly diminishes as $a$ decreases below $a_\mathrm{div}$, which coincides with the rapid increase of $e_\mathrm{min}$ or, equivalently, rapid decrease of $j_\mathrm{max}$, which is now in the low-$j_\mathrm{max}$ limit. This is because for high-$\jmax$ 
circulating trajectories, a smaller $a$ leads to a smaller $\jmax$, which in turn means that the binary spends more time at `high' eccentricities (say with $e$ above $0.9$). The cluster tide-driven secular evolution is faster at high $e$ than at $e\sim 0$ because, even though the torque on a binary with, say, $e=0.1$ is comparable to that on a binary with $e=0.9$, the angular momentum of the latter is significantly smaller, so the relative change in angular momentum
occurs over a much shorter timescale. For that reason $t_\mathrm{sec}$ decreases with decreasing $a$, very rapidly (in a non-power law fashion) for $a \lesssim a_\mathrm{div}$. We call this interval of rapid $t_\mathrm{sec}$ decay regime $\mathcal{C}$.   

In Appendix \ref{sec:tsec_type2_appendix} we use equations (\ref{eqn:tsec_exact}), (\ref{eq:djdtGR}) to derive an approximation (\ref{eqn:tsecCD}) for $\tsec$ valid in regime $\mathcal{C}$. Depending on the relationship between $a_\mathrm{div}$ and $a_\mathrm{weak}$, there are two possibilities. For $a_\mathrm{weak}\ll a\lesssim a_\mathrm{div}$ (as in Figure \ref{fig:Numerical_LK}) equation (\ref{eqn:tsecCD}) gives 
%%%%%%
\begin{align}
    \tsec^\mathcal{C} &\approx 
    \frac{2\pi}{15 \Gamma A} \frac{\sqrt{2G(m_1+m_2)p_\mathrm{min}}}{d^2}\cos\imin
    \nonumber \\
    &\times \left(\frac{a}{d}\right)^{13/4}\left|1-\left(\frac{a}{a_\mathrm{div}} \right)^{7/2}\right|^{-3/2}.
\label{eqn:t_sec_C-1}
\end{align}
%%%%%%
On the other hand, for $a\lesssim a_\mathrm{div}\ll a_\mathrm{weak}$ (as in Figure \ref{fig:Numerical_Initially_ModLib}) we find
%%%%%%
\begin{align}
    \tsec^\mathcal{C} &\approx 
    \frac{2\pi}{15 \Gamma A} \frac{\sqrt{2G(m_1+m_2)p_\mathrm{min}}}{d^2}
    \nonumber \\
    &\times \left(\frac{a}{d}\right)^{3/2}\frac{2-\left(a/a_\mathrm{div} \right)^{7/2}}{\left|1-\left(a/a_\mathrm{div} \right)^{7/2}\right|^{3/2}}.
\label{eqn:t_sec_C-2}
\end{align}
%%%%%%
Both of these expressions show a rapid decay of $\tsec$ as $a$ drops even slightly below $a_\mathrm{div}$ because of the $a$-dependent term in the denominator. However, this term rapidly becomes constant as $a$ decreases further, switching again to a power-law behavior of $\tsec(a)$, which we cover next.

\subsubsection{Regime $\mathcal{D}$: Moderate GR, $a \ll a_\mathrm{div}, a_\mathrm{weak}$}
\label{sec:RegD}

As $a$ becomes substantially smaller than both $a_\mathrm{div}$ and $a_\mathrm{weak}$, and the binary is in the moderate-GR regime, GR precession plays an even more important role in determining $\tsec$. We call this situation regime $\mathcal{D}$. In this case we can still use equation (\ref{eqn:tsecCD}) to find in the appropriate limit that
%%%%%%
\begin{align}
    \tsec^\mathcal{D} &\approx 
    \frac{4\pi}{15 \Gamma A} \frac{\sqrt{2G(m_1+m_2)p_\mathrm{min}}}{d^2}
    \left(\frac{a}{d}\right)^{3/2}.
\label{eqn:t_sec_D}
\end{align}
%%%%%%
The predicted scaling $ \tsec^\mathcal{D}\propto a^{3/2}$ matches what we observed at the low-$a$ end of Figure \ref{fig:Numerical_LK}i. Note that this expression is valid regardless of the relationship between $a_\mathrm{div}$ and $a_\mathrm{weak}$. Again the secular timescale decreases (rather than increasing like one would naively expect) as the semimajor axis shrinks, although not as rapidly as in regime $\mathcal{C}$.  

%%%%%%%%%%%%%%%%%%%%%%%%%%%%%%%%%%%%%%%%
%%%%%%%%%%%%%%%%%%%%%%%%%%%%%%%%%%%%%%%%
%%%%%%%%%%%%%%%%%%%%%%%%%%%%%%%%%%%%%%%%
%%%%%%%%%%%%%%%%%%%%%%%%%%%%%%%%%%%%%%%%
\subsection{The evolution of semimajor axis}
\label{sec:SMA_Evolution}
%%%%%%%%%%%%%%%%%%%%%%%%%%%%%%%%%%%%%%%%
%%%%%%%%%%%%%%%%%%%%%%%%%%%%%%%%%%%%%%%%
%%%%%%%%%%%%%%%%%%%%%%%%%%%%%%%%%%%%%%%%
%%%%%%%%%%%%%%%%%%%%%%%%%%%%%%%%%%%%%%%%
%Moreover, except when we consider
%circulating trajectories in the moderate GR regime
%(\S\ref{sec:Gamma_Regime_I_Moderate}), when evaluating the right hand side of
%this equation we can take the expression for $\tsec(a')$ directly from equation
%\eqref{eqn:t_sec}.

In this section we aim to understand how the semimajor axis of a binary decays with time in certain asymptotic regimes.
To achieve this we first
write down expressions for the decay in semimajor axis over one secular
cycle, $\Delta a$, in terms of $a$. This allows us to understand the behavior we saw in panel (j) of Figure \ref{fig:Numerical_LK}. Then we plug our expressions for $\tsec(a)$ and $\Delta a (a)$ into the right hand side of \eqref{eqn:t_of_a_WTM} to calculate (very approximate) expressions for $\tau_a$. 
%We now show how to calculate $\Delta a$. 

To begin, we integrate equation \eqref{eqn:dadt_GW} over one secular
cycle, approximating $a$ as constant to lowest order (which is valid since
$\vert \Delta a \vert \ll a$ by assumption for a slow merger). The result
is\footnote{Note that equation \eqref{eqn:Delta_a_sec_cycle} is
essentially identical to the first line in equation (55) of
\citet{Randall2018-uq} --- see Appendix \ref{sec:RX18}.}
%%%%%%%%%%%%%
\begin{align}
\label{eqn:Delta_a_sec_cycle}
\Delta a 
&\approx 
-\frac{\lambda_0}{a^3}\int_{\mathrm{sec. \, cycle}} \frac{\md t}{(1-e^2)^{7/2}} \left( 1+\frac{73}{24}e^2 + \frac{37}{96}e^4\right), 
\end{align}
%%%%%%%%%%%%%
where $\lambda_0 \equiv (64/5)G^3 c^{-5} m_1m_2(m_1+m_2)$ is independent of $a$.
Assuming the binary reaches very high maximum eccentricity $e_\mathrm{max} \to
1$, we show in Appendix \ref{sec:Delta_a_appendix} that we can ultimately approximate this as (equation \eqref{eqn:Delta_a_of_xmax}):
%%%%%%%%%%%%%%%
\begin{align}
\Delta a \approx &-\lambda_2
\times \frac{\xi}{a^{3/2}\vert j_+ j_0\vert},
\label{eqn:Delta_a_of_xmax_simple}
\end{align}
%%%%%%%%%%%%%%%
where $\lambda_2 \equiv
1360G^{7/2}m_1m_2(m_1+m_2)^{3/2}/[9c^5A(2\pmin)^3\sqrt{\vert 25\Gamma^2-1
\vert}]$
is independent of $a$, and $\xi$ is a complicated function of $a$ and other parameters --- see equation \eqref{eqn:dimensionless_integral}. 

To simplify the expressions for $\xi$, $j_+$ and $j_0$ in equation \eqref{eqn:Delta_a_of_xmax_simple} we need to specify the strength of GR precession.
The asymptotic regimes of interest for evaluating $\Delta a$ are therefore the weak, moderate and strong GR regimes. Unfortunately these 
do not map precisely onto the phase space regimes $\mathcal{A}$-$\mathcal{D}$ that we used to understand $\tsec(a)$ behavior in \S\ref{sec:secular_timescale}.  Nevertheless, there are situations where a binary sits in, for instance, both the weak GR regime and phase space regime $\mathcal{A}$, and in such situations clean analytic results for $\tau_a$ are possible, as we will see.

%%%%%%%%%%%%%%%%%%%%%%%%%%%%%%%%%%%%%%%%%%%%%
\subsubsection{Weak GR}
%%%%%%%%%%%%%%%%%%%%%%%%%%%%%%%%%%%%%%%%%%%%%

In Appendix \ref{sec:Delta_a_appendix} we show that for weak GR ($a\lesssim a_\mathrm{weak}$),
%%%%%%%%%%%%%%%
\begin{align}
\xi  \approx \frac{8}{15} 
    \frac{1}{\sqrt{1+\vert \sigma\vert }}.
 \label{eqn:dimensionless_integral_not_II}
\end{align}
%%%%%%%%%%%%%%%
From Figures \ref{fig:jPlus_of_a_Gamma_Regime_I} and
\ref{fig:jPlus_of_a_Gamma_Regime_II} (or by inspection of equation \eqref{eqn:sigma_of_a}) we know that for weak GR we normally have
$\sigma \lesssim 1$; thus $\xi$ will also be $\mathcal{O}(1)$ and scale weakly
with $a$ in the weak GR regime. Since  $j_+$ and $j_0$ also scale weakly with
$a$ in this regime (see equations \eqref{eqn:jplus_of_a}, \eqref{eqn:j0_of_a}), we find from \eqref{eqn:Delta_a_of_xmax_simple} that
%%%%%%%%%%%%%%%
\begin{align}
\Delta a \propto a^{-3/2}. 
\label{eqn:Delta_a_scaling_weak}
\end{align}
%%%%%%%%%%%%%%%
This matches what we saw at the high-$a$ end of Figure \ref{fig:Numerical_LK}j.
Physically, the scaling \eqref{eqn:Delta_a_scaling_weak} just reflects the fact that the time spent in the high
eccentricity state is proportional to the secular timescale, and that in the
weak GR regime $ \tsec \propto a^{-3/2}$ (equation
\eqref{eqn:tsec_scaling_A}). Indeed, one might have guessed the
result \eqref{eqn:Delta_a_scaling_weak} a priori by noting from
\eqref{eqn:dadt_GW} that for $j\approx \jmin$ we have $(\md a/\md
t)_\mathrm{GW} \propto a^{-3}\jmin^{-7}$. Since the time spent
at high eccentricity is $t_\mathrm{min} \sim \jmin \tsec$
(Paper III), we get $\Delta a \sim (\md a/\md t)_\mathrm{GW}
\times \tmin \propto a^{-3}\jmin^{-6} \tsec$. Plugging in
\eqref{eqn:jmin_of_a} for $\jmin$ we simply get $\Delta a \propto
\tsec \propto a^{-3/2}$.
%of GW emission is roughly (power of GW emission) $\times$ (time spent in high
%%$e$ state). The first of these depends mainly on $\jmin$ while the second is
%$\jmin \tsec \propto \jmin a^{-3/2}$.
%Since $

%Also, and perhaps more importantly, the
%transition between weak and moderate GR behaviors around $a_\mathrm{weak}$ is
%typically rather blurred, so in truth $a=a_\mathrm{weak}$ rarely signifies a
%precise boundary between these different dynamical regimes, as we see in all
%numerical examples in this section.

We can now evaluate the characteristic decay timescale $\tau_a$ if we assume not only that the binary is in the weak GR regime but also that it is on a librating phase space trajectory (i.e. that we are in regime $\mathcal{A}$).  In this case we can plug equations \eqref{eqn:t_sec_weak}
and \eqref{eqn:Delta_a_of_xmax_simple},
\eqref{eqn:dimensionless_integral_not_II} into \eqref{eqn:decay_timescale} to find
\begin{align}
    \tau_a =  \Lambda_\mathrm{weak} U^{-1}  a,
\label{eqn:decay_time_weak}
\end{align}
where 
\begin{align}
    \Lambda_\mathrm{weak} \equiv \sqrt{1+\sigma } \, \frac{\vert j_+j_0\vert}{\sqrt{\Delta}}K\left( \frac{j_+}{\sqrt{\Delta}}\right),
    \label{eqn:dimensionless_quantity_weak}
\end{align}
is dimensionless and typically $\mathcal{O}(1)$ away from separatrices, and we defined a `decay rate' 
\begin{align}
    U &\equiv \frac{272 G^3m_1m_2(m_1+m_2)}{3(2\pmin)^3c^5}
    \label{eqn:decay_speed_characteristic}
    \\
    &=1.4\, \mathrm{AU} \,\mathrm{Gyr}^{-1}\times \left( \frac{\pmin}{10^{-2}\mathrm{AU}} \right)^{-3}
\left( \frac{m}{10 M_\odot} \right)^{3}.
\label{eqn:decay_speed_numerical}
\end{align}
(In the numerical estimate \eqref{eqn:decay_speed_numerical} we put $m_1=m_2=m$).
The scaling $\tau_a \propto a$ predicted by equation \eqref{eqn:decay_time_weak} is exactly what we see at the high-$a$ end in Figure \ref{fig:Numerical_LK}k.

\subsubsection{Moderate GR}
%%%%%%%%%%%%%%%%%%%%%%%%%%%%%%%%%%%%%%%%%%%%%

\begin{figure}
\centering
\includegraphics[width=0.7\linewidth]{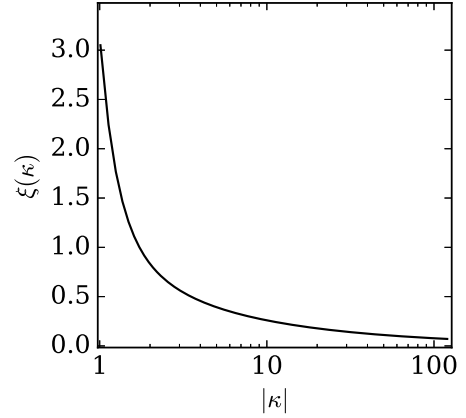}
\caption{Plot of the function $\xi$ as a function of $\vert \kappa \vert$
 according to \eqref{eqn:dimensionless_integral_type_II}, which
 is valid for low-$\jmax$ circulating trajectories in the moderate GR regime. Note that the horizontal axis is logarithmic.}
\label{fig:dimensionless_integral_type_II}
\end{figure}

In the moderate GR regime we can get a scaling for $\Delta a$ if we 
assume the $(d/a)^{7/2}$ terms 
dominate in equations \eqref{eqn:jplus_of_a},
\eqref{eqn:j0_of_a}.  Then from \eqref{eqn:Delta_a_of_xmax_simple}:
%%%%%%%%%%%%%%%
\begin{align}
\Delta a \approx -\frac{\lambda_2 \xi}{d^{7/2}}\frac{\sqrt{\vert
25\Gamma^2-1 \vert}}{10\Gamma} \times a^{2}.
\label{eqn:Delta_a_II_moderate}
\end{align}
%%%%%%%%%%%%%%%
We can simplify the expression for $\xi$ (equation \eqref{eqn:dimensionless_integral}) if we further assume the binary is on a low-$\jmax$ circulating trajectory (which is inevitably true at some point before the strong GR regime is reached).
In this case $\jmax = -\sigma j_-$ (equation \eqref{eqn:jmax_type_II}) so that 
$x_\mathrm{max} = x_\sigma
\approx (-\sigma j_-)/(\gamma j_-) = -\sigma/\gamma \equiv - \kappa$. 
For moderate GR it is 
also
easy to show that $\vert x_\alpha \vert \approx \gamma^{-2} \ll 1$, so we can
ignore $\vert x_\alpha \vert$ compared to $x$ in
\eqref{eqn:dimensionless_integral}.  As a result we find
%%%%%%%%%%%%%%%
\begin{align}
\xi(\kappa) \approx \int_1^{\vert \kappa \vert} \frac{x^{-6}\md x}{\sqrt{(x-1)(\vert \kappa \vert -x)}}.
\label{eqn:dimensionless_integral_type_II}
\end{align}
%%%%%%%%%%%%%%%
In Figure \ref{fig:dimensionless_integral_type_II} we plot $\xi$ as a
function of $\vert \kappa \vert $ according to
\eqref{eqn:dimensionless_integral_type_II}; in particular we see that
$\xi \sim 1$ except for very large $\vert \kappa \vert \gg 1$. 
%%%%%%%%%%%%%%%  
Also, away from
$a\approx a_\mathrm{div}$ (i.e. in regime $\mathcal{D}$) we know $\kappa$ is almost never large compared to
unity, and scales weakly with $a$ (Figures
\ref{fig:jPlus_of_a_Gamma_Regime_I}-\ref{fig:jPlus_of_a_Gamma_Regime_II}),
so we can treat $\xi$ as an order-unity constant for a rough analysis.  As a
result, \eqref{eqn:Delta_a_II_moderate} predicts a scaling
%%%%%%%%%%%%%%%
\begin{align}
\Delta a \propto a^2.
\label{eqn:Delta_a_scaling_moderate}
\end{align}
%%%%%%%%%%%%%%%
We see from \eqref{eqn:Delta_a_scaling_moderate} that, unlike for
weak GR (equation \eqref{eqn:Delta_a_scaling_weak}), here the
individual decrements in semimajor axis $\Delta a$ get \textit{smaller} as the semimajor
axis $a$ shrinks.  The scaling \eqref{eqn:Delta_a_scaling_moderate} matches what we saw in Figure
\ref{fig:Numerical_LK}j towards the low-$a$ end.

One can understand the result \eqref{eqn:Delta_a_scaling_moderate}
qualitatively as follows. Like for weak GR, at very high $e$ we again have
roughly $(\md a /\md t )_\mathrm{GW} \propto a^{-3}\jmin^{-7}$, and again using
\eqref{eqn:jmin_of_a} this is $\propto a^{1/2}$. This time, since the
binary spends a large fraction of its secular period in the vicinity of $\jmin$,
we get a rough estimate of $\Delta a$ by multiplying $(\md a /\md t
)_\mathrm{GW}$ not by $\tmin$, but by $\tsec$. Using the scaling
\eqref{eqn:t_sec_D} we get $\Delta a \propto a^{1/2}
\times a^{3/2} \propto a^{2}$. Loosely speaking, the factor of $\xi$ in equation
\eqref{eqn:Delta_a_II_moderate} accounts for the fraction of time
that the binary actually spends in the vicinity of $\jmin$ during each secular
cycle.

We can also compute the decay timescale $\tau_a$ in the moderate GR, low-$\jmax$ circulating regime.
To do so we plug \eqref{eqn:t_sec_D} and
\eqref{eqn:Delta_a_of_xmax_simple} into
\eqref{eqn:decay_timescale} to find\footnote{Note
that we do \textit{not} take $\Delta a$ from
\eqref{eqn:Delta_a_II_moderate}: instead we used the more general
equation \eqref{eqn:Delta_a_of_xmax_simple}, which allows us to take
advantage of the cancellation of the factors $\vert j_+j_0\vert$
\textit{without} having to assume the dominance of the $(d/a)^{7/2}$ terms in
$j_+^2$, $j_0^2$. Similarly, $\xi$ as given in equation
\eqref{eqn:dimensionless_integral_type_II} does not rely on this
assumption. } 
\begin{align}
    \tau_a =  \Lambda_\mathrm{mod} U^{-1} (2\pmin)^{1/2} a^{1/2},
    \label{eqn:decay_time_mod}
\end{align}
where
\begin{align}
    \Lambda_\mathrm{mod} \equiv \frac{4\pi(1-\kappa)}{5}\left[\int_1^{\vert \kappa \vert} \frac{x^{-6}\md x}{\sqrt{(x-1)(\vert \kappa \vert -x)}}\right]^{-1},
\end{align}
%%%%%%%%%%%%%%%%%%%%%%%%%%%%%%%%% 
%\begin{align}
%t-t_\mathrm{i} \approx& - \frac{3c^5(2\pmin)^{7/2}}{170G^3m_1m_2(m_1+m_2)} \nn \\&\times \frac{\pi}{2}\int_{a_\mathrm{i}}^{a(t)}
%\md a \, \frac{(1-\kappa)}{a^{1/2}\xi},
%\label{eqn:t_of_a_Type_II}
%\end{align}
%%%%%%%%%%%%%%%%%%%%%%%%%%%%%%%%%
which is typically $\mathcal{O}(1)$.
The scaling $\tau_a \propto a^{1/2}$ is confirmed at the low-$a$ end of Figure \ref{fig:Numerical_LK}k.
The moderate GR decay time \eqref{eqn:decay_time_mod} is shorter than the analogous
weak GR result decay time \eqref{eqn:decay_time_weak} by a factor $\sim \sqrt{2\pmin/a} \sim \jmin \ll 1$.

\subsubsection{Strong GR}
\label{sec:a_of_t_strong}
%%%%%%%%%%%%%%%%%%%%%%%%%%%%%%%%%

Once the strong GR regime is reached, equation
\eqref{eqn:t_of_a_WTM} ceases to be valid because the binary
decouples from cluster tides and so no longer undergoes secular eccentricity
oscillations (Paper III). %Precisely, in the
%asymptotic limit $\epsGR \gg \epsstr$, the binary undergoes pure precession at a rate 
%%%%%%%%%%%%
%\begin{align}
%\dot{\omega}_\mathrm{GR} 
%=\frac{3[G(m_1+m_2)]^{3/2}}{c^2 a^{5/2} j^2},
%\label{eqn:omega_dot_GR}
%\end{align}
%%%%%%%%%%%%
The evolution of $a$,
$e$ is then dictated purely by equations \eqref{eqn:dadt_GW},
\eqref{eqn:dedt_GW}. Supposing the transition to the strong GR regime
happens at some reference time
$t_\mathrm{i}$, we know from \S\ref{sec:conservation_laws} that for $t>t_\mathrm{i}$
the binary conserves its value of $ p=a(1-e)=\pmin$, meaning $j =
(2\pmin)^{1/2}a^{-1/2}$ (although $\pmin$ can be a factor of $\sim 2$ larger than its initial value $\pmin(t=0)$ --- see Figure \ref{fig:Numerical_Initially_ModLib_zoom2}d and the final paragraph of \S\ref{sec:Numerical_Example_3}.). With this we can eliminate eccentricity from equation \eqref{eqn:dadt_GW}, resulting in
%\cmtrr{Need to stress that not using second equality in definition (\ref{eqn:decay_timescale}) here, only the first one. Also, note somewhere that $\pmin$ is not its initial value, but is different by up to $\sim 2$.}
%%%%%%%%%%%%%
\begin{align}
\frac{\md a}{\md t} \approx -\frac{\lambda_1}{(2\pmin)^{7/2}} a^{1/2}.
\label{eqn:dadt_strong}
\end{align}
%%%%%%%%%%%%%
We now use the definition $\tau_a \equiv \vert \md \ln a/\md t \vert^{-1}$ (equation \eqref{eqn:decay_timescale}) to calculate $\tau_a$ directly from \eqref{eqn:dadt_strong} as
\begin{align}
    \tau_a = \frac{8}{5}U^{-1} (2\pmin)^{1/2} a^{1/2}.
    \label{eqn:decay_time_strong}
\end{align}
%%%%%%%%%%%%%
%\begin{align}
%a(t) \approx a_\mathrm{i} \left( 1-\frac{t-t_\mathrm{i}}{\tau_\mathrm{strong}}\right)^2,
%    \label{eqn:a_of_t_strong}
%\end{align}
%%%%%%%%%%%%%
%where
%%%%%%%%%%%%%
%\begin{align}
%%\tau_\mathrm{strong} \approx \frac{6c^5(2\pmin)^{7/2}a_\mathrm{i}^{1/2}}{170G^3m_1 m_2(m_1 + m_2)}= \frac{2}{\pi Q_2} \times \tau_2,
%\label{eqn:taustrong}
%\end{align}
%%%%%%%%%%%%%
%and $\tau_2$ is given in equation
%\eqref{eqn:tau2}. 
Thus the only difference
between the characteristic decay timescale in the moderate GR regime (equation \eqref{eqn:decay_time_mod})
 and that during the strong GR
regime (equation \eqref{eqn:decay_time_strong}) is a factor of  order unity which depends very weakly on $a$.  This explains why the behavior of $\tau_a \propto a^{1/2}$ is barely modified in Figure \ref{fig:Numerical_LK}k once the binary enters the strong GR regime ($a\lesssim 10^{0.45}$AU).

\section{Numerical examples}
\label{sec:Numerical_Examples}
%%%%%%%%%%%%%%%%%%%%%%%%%%%%%%%%%%%%%%%%%%%%%%%%%%%%%%%%%%%%%%%%%
%%%%%%%%%%%%%%%%%%%%%%%%%%%%%%%%%%%%%%%%%%%%%%%%%%%%%%%%%%%%%%%%%

In this section we will provide further numerical examples akin to Figure \ref{fig:Numerical_LK}, in particular for binaries moving in non-Keplerian potentials. Our aim is to
verify and elucidate the approximate analytical results derived in
\S\S\ref{sec:GW_emission}-\ref{sec:phase_space_evolution}. We do this by direct numerical integration
of the DA equations of motion, including both GR precession and GW emission, for
various binaries that undergo slow mergers. First we give two Examples with
$\Gamma > 1/5$
(\S\S\ref{sec:Numerical_Example_2}-\ref{sec:Numerical_Example_3}),
the first of which exhibits all the hallmark behavior of a slow merger
beginning in the weak GR regime, and the second of which allows us to focus on
the late-stage (moderate and strong GR) evolution. We then provide one further
Example, this time for a binary with $0< \Gamma \leq 1/5$
(\S\ref{sec:Numerical_Example_4}). Note that we also provide one
additional numerical example in the LK limit in Appendix
\ref{sec:RX18}, when comparing our work with that of
\citet{Randall2018-uq}.

{To be clear, we note that we ran many more numerical experiments of
slow mergers than those shown here. We have chosen to present here the minimal
number of examples that still capture qualitatively all the possible interesting
 evolutionary scenarios.  (There are of course non-interesting cases, such
as binaries that are so tightly bound the cluster essentially plays no role
in their evolution, but we do not include them here).}

\subsection{Example 2: $\Gamma =0.42>1/5$. An initially librating trajectory in the weak GR regime}
\label{sec:Numerical_Example_2}
%%%%%%%%%%%%%%%%%%%%%%%%%%%%%%%%%%%%%%%%%%%%%%%%

%%%%%%%%%%%%%%%%%%%%%%%%%%%%%%%%%%%%%%%%%%%%%%%%
\begin{figure*}
\centering
   \includegraphics[width=0.99\linewidth]{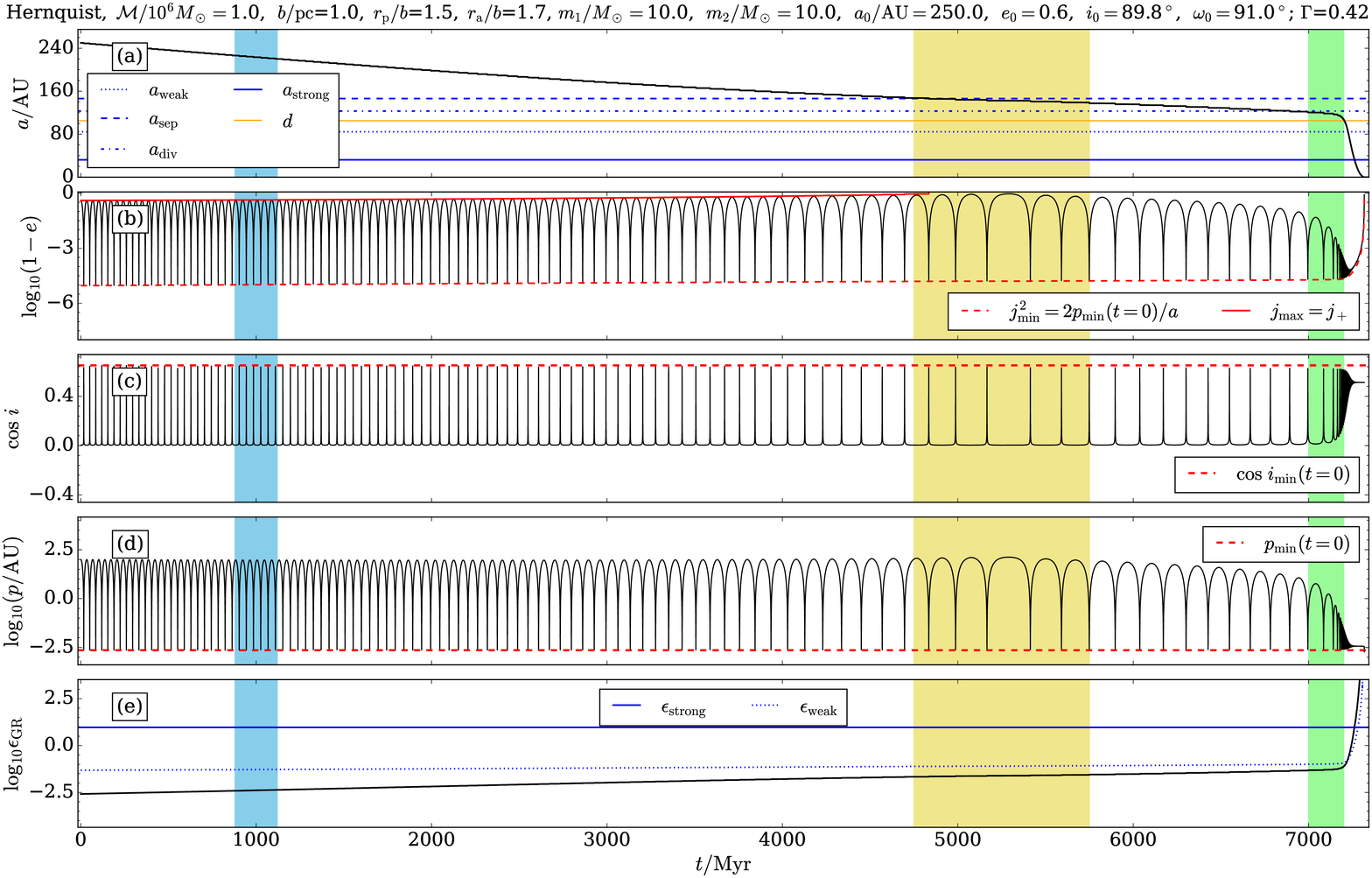}
    \includegraphics[width=0.99\linewidth]{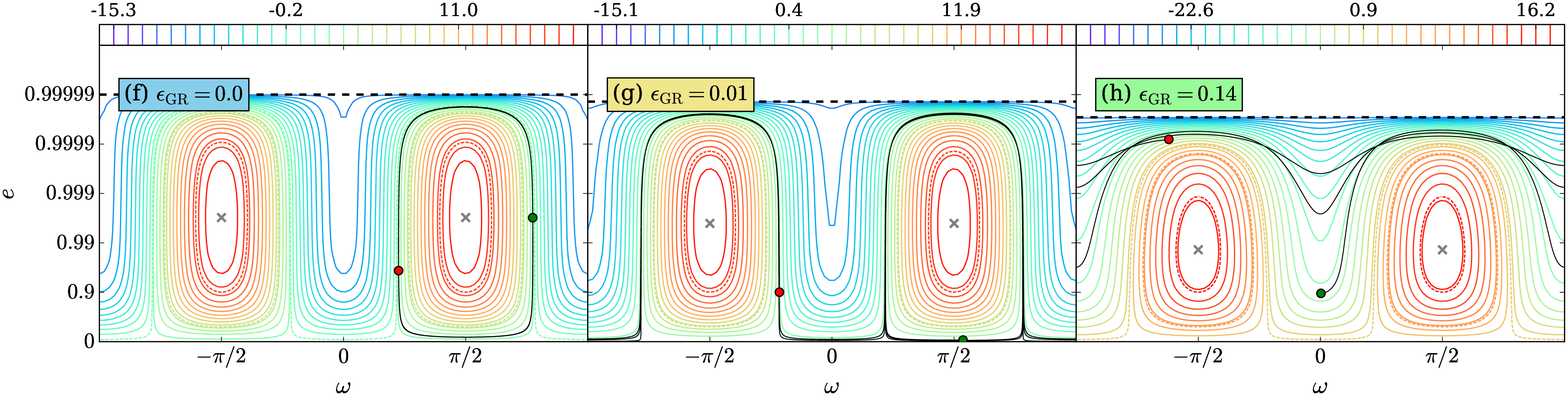}
    \includegraphics[width=0.99\linewidth]{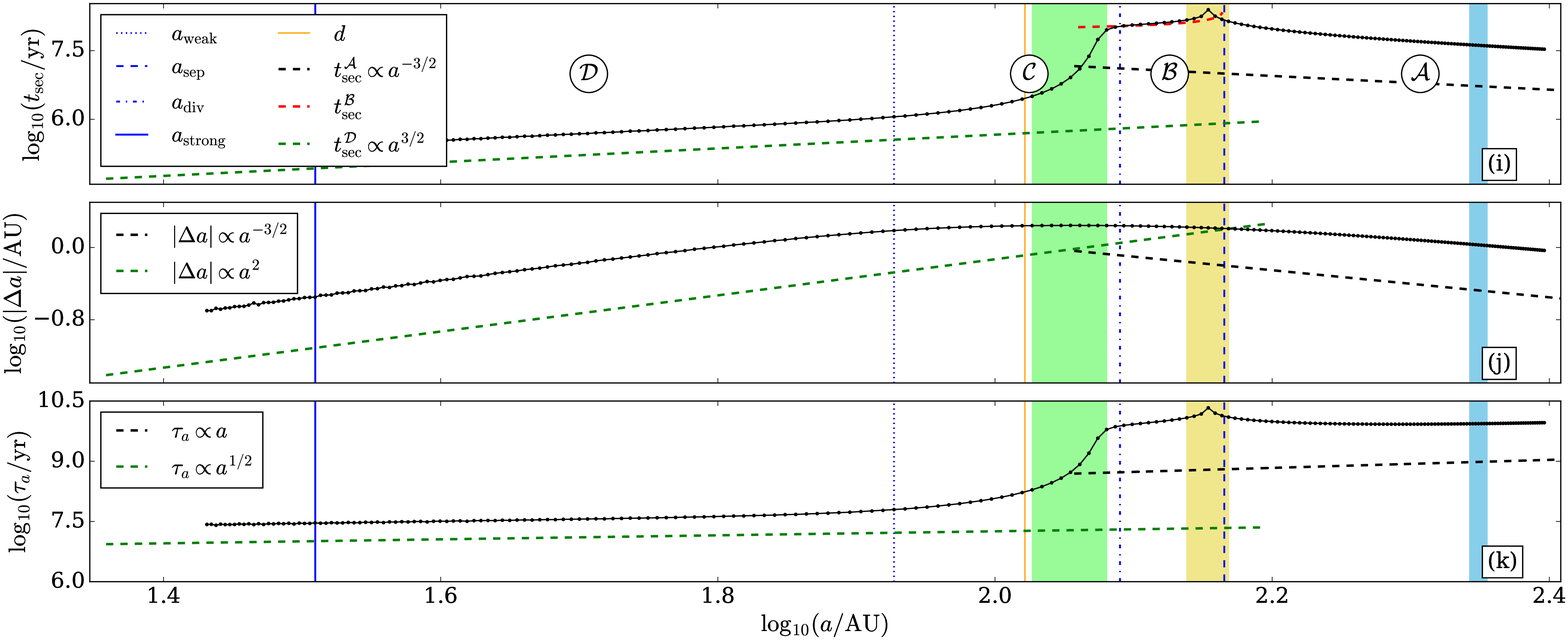}
\caption{\textit{Example 2}. The binary orbits a Hernquist cluster $\Phi(r) = -G\mathcal{M}/(b+r)$ with $\mathcal{M}=10^6M_\odot$ and $b=1\mathrm{pc}$, and the outer orbital peri/apocentre is chosen to be $(\rp/b, \ra/b)=(1.5,1.7)$. See
\S\ref{sec:Numerical_Example_2} for a detailed discussion.}
    \label{fig:Numerical_Initially_WeakLib}
\end{figure*}
%%%%%%%%%%%%%%%%%%%%%%%%%%%%%%%%%%%%%%%%%%%%%%%%

In Figure \ref{fig:Numerical_Initially_WeakLib} we show the result of integrating the DA equations of motion for a binary that orbits the spherical Hernquist potential $\Phi(r)=-G\mathcal{M}/(b+r)$, where $\mathcal{M}=10^6M_\odot$ and $b=1\mathrm{pc}$.  The outer orbital peri/apocentre is chosen to be $(\rp/b, \ra/b)=(1.5,1.7)$. The resulting $\Gamma$ value is $0.42 > 1/5$. The figure is set up in precisely the same way as Figure \ref{fig:Numerical_LK}.

Let us first focus on the initial $\sim 3000$ Myr.
From panel (a) we see that at $t=0$ the binary has $a_0=250$ AU, and that for the
first $\sim 3000$ Myr, $a(t)$ exceeds significantly each of the four critical
values $a_\mathrm{weak}$, $a_\mathrm{sep}$, $a_\mathrm{div}$,
$a_\mathrm{strong}$, which were defined in
\S\ref{sec:lengthscales} and which we show with horizontal
blue lines (see legend). It follows that during this time, the binary resides in
the weak GR regime: and indeed, we see from panel (e) that $\epsGR$ is initially
far smaller than $\epsweak$ (blue horizontal dotted line). Moreover, panel (b)
shows that the binary undergoes the expected secular eccentricity oscillations
(initially on a timescale of $\sim 30$Myr), and reaches a very high maximum
eccentricity of $1-e_\mathrm{max} \approx 10^{-5}$. Concomitantly there are
secular oscillations in inclination $i$ (panel (c)) and pericentre distance $p$
(panel (d)), though as predicted in \S\ref{sec:conservation_laws} the
values of $\cos i_\mathrm{min}$ and $\pmin$ reached at the peak of each secular
eccentricity cycle are very nearly conserved (see the dashed red horizontal
lines in these panels). Similarly, from panel (b) we see that the maximum
eccentricity of the binary is well described by $\jmin^2 =2\pmin/a$ (equation
\eqref{eqn:jmin_of_a}), while its minimum is well described by $\jmax
= j_+$. The latter fact implies that the binary is initially on a librating
phase space trajectory (equation \eqref{eqn:jmax_librating}), and this is
confirmed by panel (f), in which we show the phase space evolution during the
time interval denoted by the blue shaded stripe. 

%: returning to panel (a) we see that $a(t)$ decays
%roughly linearly while the binary is in the weak GR regime, which is exactly
%what we predicted in \S\ref{sec:SMA_Evolution} (equation
%\eqref{eqn:a_of_t_weak}).

So, we have a binary on a librating trajectory (regime $\mathcal{A}$) in the weak GR regime, whose semimajor
axis is slowly decaying with time. From panels (i), (j) and (k) we see that the binary obeys all the expected scalings for $\tsec$, $\Delta a$ and $\tau_a$ in this regime, namely equations \eqref{eqn:t_sec_weak}, \eqref{eqn:Delta_a_scaling_weak} and \eqref{eqn:decay_time_weak} respectively.
As we know from
\S\ref{sec:phase_space_evolution} there are two key things that
happen next to such a binary: one is that it enters the moderate GR regime, and
the other is that its $(\omega,e)$ phase space trajectory crosses the separatrix
and becomes circulating (ultimately a low-$\jmax$ circulating trajectory). We also know
from Figure \ref{fig:jPlus_of_a_Gamma_Regime_I} that these two
occurrences can happen in any order.
In this particular case the binary crosses the separatrix first: we see from
panel (a) that $a$ crosses $a_\mathrm{sep}$ around $t=4800$ Myr, and from
panel (b) that around this time the minimum eccentricity gets very close to zero
and then starts to increase and ceases to be well described by $\jmax=j_+$.
This inference is confirmed in panel (g), in which we see explicitly the
evolution from libration to circulation that occurs during the time interval
denoted by the yellow shaded stripe.  
Panel (i) confirms the expected scalings of $\tsec$, $\Delta a$ and $\tau_a$ as the binary moves through regime $\mathcal{B}$.

At around $t=7100$ Myr, the binary's dynamical evolution changes
dramatically: the semimajor axis $a$ approaches $a_\mathrm{div}$ and its decay
accelerates, whereas the secular timescale becomes very short (decaying in the non-power law fashion expected of regime $\mathcal{C}$ --- see \S\ref{sec:RegC}). Furthermore we see
that $\jmin^2=2\pmin/a$ is still a good approximation for the maximum
eccentricity and, though we do not show it here, the minimum eccentricity at
this stage is fairly well-described by $j_\mathrm{max} = -\sigma j_-$ (we defer
a more careful, `zoomed in' discussion of the late stages of a slow merger to \S\ref{sec:Numerical_Example_3}). These characteristics
are the hallmark of low-$\jmax$ circulating trajectories in the moderate GR regime.  To
confirm this, we look at panel (h), which shows the phase space evolution during
the green striped time interval. We see clearly that the binary rapidly evolves towards
a purely high-eccentricity circulating trajectory (into regime $\mathcal{D}$), whereafter it soon enters
the strong GR regime and then merges.

\subsection{Example 3: $\Gamma =0.42>1/5$. A binary initially in the moderate GR regime}
\label{sec:Numerical_Example_3}
%%%%%%%%%%%%%%%%%%%%%%%%%%%%%%%%%%%%%%%%%%%%%%%%

%%%%%%%%%%%%%%%%%%%%%%%%%%%%%%%%%%%%%%%%%%%%%%%%
\begin{figure*}
\centering
   \includegraphics[width=0.99\linewidth]{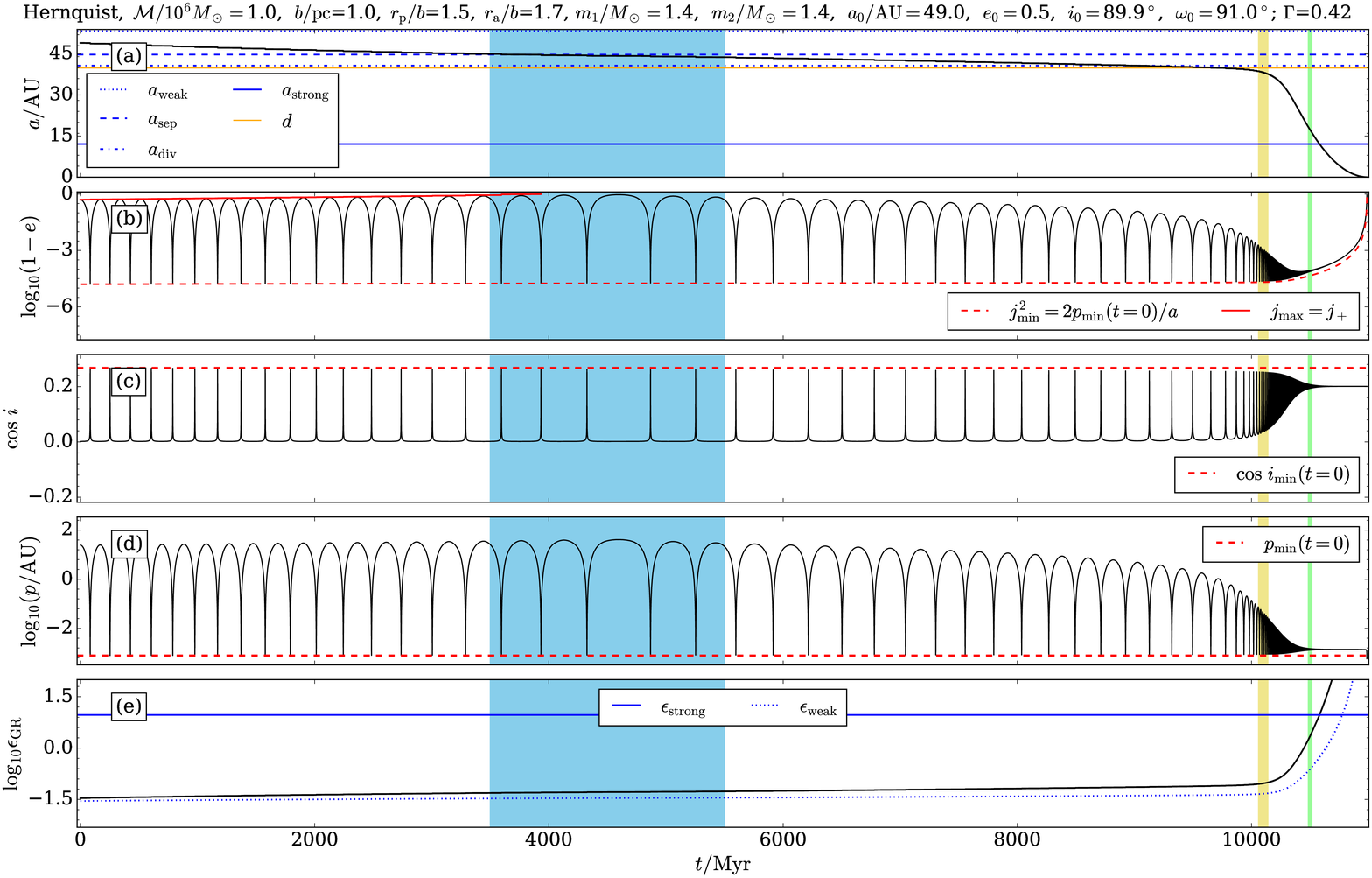}
    \includegraphics[width=0.99\linewidth]{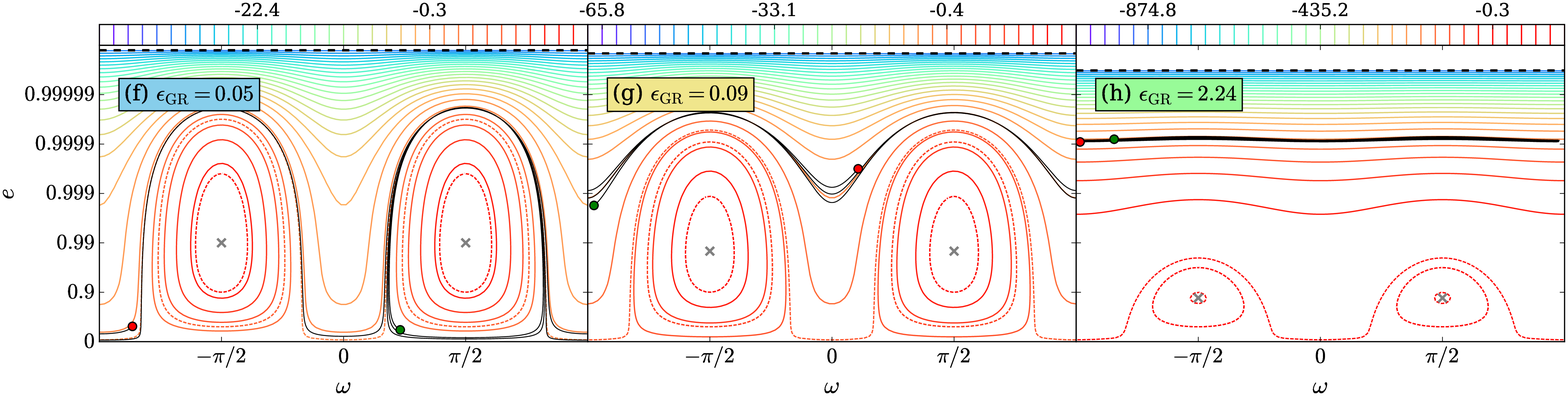}
    \includegraphics[width=0.99\linewidth]{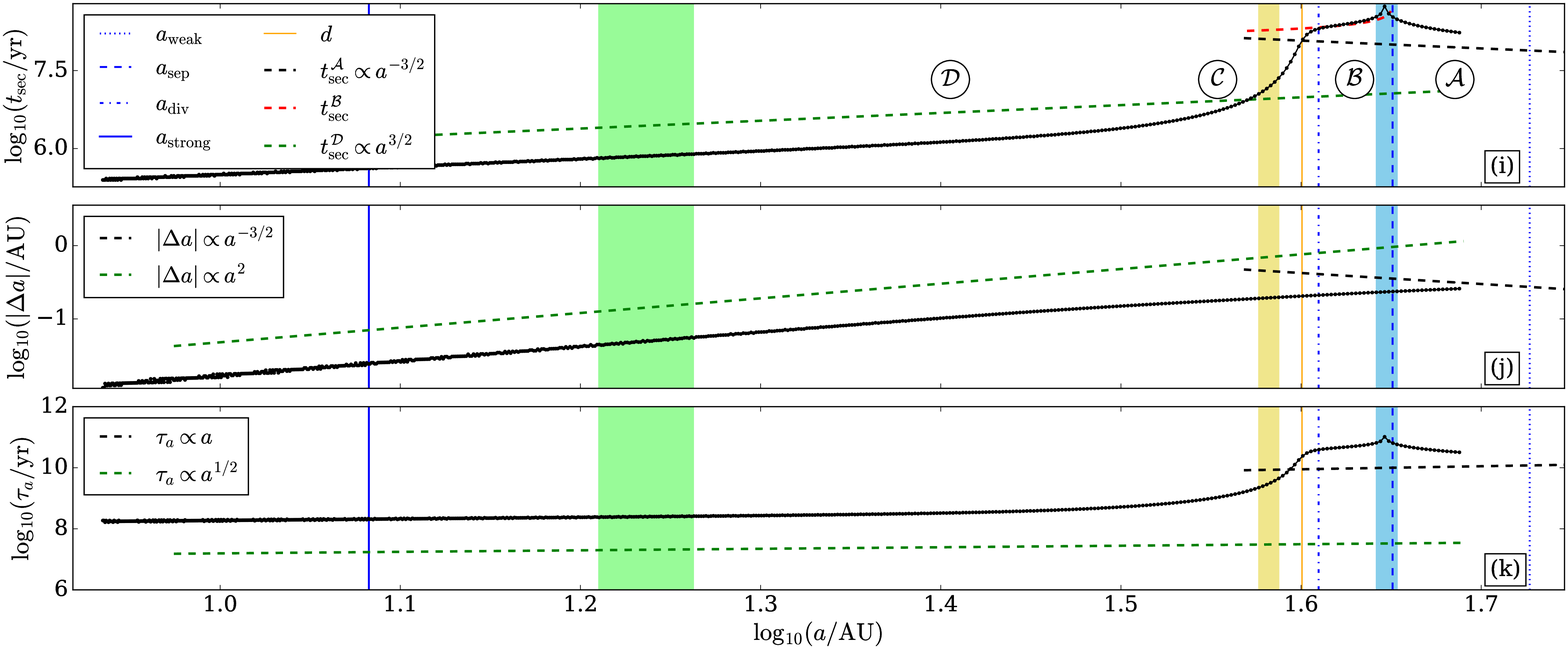}
\caption{\textit{Example 3}. The potential and outer orbit are
 the same as in Example 2 (Figure
\ref{fig:Numerical_Initially_WeakLib}), so that again $\Gamma=0.42$, 
but the binary constituent masses and inner orbit initial conditions are different.
In this case the binary begins in the moderate GR regime on a librating phase space trajectory.}
    \label{fig:Numerical_Initially_ModLib}
\end{figure*}
%%%%%%%%%%%%%%%%%%%%%%%%%%%%%%%%%%%%%%%%%%%%%%%%

In Figure \ref{fig:Numerical_Initially_ModLib} we show the evolution
of a lower mass binary ($m_1=m_2=1.4M_\odot$), but on the same outer orbit in the
same cluster as in Example 2; thus we again have $\Gamma=0.42$.  We choose
different initial conditions for the inner orbit, in particular $a_0 = 49$ AU.
We see from panel (a) that because of this choice, $a < a_\mathrm{weak}$ initially, putting
the binary just inside the moderate GR regime  (see also panel (e)). On the
other hand, initially $a > a_\mathrm{sep}$ so the phase space trajectory  librates. This is different to Example 1 since in that case, by
the time the binary reached the moderate GR regime, its phase space trajectory was
already circulating. Thus with Example 3 we will not only be able to focus on
the `late-time' behavior of a slow merger, i.e. its evolution through the
moderate and strong GR regimes (as promised in
\S\ref{sec:Numerical_Example_2}), but also to see a phase space
transition from librating to circulating within the moderate GR regime.

%Let us note here that the decay of $a(t)$ shown in panel (a) is qualitatively
%very similar to that in Example 2 (Figure
%\ref{fig:Numerical_Initially_WeakLib}a). Indeed, this turns out to be
%the case for all slow mergers: their semimajor axes decay approximately linearly to
%begin with, then more rapidly (roughly following the quadratic law of equation
%\eqref{eqn:a_of_t_moderate}) once they get well into the low-$\jmax$
%circulating regime, and then like equation \eqref{eqn:a_of_t_strong}
%in the strong GR regime.  What we want to focus on here is the behavior of $e$,
%$p$ and $i$ in these latter stages of moderate and strong GR.

%%%%%%%%%%%%%%%%%%%%%%%%%%%%%%%%%%%%%%%%%%%%%%%%
\begin{figure*}\centering
\includegraphics[width=0.99\linewidth]{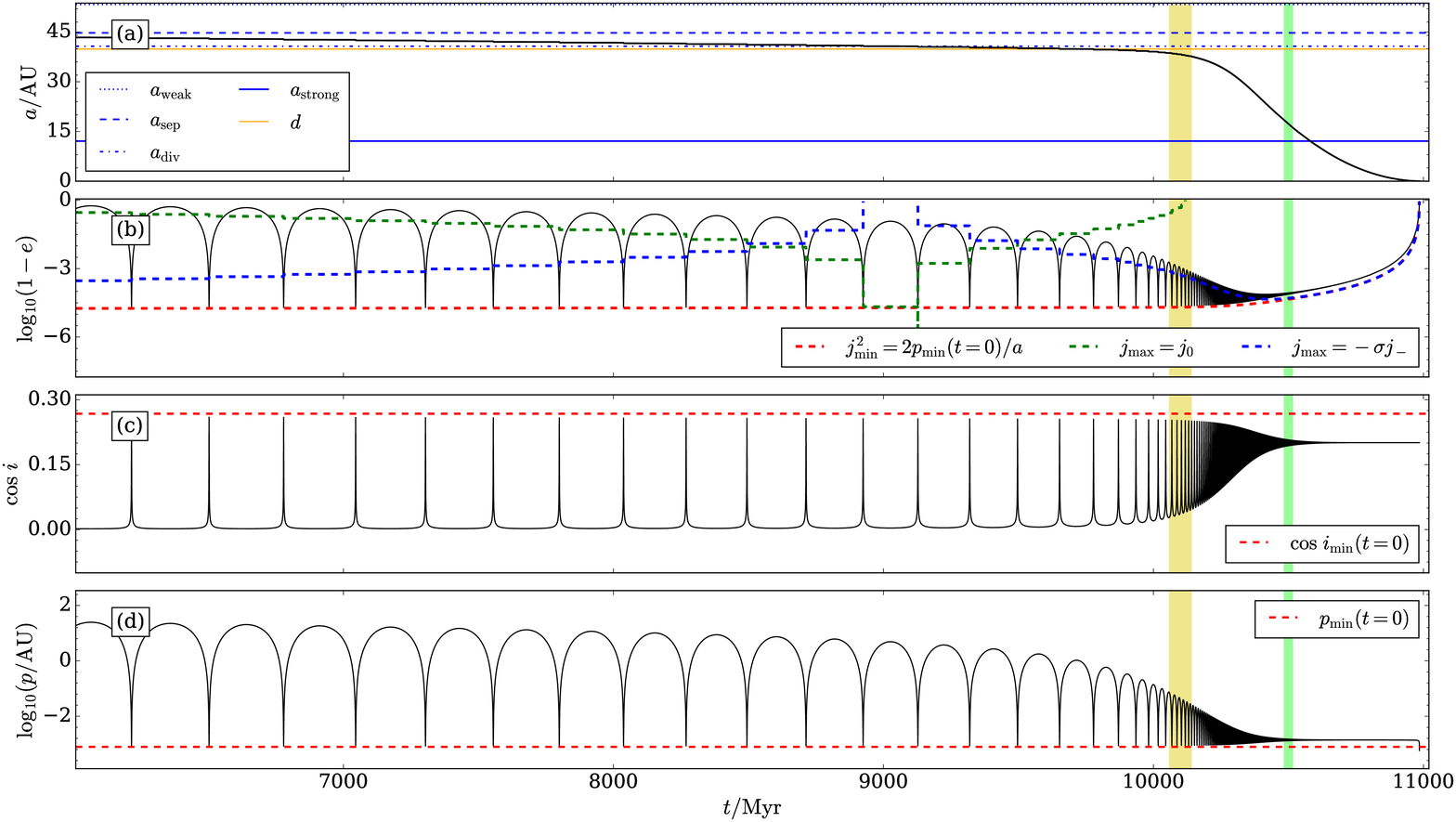}
\caption{Zoomed-in version of panels (a)-(d) from Figure
\ref{fig:Numerical_Initially_ModLib}, focusing on $t > 6000$ Myr.}
    \label{fig:Numerical_Initially_ModLib_zoom}
\end{figure*}%
%%%%%%%%%%%%%%%%%%%%%%%%%%%%%%%%%%%%%%%%%%%%%%%%
%\hfill <-- it is superfluous 
\begin{figure*}\centering
 \includegraphics[width=0.99\linewidth]{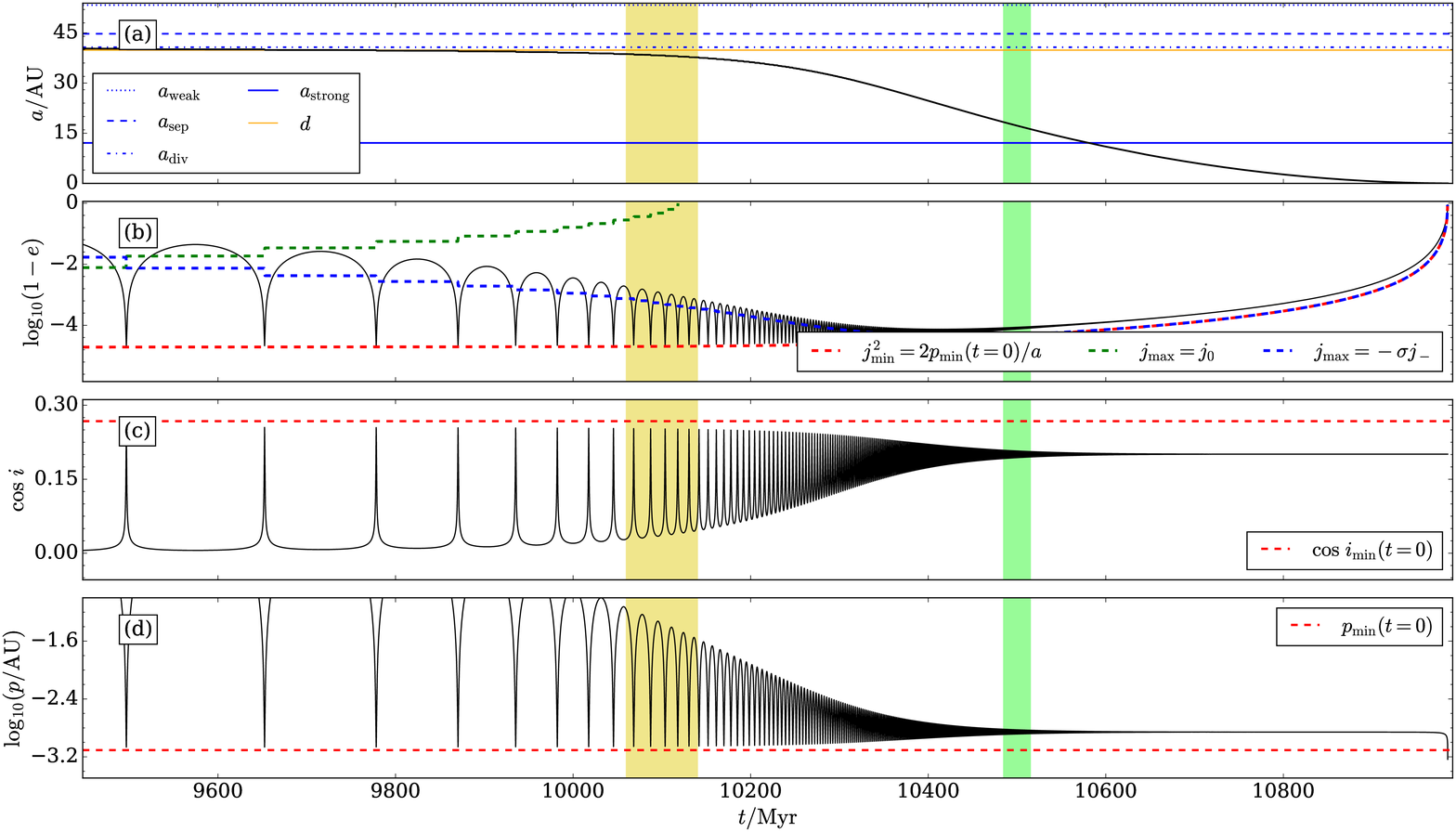}
 \caption{Further zoomed-in version of panels (a)-(d) from Figure
 \ref{fig:Numerical_Initially_ModLib}, this time showing from $t\approx 9400$ Myr to merger.}
     \label{fig:Numerical_Initially_ModLib_zoom2}
\end{figure*}%
%%%%%%%%%%%%%%%%%%%%%%%%%%%%%%%%%%%%%%%%%%%%%%%%

From panel (b) we see that while the binary is on a librating phase space trajectory
($t\lesssim 4500$ Myr) its maximum eccentricity is well described by $\jmax =
j_+$ and its secular period increases with time.  Once it enters the circulating
region its secular period begins to decrease with time. 
Panel (i) shows that the $\tsec(a)$ behavior is again split cleanly into four regimes $\mathcal{A}-\mathcal{D}$.
%Of course this was also
%the case in Example 2, though in that case the transition to circulation occurred in the weak
%GR regime. 
%We note from panel (k) that until $t\sim 10000$ Myr, $\tau_a \propto $, remains a gentle, near-linear one, by virtue of the
%fact that the binary is on a high-$\jmax$ circulating trajectory (e.g. panel (f)) despite belonging
%to the moderate
%GR regime. The
%fast, nonlinear decay in semimajor axis does not occur until $t\gtrsim 10000$
%Myr, because it is not until then that the circulating trajectory can be
%well-described as low-$\jmax$ (panel (g)). 
%In a moment we will zoom in further on this very late-stage evolution and look
%at panels (a)-(d) more closely. For %now, to conclude the discussion of Figure
%\ref{fig:Numerical_Initially_ModLib} we consider 
Panels (j) and (k) show that the binary also exhibits the expected asymptotic behavior for $ \Delta a(a)$ and $\tsec(a)$ at small semimajor axes (i.e. in regime $\mathcal{D}$). This is unsurprising since we know from panel (g) that the
binary has reached low-$\jmax$ circulation by this stage. 
On the other hand, the large-$a$ scalings $\vert \Delta a \vert \propto
a^{-3/2}$ (equation \eqref{eqn:Delta_a_scaling_weak}) and $\tau_a \propto a$ (equation \eqref{eqn:decay_time_weak}) are never cleanly realized,
because the binary
does not begin its life in the weak GR regime.

%%%%%%%%%%%%%%

Let us now turn to Figure \ref{fig:Numerical_Initially_ModLib_zoom},
in which we zoom in on panels (a)-(d) of Figure
\ref{fig:Numerical_Initially_ModLib}, focusing on $t \gtrsim 6000$
Myr. In panel (b) we no longer show the $\jmax = j_+$ solid red line, since we
know that for this time range the binary is certainly on a circulating trajectory.
However, we have added a green dashed line that shows the minimum eccentricity
that would be obtained if the binary was on a high-$\jmax$ circulating trajectory, i.e.
with $\jmax = j_0$ (equation \eqref{eqn:jmax_type_I}).  We have also
added a blue dashed line showing the low-$\jmax$ solution $\jmax = -\sigma j_-$
(equation \eqref{eqn:jmax_type_II}). We see that until around $8000$
Myr the evolution is best described as a high-$\jmax$ circulating trajectory with $\jmax =
j_0$. There is then a transitional stage around $t \approx 9000$ Myr wherein neither
high-$\jmax$ nor low-$\jmax$ is a good description (this corresponds to $j_0^2$ approaching
and then crossing zero from above in Figures
\ref{fig:jmax_circulating} and
\ref{fig:jPlus_of_a_Gamma_Regime_I}). After $t \approx 9400$ Myr the
evolution is quite well-described as a low-$\jmax$ circulating trajectory,
$j_\mathrm{max}=-\sigma j_-$. On the other hand there is small systematic error
in this prediction, which we will explain momentarily.

%%%%%%%%%%%%%%

We finish our discussion of Example 3 by zooming in on the very latest stage of
the evolution, $t \gtrsim 9400$ Myr, which we plot in Figure
\ref{fig:Numerical_Initially_ModLib_zoom2}. It is clear from this
figure that at these late times the conservation of $\pmin$ (panel (d)) begins
to fail, evolving from its value on the red dashed line ($\pmin(t=0) \approx
10^{-3.11}$ AU) towards a slightly larger value. The reason for the evolution of $\pmin$ is that 
by this stage $\emin$ has got so large that one cannot think of GW emission as being confined to just the peak-eccentricity portion of a secular cycle and negligible elsewhere.  Instead there is a non-negligible amount of emission throughout the whole cycle, so $a$ decays continuously rather than in a step-like fashion, as can be confirmed by zooming in on panel (a).
As a result, the binary's semimajor axis upon entering the peak of its next secular cycle is slightly smaller than when it left the peak of the previous cycle, so $\epsGR$ is slightly larger and hence the binary achieves a larger value of $\pmin \approx a\jmin^2/2$ (see also \S3.3 of
\citet{Wen2003-jf}). 
%\footnote{There is a different way to think about this in the context of
%\S\ref{sec:conservation_laws}: as $\emax$ is diminished from one
%secular cycle to the next, the approximation $\md p/\md a \approx 0$ --- which
%was derived in the limit $e\to 1$ --- gets gradually worse.}. 
The magnitude of
the oscillations in $p$ also diminish with time until, after around $t=10600$
Myr, the binary reaches the strong GR regime and the oscillations are
quenched. As predicted in \S\ref{sec:conservation_laws} the value of
$p$ itself then remains effectively constant almost all the way to merger,
taking a value\footnote{\citet{Wen2003-jf} estimated that $p_\mathrm{strong}/\pmin$
should lie in the approximate range $(1, 3)$ --- see her equation (31) and the
surrounding discussion.  In fact, the following simple physical argument
suggests the value ought to be $\approx 2$ \citep{Ford2006-qj}. Since GW
emission is very poor at dissipating angular momentum, $J\equiv \sqrt{G(m_1+m_2)
a(1-e^2)} = \sqrt{G(m_1+m_2) p(1+e)}$ is roughly constant during this phase. When
$p=\pmin$ we have $e\approx 1$, whereas upon circularisation we have
$p=p_\mathrm{strong}$ and $e\sim 0$; thus $2\pmin \approx
p_\mathrm{strong}$, i.e. $p_\mathrm{strong}/\pmin \approx 2$. For all numerical
examples presented in this paper, $p_\mathrm{strong}/\pmin \in (1.4, 1.8)$.}
$p_\mathrm{strong} \approx 10^{-2.86} \mathrm{AU}\approx 1.8 \pmin(t=0)$. We see from panel
(b) that the resulting underestimate of $\pmin$ at these late times leads to a
slight overestimate of both the maximum and minimum eccentricities (the blue and
red dashed lines each sit slightly too low in this panel) --- hence the
systematic error in the prediction $\jmax = -\sigma j_-$ which was computed
using the $\pmin$ value from $t=0$. Finally, we note that the conservation of
$\imin$ also fails in these latter stages of the evolution (panel (c)), with
$\cos i_\mathrm{min}$ undergoing a decrease from $\cos i_\mathrm{min}(t=0) =
0.265$ to roughly $\cos i_\mathrm{strong} \approx 0.20$. This change --- which
occurs for the same reason as that in $\pmin$, and is also discussed in \S3.3
of \citet{Wen2003-jf} --- does not make a significant difference to our analysis
because the $\cos^2\imin$, $\sin^2 \imin$ terms in e.g. equations
\eqref{eqn:jplus_of_a}-\eqref{eqn:j0_of_a} are already
dominated by the $(d/a)^{7/2}$ terms by this stage.
%%%%%%%%%%%%%%%%%%%%%%%%%%%%%%%%%%%%%%%%%%%%%%%%
%%%%%%%%%%%%%%%%%%%%%%%%%%%%%%%%%%%%%%%%%%%%%%%%

\subsection{Example 4: $\Gamma =0.176 < 1/5$.}
\label{sec:Numerical_Example_4}
%%%%%%%%%%%%%%%%%%%%%%%%%%%%%%%%%%%%%%%%%%%%%%%%
%%%%%%%%%%%%%%%%%%%%%%%%%%%%%%%%%%%%%%%%%%%%%%%%

%%%%%%%%%%%%%%%%%%%%%%%%%%%%%%%%%%%%%%%%%%%%%%%%
\begin{figure*}
\centering
   \includegraphics[width=0.99\linewidth]{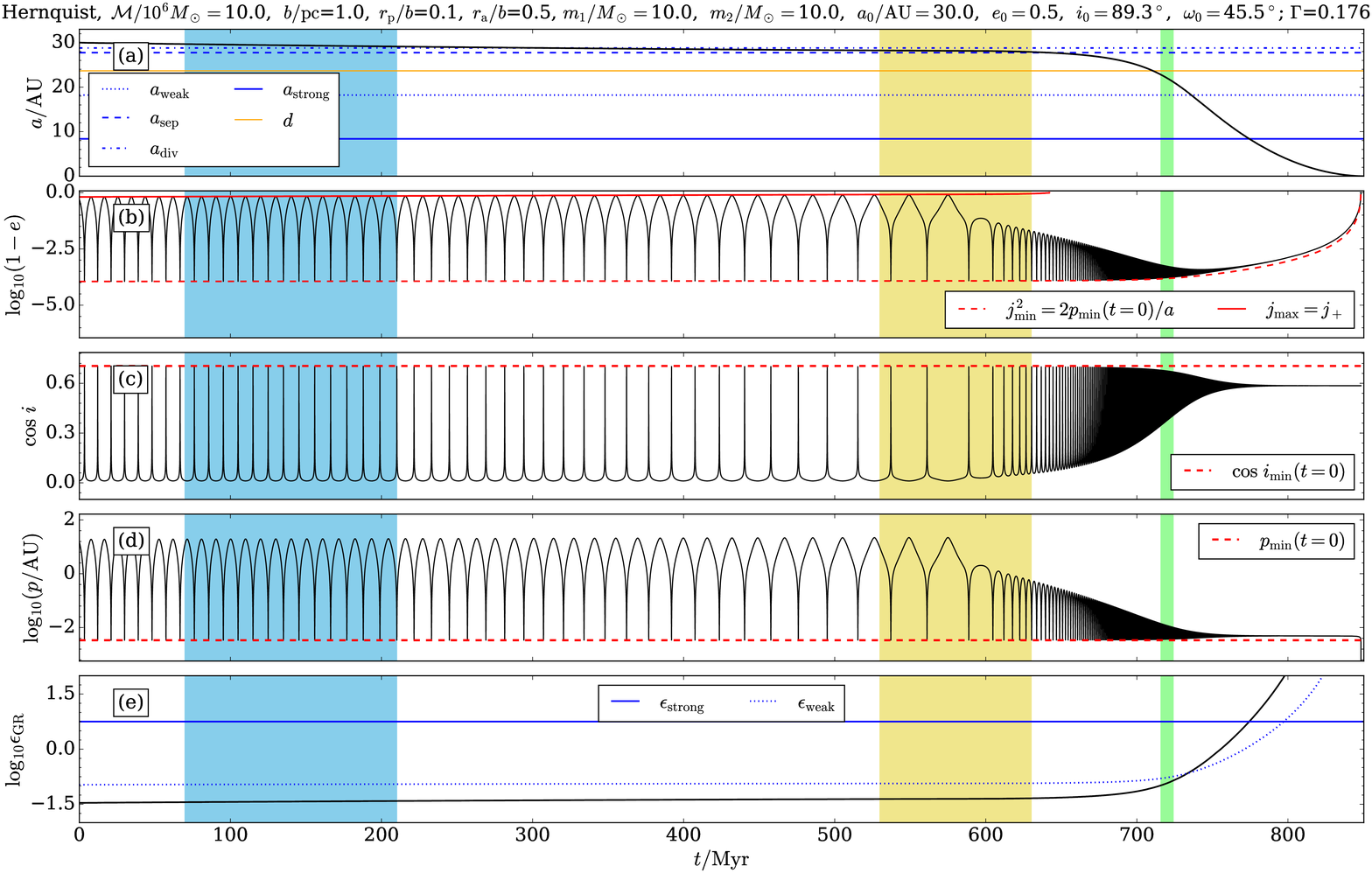}
    \includegraphics[width=0.99\linewidth]{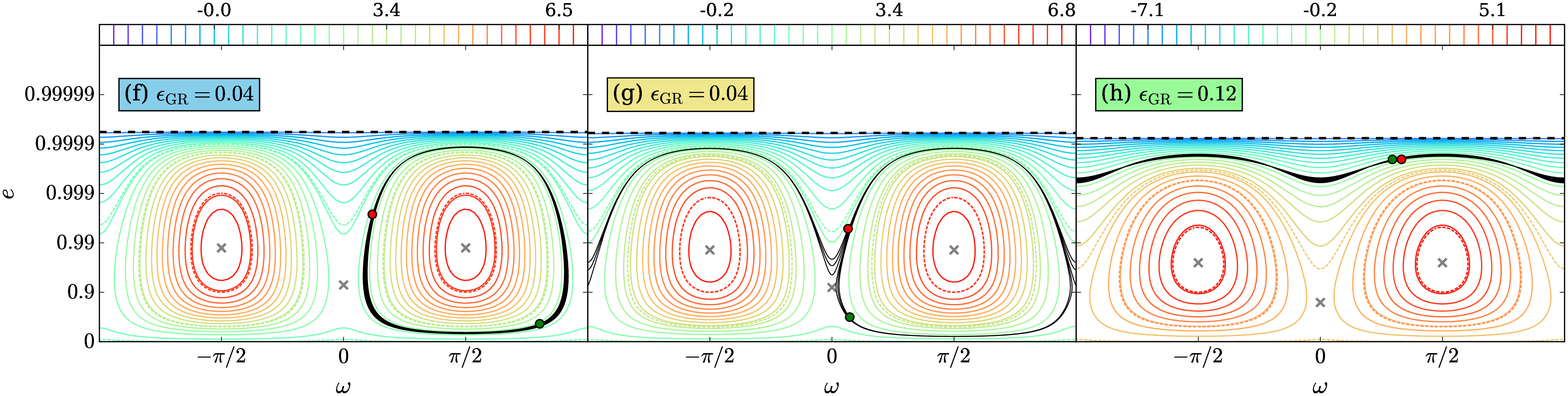}
    \includegraphics[width=0.99\linewidth]{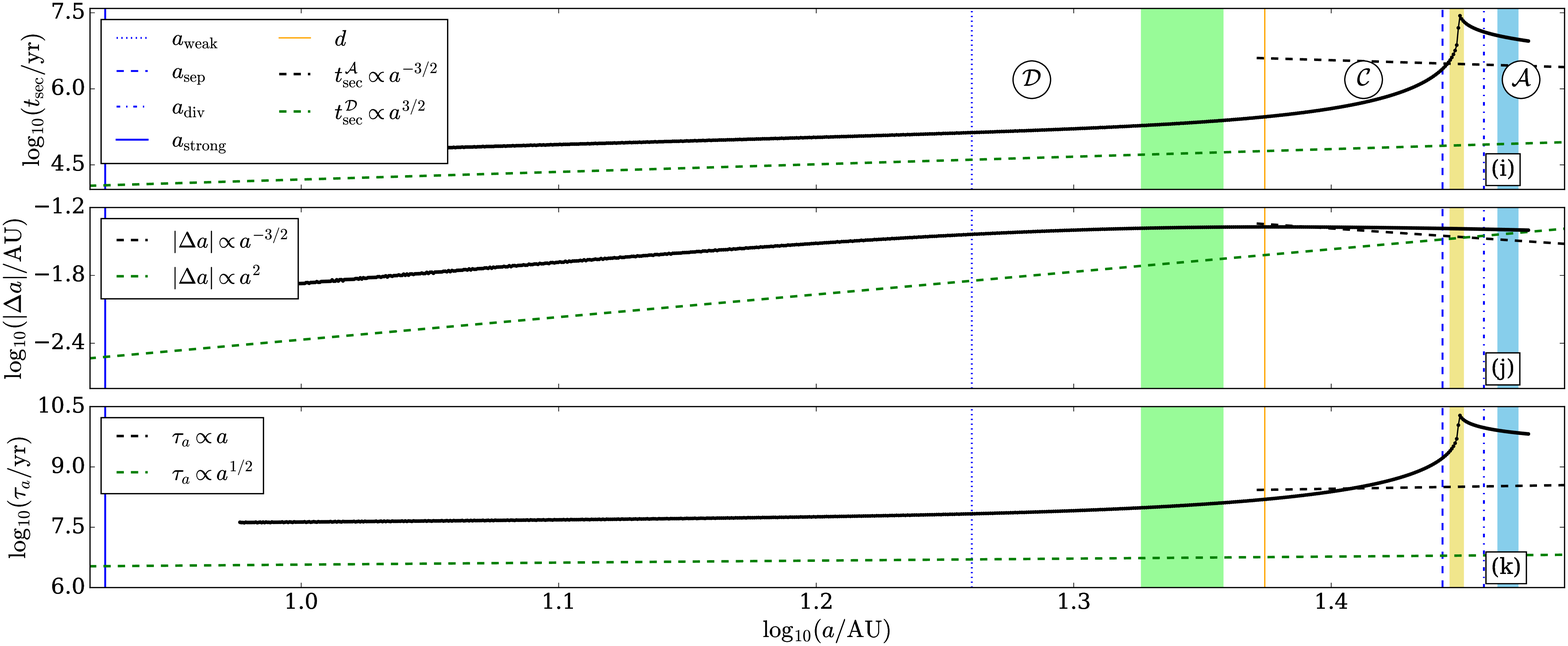}
\caption{\textit{Example 4}. Slow merger in the $0< \Gamma \leq 1/5$ regime.
The binary orbits an $\mathcal{M}=10^7 M_\odot$ Hernquist cluster.
The outer orbit is much smaller than in Examples 2 and 3, giving $\Gamma=0.176 < 1/5$.}
\label{fig:Numerical_Gampt176}
\end{figure*}
%%%%%%%%%%%%%%%%%%%%%%%%%%%%%%%%%%%%%%%%%%%%%%%%
In Figure \ref{fig:Numerical_Gampt176} we provide one more Example,
this time of a $m_1=m_2=10M_\odot$ binary orbiting a $\mathcal{M}=10^7M_\odot$
Hernquist cluster. We take new initial parameters for the inner orbit, as well
as a much smaller outer orbit $(\rp/b, \ra/b) = (0.1, 0.5)$, resulting in
$\Gamma = 0.176 < 1/5$.

%the `fencepost' pattern to the $e$ and $t$ evolution in panels (b) and (d) is
%just symptomatic of the phase space structure for $0 < \Gamma \leq 1/5$:
%indeed the shape of the contours is inverted (see XX).
This binary begins its life in the weak GR regime
($a>a_\mathrm{weak}$), and is initially on a librating phase space trajectory ($a>
a_\mathrm{sep}$). It moves into the circulating regime ($a< a_\mathrm{sep}$) at
around $t=600$ Myr and then into the moderate regime ($a< a_\mathrm{weak}$) at
around $720$ Myr, and by around $860$ Myr it has merged.
%By now it will come as
%no surprise that the semimajor axis evolution is qualitatively very similar to
%that in Example 2 (for $\Gamma > 1/5$), with a roughly linear decrease in $a(t)$
%in the weak GR regime and a steeper, nonlinear decay thereafter.  Also
Unsurprisingly, the secular period increases with time while the phase space
orbit librates (regime $\mathcal{A}$ in panel (i)), and decreases with time once it begins to circulate (regimes $\mathcal{C}$ and $\mathcal{D}$). There is no regime $\mathcal{B}$ in this example, which follows from the fact that in this case $a_\mathrm{sep}<a_\mathrm{div}$, so the divergence at $a=a_\mathrm{div}$ does not occur while the binary is on a circulating trajectory.
In fact we always have $a_\mathrm{sep}<a_\mathrm{div}$ for $0 < \Gamma < 1/5$, as guaranteed by equation \eqref{eqn:asep_over_adiv}.
However we emphasize that this behavior is not limited to $\Gamma < 1/5$, and in fact it is easy to find examples of binaries with $\Gamma > 1/5$ that 
have $a_\mathrm{sep}/a_\mathrm{div} < 1$ and hence also have no $\mathcal{B}$ regime.

It is notable in this Example that compared to Examples 1-3, the transition in the behavior of $e, i$ and $p$ around $600$Myr is very abrupt. At
the beginning of the yellow shaded time interval the binary is on a librating
orbit (starting at the green dot in panel (g)); then at around $600$ Myr it
drifts through the separatrix and joins the family of circulating trajectories; and by the end of the yellow interval it is on a very
high eccentricity circulating trajectory (ending at the red dot).  
In other words, at some point around the middle of the yellow time interval
the binary effectively `jumps' from libration to low-$\jmax$ circulation.
The reason for this jump is that like in other examples, 
the binary inevitably ends up on a high-eccentricity circulating trajectory, and for $0< \Gamma< 1/5$ the 
eccentricity of such a trajectory is forced to be larger than that of the saddle point at $\omega=0$,
namely $e_{\mathrm{f},0}$ --- see Paper III. % (Also, it is
%perhaps not surprising that low-$\jmax$ circulation is reached quickly given the lack
%of high-$\jmax$ solutions for $\Gamma<1/5$ shown in Figure
%\ref{fig:jmax_circulating}). 
%One result of this sharp transition is
%that the change in minimum eccentricity behavior (panel (b)) is very abrupt
%around $600$ Myr, rather than smooth as it was for $\Gamma > 1/5$. 
Thereafter
the evolution matches the usual low-$\jmax$ behavior as seen for
$\Gamma > 1/5$, followed by a merger.

Finally, at the extremes of panels (i)-(k) we see the expected asymptotic
behavior for regimes $\mathcal{A}$ and $\mathcal{D}$ taking shape. However the proper scalings are never fully
developed at large $a$, simply because at $t=0$ the binary is
too close to the separatrix and to $a_\mathrm{div}$ for $j_+$, $j_0$ etc. to be
considered near-constant.

%What is remarkable in this case is that these vanishingly tiny
%%eccentricity oscillations, normally symptomatic of the strong GR regime, happen
%well before $\epsGR$ reaches $\epsstr$. Again this is a consequence of the
%distinct phase space structure for $0<\Gamma\leq 1/5$ as compared to $\Gamma
%>1/5$. 

%%%%%%%%%%%%%%%%%%%%%%%%%%%%%%%%%%%%%%%%%%%%%%%%
%%%%%%%%%%%%%%%%%%%%%%%%%%%%%%%%%%%%%%%%%%%%%%%%
\section{Discussion}
\label{sec:Discussion}
%%%%%%%%%%%%%%%%%%%%%%%%%%%%%%%%%%%%%%%%%%%%%%%%
%%%%%%%%%%%%%%%%%%%%%%%%%%%%%%%%%%%%%%%%%%%%%%%%
%%%%%%%%%%%%%%%%%%%%%%%%%%%%%%%%%%%%%%

In this paper, we have extended our theory of secular dynamics of binaries
in stellar clusters by accounting for the effect of GW emission.  We have
demonstrated that cluster tides are capable of driving binaries to very high
eccentricity, where they can emit GW bursts, shrink in semimajor axis, 
and ultimately merge.  
Our results also encompass --- and in several ways extend --- the theory of (test-particle, quadrupole, doubly-averaged) LK-driven compact object binary
mergers, which is recovered exactly in the limit $\Gamma=1$.

Throughout this paper we have focused on understanding the physics of `slow
mergers', i.e. those mergers that require many secular periods, but would not have
occurred within a Hubble time had the cluster tidal perturbation not been present. 
This meant focusing on initially high-inclination systems for which eccentricity excitation is possible.
We have also ignored octupole effects, short
timescale fluctuations, stellar flybys and the like (see Paper II for discussion). Yet even in this relatively
simple setting we have seen that the evolution of a binary from `birth' to
merger can be rather complex and is, in general, analytically intractable. Key to
making analytical progress was our identification of several asymptotic
regimes both in GR strength (weak, moderate, strong) and in 
phase space trajectory (librating, high-$\jmax$ circulating, low-$\jmax$ circulating). We
emphasize that the analytical results derived in these regimes are only approximations.
In practice
the boundaries between asymptotic regimes are blurry and poorly separated,
especially in $a$-space. Nevertheless, they have been sufficient for our
purpose, which was to gain analytical --- and consequently, physical --- insight
into an important class of problems that have traditionally been outsourced to a
computer. In Table \ref{tab:summary} we summarize our approximate asymptotic results.
%%%%%%%%%%%%%%%%%%%%%%%%%%%%%%%%%%%%%%
\begin{table*}
	\centering
	\caption{Summary of key asymptotic results}
	\label{tab:summary}
	\begin{tabular}{lcccc} % 4 columns, alignment for each
		\hline
		  &  Weak GR \& librating (regime $\mathcal{A})$ & Moderate GR \& low-$\jmax$ circulating (regime $\mathcal{D}$) & Strong GR &\\
		  \hline
Conserved quantities & $\pmin, \imin$ & $\pmin, \imin$ & $p$, $i$ &\\
$\tsec$ & $\propto a^{-3/2}$ & $\propto a^{1/2}$ & N/A &\\
	$\Delta a$ & $\propto a^{-3/2}$ & $\propto a^2$ & N/A &\\
		$\tau_a$ & 
		$\Lambda_\mathrm{weak} U^{-1}a$& 
		$\Lambda_\mathrm{mod} U^{-1}(2\pmin)^{1/2}a^{1/2}$
		& 
		$(8/5) U^{-1}(2\pmin)^{1/2}a^{1/2}$ &\\
		\hline
	\end{tabular}
\end{table*}
%%%%%%%%%%%%%%%%%%%%%%%%%%%%%%%%%%%%%%

To conclude this paper, we first discuss in
\S\ref{sec:merger_timescale} the implications of our results for the
calculation of the total merger timescale.  Finally in \S\ref{sec:LK} we
discuss our work more broadly in the context of previous studies of LK-driven
mergers.

%GW inspirals are possible because there is a window in between $\epsGR^2$ and
%%$\epsGR$ where the GW emission is possible and eccentricity maxima are not be
%strongly quashed by GR precession.  That's because $\emax$ is quadratic in
%$\epsGR$ not linear, which allows the 2.5PN terms to kick in before the max
%eccentricity is killed off for good. I think this is discussed in
%\citet{Wen2003-jf}.

%Eccentricity `trapping' a very common phenomenon that is not exclusive to GW
%inspirals --- see also Anderson work on HJs. Trapping is discussed by
%\citet{Liu2018}.
%%%%%%%%%%%%%%%%%%%%%%%%%%%%%%%%%%%%%%%
\subsection{Merger timescale}
\label{sec:merger_timescale}
%%%%%%%%%%%%%%%%%%%%%%%%%%%%%%%%%%%%%%%

Secular dynamics of binaries including GW emission is a problem that has been
considered many times in the LK context for hierarchical triple systems.
Many LK studies that include GW emission 
are focused on the resulting observable merger rate,
i.e. the number of binaries that merge per cubic Gpc per year in the local universe. To compute
such a rate --- as we did for cluster tide-driven compact object mergers in
\cite{Hamilton2019-mq} --- one needs to know the time it
takes for a given binary to merge as a function of its initial conditions. There
are basically two ways to approach this problem. One can either integrate the
equations of motion (DA, SA or N-body) directly and read off the merger time
from the simulation, or one can seek an approximate (semi-)analytic formula that
parameterizes the merger time in terms of those initial conditions (which can be checked using direct
numerical integration for a small number of cases). The latter
approach is obviously much faster when one is dealing with millions or billions
of binary initial conditions in a Monte-Carlo population synthesis.

The merger time formula usually used in compact object merger
calculations in the LK literature is
%%%%%%%%%%%%%%%%%%%
\begin{align}
    T_\mathrm{m} &\equiv
    T_\mathrm{m}^\mathrm{iso}(a(0),e_\mathrm{max}(0))
    \times (1-e^2_\mathrm{max}(0))^{-1/2} 
   \nn  \\ &= \frac{3c^5a(0)^4}{85G^3(m_1+m_2)m_1m_2} (1-e_\mathrm{max}(0))^{3},
   \label{eqn:Tm}
\end{align}
%%%%%%%%%%%%%%%%%%%
where
%%%%%%%%%%%%%%%%%%%
\begin{align} 
    T_\mathrm{m}^\mathrm{iso}(a,e) &= \frac{3c^5a^4}{85G^3(m_1+m_2)m_1m_2} (1-e)^{7/2},
    \label{eqn:Tm_iso}
    \end{align}
%%%%%%%%%%%%%%%%%%%
is simply the merger time of an isolated binary with initial semimajor axis $a$
and very high initial eccentricity $e \approx 1$ \citep{Peters1964-fb}. The
formula \eqref{eqn:Tm} is typically justified via the following
heuristic argument \citep{Miller2002-co, Thompson2011-wv, Liu2018-kg,Randall2018-uq}. First, one assumes
that the GW emission is negligible except around $e\approx e_\mathrm{max}$, and
so the total amount of time that needs to be spent at $e\approx e_\mathrm{max}$
before the binary merges is $\approx T_\mathrm{m}^\mathrm{iso}(e_\mathrm{max})$.
But the amount of time that the binary actually spends in the vicinity of
$e_\mathrm{max}$ in each secular cycle is  $\approx \jmin \tsec \equiv
(1-\emax^2)^{1/2} \tsec$; thus the number of secular cycles required until the
time spent around $e_\mathrm{max}$ accumulates to
$T_\mathrm{m}^\mathrm{iso}(e_\mathrm{max})$ is $
T_\mathrm{m}^\mathrm{iso}(e_\mathrm{max})/[(1-\emax^2)^{1/2}\tsec]$. To get the
total merger time we multiply this by $\tsec$.  Finally, evaluating everything at
$t=0$ we get the formula \eqref{eqn:Tm}.

Of course, this heuristic derivation can be criticized on several levels. For instance,
it makes no distinction between the values of $e_\mathrm{max}$, $\tsec$ at $t=0$
and their values at later times, even though we know
(\S\S\ref{sec:GW_emission}-\ref{sec:phase_space_evolution}) that both of these quantities vary with
$a$.  Also, it does not accurately treat the behavior of $e$ around
$e_\mathrm{max}$, instead assuming that $e$ is precisely equal to
$e_\mathrm{max}$ within a discrete time window which lasts for
$(1-e_\mathrm{max}^2)^{1/2}\tsec$, and that GW emission is negligible outside that window.

In reality, we know from Paper II that even in the absence of GR precession, the fraction of each secular period spent in the
vicinity of high eccentricity is not precisely proportional to $\tsec$. Indeed, combining equations (34) and (59) of Paper II with equations \eqref{eqn:t_sec_weak} and \eqref{eqn:dimensionless_quantity_weak} of the present paper, 
we find that the time for $j$ to change from $\jmin$ to $\sqrt{2}\jmin$ in the $\epsGR \to 0$ limit (in regime $\mathcal{A}$) is
\begin{equation}
    t_\mathrm{min} \approx  \tsec^\mathcal{A} \jmin \times (2\Lambda_\mathrm{weak})^{-1}.
    \label{eqn:tmin_tsec_relation}
\end{equation}
The factor $(2\Lambda_\mathrm{weak})^{-1}$ can be significantly different from unity if the binary is near a separatrix --- see Paper II. Finally, the expression for $\tmin$ becomes even more complicated when we do include GR precession, especially for large values of $\sigma$ and/or $\kappa$. 
%Finally, it does not contain the information that $\tsec$ is not constant but changes dramatically throughout the evolution (although this turns out not to matter too much, since the crucial quantity is the \textit{fraction} of $\tsec$ spent near $\emax$ rather than $\tsec$ itself).

Despite these shortcomings, the formula \eqref{eqn:Tm} actually works
reasonably well in practice (to within a factor of order unity) when compared to direct
numerical integration of the (DA, test-particle quadrupole) equations of motion for triple systems
\citep{Thompson2011-wv,Liu2018-kg,Randall2018-uq}. To see why this might be the case, we
now show that one can actually derive the formula \eqref{eqn:Tm} in a slightly
less hand-waving fashion using the results of this paper. 
For slow mergers, a rather general formula for the merger time $t_\mathrm{m}$ is found by integrating equation
\eqref{eqn:t_of_a_WTM} from $t=0$ to $t= t_\mathrm{m}$:
\begin{align}
    t_\mathrm{m} \approx \int_{a(0)}^0 \md a' \frac{\tsec(a')}{\Delta a(a')} = \int_{a(0)}^0 \md a' \frac{\tau_a(a')}{a'}
\label{eqn:Tm_general}
\end{align}
(see equation (57) of \citet{Randall2018-uq}). 
% Of course we cannot claim that this
%will be completely accurate because it does not account for the brief time the
%binary spends in the strong GR regime.
Of course, as it stands \eqref{eqn:Tm_general} is
an entirely impractical formula given the
complexity of the general analytic expression  for $\tau_a$
that must then be integrated over.
To make progress we assume
that the majority of a slow merger is spent in the weak GR regime,
and that by
ignoring the time spent in the moderate and strong GR regimes we do not impart
any major error (though we note that this approximation would fail in Figure \ref{fig:Numerical_Initially_ModLib}, for example). Then a decent approximation to $\tau_a$ is given by equation \eqref{eqn:decay_time_weak}. Plugging this into \eqref{eqn:Tm_general} gives 
\begin{align}
    t_\mathrm{m} \approx \Lambda_\mathrm{weak} {U}^{-1} a(0),
    \label{eq:Tm_weak}
\end{align}
with $\Lambda_\mathrm{weak}$ and $U$ given in equations \eqref{eqn:dimensionless_quantity_weak} and \eqref{eqn:decay_speed_characteristic} respectively. Of course, $\Lambda_\mathrm{weak}$ accounts for the fact that
the time spent at highest eccentricity is not precisely proportional to $\tsec$ --- see equation \eqref{eqn:tmin_tsec_relation}.
Since $\pmin(t)
\equiv a(t) \times (1-e_\mathrm{max}(t))$ is conserved throughout a slow merger,
we can substitute in equation \eqref{eqn:decay_speed_characteristic} the expression
\begin{align}
    (2\pmin)^3 &=\nn 8(1-\emax(0))^3 a(0)^3 
    \\
    & \approx (1-\emax^2(0))^3 a(0)^3,
    \label{eqn:2pminapprox}
\end{align}
where in the second line we assumed $e_\mathrm{max}(0) \approx 1$. 
Comparing
the result to \eqref{eqn:Tm}, we find that
in this approximation the merger occurs at time
\begin{align}
t_\mathrm{m} \approx (5\Lambda_\mathrm{weak}/16) \times T_\mathrm{m},
\end{align}
Thus provided $\Lambda_\mathrm{weak} \sim 1$, we recover the standard estimate of the merger
 timescale \eqref{eqn:Tm} to within a factor of order unity.
%Indeed, suppose that instead of doing the integral in \eqref{eqn:Delta_a_formal} properly, 
%we just evaluated the integrand at $\jmin$ and multiplied by the naive timescale $(\jmin \tsec)/2$.  
%Then we would find $\Delta a = -2 \lambda_1 a^{-3} \tsec \jmin^{-6} = -2 \lambda_1 (2\pmin)^3 \tsec$, where we used the approximation \eqref{eqn:jmin_of_a}.  We could then get $t_\mathrm{m}$ by plugging this expression for $\Delta a$ into the first equality in \eqref{eqn:Tm_general}, \textit{causing the factor of $\tsec$ to cancel out as expected}; 
%performing the resulting integral over $a$ we would then find $t_\mathrm{m} = (16/5) U^{-1} a(0) = T_\mathrm{m}$.
%In the more nuanced calculation that gave equation \eqref{eq:Tm_weak}, 
%the quantity $\Lambda_\mathrm{weak}$ accounts for the detailed evolution of $j(t)$ near peak eccentricity; in our 
%dirty version, $\Lambda_\mathrm{weak}$ is just replaced with a fudge factor $16/5$. In other words there is always some factor multiplying $U^{-1}a(0)$ even after the $\tsec$ values have canceled out; $\Lambda_\mathrm{weak}$ is just a more accurate one than $16/5$. }

If anything, one might expect that $T_\mathrm{m}$ will be an overestimate of the
`true' merger time (even if one calculates this `true' time by integrating the
DA quadrupolar equations, i.e. ignoring SA effects, octupolar terms, and so on).
That is because, as we saw in \S\ref{sec:secular_timescale} and
\S\ref{sec:Numerical_Examples}, the decay of $a(t)$ speeds up
substantially once the binary reaches its low-$\jmax$ circulating phase in the
moderate GR regime (see equation \eqref{eqn:decay_time_mod}). Thus,
approximating the entire decay using the weak GR equation
\eqref{eqn:decay_time_weak}) may seem overly conservative. It is
therefore surprising to note Figure 8 of \citet{Thompson2011-wv} and 
Figure 5 of \citet{Randall2018-uq}, both of which suggest
that $T_\mathrm{m}$ typically \textit{underestimates} the true (DA) merger
time by a factor $\sim 2$ for compact object binaries in hierarchical triple
systems. In future work it might be interesting to understand more deeply the
reason for this trend.  It may also be profitable to try to use the results of
this paper to calibrate a merger timescale formula that is more accurate than
\eqref{eqn:Tm} --- even an estimate with typical in error at the
level of only a few tens of percent would be a significant improvement. On the
other hand, for realistic calculations such a formula may be of limited interest,
since the true merger time can be greatly shortened when one includes
sub-secular (e.g. `singly-averaged') effects, octupolar terms, and so on --- see
e.g. \citet{Antonini2014-oa,Grishin2018-da}.

%%%%%%%%%%%%%%%%%%%%%%%%%%%%%%%%%%%%%%%%%
\subsection{Relation to studies of LK-driven mergers}
\label{sec:LK}
%%%%%%%%%%%%%%%%%%%%%%%%%%%%%%%%%%%%%%%%%

As discussed in \S\ref{sec:merger_timescale}, most LK studies `solve'
the problem of GW-assisted mergers either by direct numerical integration or by
stating and then evaluating the merger time formula \eqref{eqn:Tm}
after calculating $e_\mathrm{max}$ from simple theory.  
There does not exist much in the literature that lies in between these extremes,
in which an attempt is made to understand in detail the physics of each stage of
the merger or to derive analytic results in specific asymptotic regimes as we
have done here. Nevertheless, some of the key ideas covered in this paper have been
considered by other authors, as we now describe. 

A central result of \S\ref{sec:GW_emission} was the approximate
conservation of $\pmin$ and $\imin$ during slow mergers:  this was ultimately
what allowed us to express various important quantities ($\jmin, \tsec$, etc.) as
functions of $a$. These conservation laws (as well as their breakdown in the
late stages of a slow merger) seem to have first been described in the LK limit by
\citet{Wen2003-jf}. The behavior of $p$ during slow mergers has subsequently been
appreciated as an important diagnostic of different regimes; for instance,
Antonini has followed the $p$ evolution in order to distinguish between `LK
dominated' and `GW dominated' regimes
\citep{Antonini2012-fv,Antonini2014-oa,Antonini2017-od}. Some basic scalings of
$\jmin$, $\tsec$, etc. with $a$ were also written down by e.g.
\citet{Miller2002-co}, \citet{Wen2003-jf}, \citet{Thompson2011-wv}, although none of these
authors venture beyond the weak GR
regime in their analytical efforts, and so did not derive the peculiar results in the moderate GR regime
that we have found here. No other studies have progressed beyond these simple scaling relations, to write down
explicit formulae like we did in \S\ref{sec:phase_space_evolution}.

Another main achievement of the present paper has been to understand the interplay between
the time-evolution of key dynamical quanties like ($a,e$), and the underlying
phase space structure. The fact that a binary initially on a librating phase
space trajectory necessarily transitions into the circulating regime as it
shrinks was first mentioned by \citet{Blaes2002-rx} (although they did not note the
accompanying qualitative change in $\tsec$ behavior). Of course, since LK
theory corresponds to $\Gamma=1>1/5$, no previous authors have noted
the new behavior that arises in the $0<\Gamma\leq 1/5$ regime, e.g. the abrupt changes in phase space trajectory and the associated sharp `kink' in $\tsec(a)$ --- see \S\ref{sec:Numerical_Example_4}.  Furthermore, to our knowledge no LK study has distinguished between high-$\jmax$ and low-$\jmax$ circulating trajectories.

The only LK study we know of to have written down a formula for the decay in
 semimajor axis $\Delta a$ over one secular cycle is \citet{Randall2018-uq} --- see
 their equation (55).  These authors also wrote down an expression (their
 equation (57)) that is essentially the same as our equation
 \eqref{eqn:t_of_a_WTM}, pertaining to the slow evolution of $a$.
In addition, \citet{Randall2018-uq} seem to be the only authors who mentioned that $\tsec$ can sometimes decrease as $a$ shrinks,  
 even though every author who has integrated the equations of motion numerically
 must have encountered this phenomenon. In Appendix \ref{sec:RX18} we look in detail at some of the calculations of \citet{Randall2018-uq}. 
 As we show
 there, \citet{Randall2018-uq} implicitly assumed weak GR and $\sigma \ll 1$ when deriving certain analytical results, so their calculations are not valid outside of this regime. %To
 %our knowledge no authors have derived explicit results in the moderate GR
 %regime until now.

%%%%%%%%%%%%%%%%%%%%%%%%%%%%%%%%%%%%%%%%%

\section{Summary}
\label{sec:Summary}

In this paper we studied the (2.5pN) GW-driven orbital decay and subsequent
merger of binary systems which are torqued to high eccentricity by cluster tides
on secular timescales. We worked in the DA, test-particle quadrupole approximation and
included the effect of (1pN) GR precession in our calculations.  Our results may
be summarized as follows.

\begin{itemize}
\item Cluster tides are capable of torquing binaries to
sufficiently high eccentricity that they emit bursts of GWs and ultimately
merge. Cluster tide-driven eccentricity excitation is therefore a viable
mechanism for producing LIGO/Virgo mergers, similar to LK-driven mergers that
have been widely explored in the past. In fact (test-particle quadrupole DA) LK-driven mergers are simply a special case of the cluster tide-driven
mergers considered here.

\item For slow mergers (those that take place over many secular periods)
there are two approximate conservation laws that hold as the semimajor axis
$a$ decays, namely conservation of the minimum pericentre distance $\pmin  =
a(1-e_\mathrm{max})$ and conservation of the minimum inclination reached
$\imin$. The evolution of a decaying binary through phase space can be
understood in terms of these conserved quantities.
\item We uncovered several asymptotic regimes both in terms of GR strength and phase space morphology.
The different regimes exhibit
different characteristic behaviors of secular timescale $\tsec(a)$, decay
in semimajor axis per cycle $\Delta a(a)$, and consequently the decay timescale $\tau_a(a)$.
\item We re-derived a formula for the merger timescale that has been much used in
LK theory, and provided a more detailed justification for it than those that have
been offered previously. 
\end{itemize}

The insights from this paper will inform future studies of LK-driven and
cluster tide-driven binary mergers. %In addition, some of the ideas developed
%here may provide qualitative insight into other problems where secular forcing,
%apsidal precession and short-range dissipation compete over the dynamics of a
%two-body system. One such problem might be the LK-driven formation of short
%period binaries and hot Jupiters in triple systems, wherein the dissipation is
%not due to GW emission but instead due to internal fluid tidal friction in the
%star(s) and/or planet \citep{Fabrycky2007-tu}.

%%%%%%%%%%%%%%%%%
%%%%%%%%%%%%%%%%%%%%%%%%%%%%%%%%%%%%%%%%%%%
\acknowledgements 

We thank Ulrich Sperhake and Bence Kocsis for comments on an earlier version of this work.
This work was supported by a grant from the Simons Foundation (816048, CH), STFC grant ST/T00049X/1 and Ambrose Monell Foundation (RRR). 

\appendix

%%%%%%%%%%%%%%%%%%%%%%%%%%%%%%%%%%%%%%%%
%%%%%%%%%%%%%%%%%%%%%%%%%%%%%%%%%%%%%%%%%
\section{High eccentricity results without gravitational wave emission}
\label{sec:high_ecc_no_GWs}
%%%%%%%%%%%%%%%%%%%%%%%%%%%%%%%%%%%%%%%%
%%%%%%%%%%%%%%%%%%%%%%%%%%%%%%%%%%%%%%%%

In this section we gather some results from Papers II-III concerning cluster
tide-driven secular dynamics \textit{without} gravitational wave emission (but
including GR precession).  
Though there is nothing strictly new here, it will be useful to have these results gathered 
in one
place and written in a form that makes their meaning transparent.

%\subsection{High eccentricity behavior}
%\label{sec:high_ecc_no_GWs}

Without GWs
the entire eccentricity evolution is dictated by equation (15) of Paper III:
%%%%%%%%%%
\begin{align}  
\frac{\md j}{\md t} =
\pm \frac{6C}{Lj^2} & \Bigg\{
 (25\Gamma^2-1)\left[(j_+^2-j^2)(j^2-j_-^2)-\frac{\epsGR}{3(1+5\Gamma)}j\right] 
\left[j^2(j_0^2-j^2)+\frac{\epsGR}{3(5\Gamma-1)}j\right] \Bigg\}^{1/2},
\label{eq:djdtGR}
\end{align}
%%%%%%%%%%
where $j_\pm^2$, $j_0^2$ are given in equations (16)-(19) of that Paper.
It will  be important that we are able to
derive simple expressions for $j_\pm$, $j_0$ etc. in the high-eccentricity
limit.  If the binary initially has $e$ not close to unity, then to reach high $\emax$ it is necessary to have both $\Theta \ll 1$ and the binary initially in the weak-to-moderate GR regime.
Making these assumptions we can use equations (46) of Paper III, which we repeat here:
%%%%%%%%%%
\begin{align} 
&j^2_+\approx \frac{2\Sigma}{1+5\Gamma} \sim 1,
\,\,\,\,\,\,\,\,\,\,\,\,\,
j^2_-\approx \frac{5\Gamma\Theta}{\Sigma} \sim \Theta \ll 1.
\label{eqn:j_plus_minus_small_Theta}
\end{align}
Moreover, evaluating equations (18)-(19) of Paper III
at $\omega = \pm\pi/2$, $j = \jmin \approx 1$
we get the following equation for $\Sigma$ which did not appear explicitly in Paper III:
\begin{equation}
    \Sigma = 5\Gamma \left( \frac{\Theta}{\jmin^2} + \frac{\epsGR}{30\Gamma \jmin}\right).
    \label{eqn:Sigma_small_Theta}
\end{equation}
Using the approximations \eqref{eqn:j_plus_minus_small_Theta}, \eqref{eqn:Sigma_small_Theta} 
 we find that
we can write $j_\pm^2$ and $j_0^2$ exclusively in terms of $\Gamma$ and the three dimensionless numbers
$\Theta,\epsGR,\jmin$, which are constants when GW emission is ignored:
%%%%%%%%%%
%%%%%%%%%%%%
\begin{align}
j_+^2 &\approx \frac{10\Gamma}{1+5\Gamma}\left( \frac{\Theta}{\jmin^2} + \frac{\epsGR}{30\Gamma\jmin} \right),
\label{eqn:jplusWTM}
\\
j_-^2 &\approx \jmin^2 \left( 1+ \frac{\epsGR \jmin}{30\Gamma \Theta}\right)^{-1},
\label{eqn:jminusWTM}
\\
j_0^2 &\approx \frac{10\Gamma}{5\Gamma -1 }\left( 1-\frac{\Theta}{\jmin^2} - \frac{\epsGR}{30\Gamma\jmin} \right).
\label{eqn:j0WTM}
\end{align}
%%%%%%%%%%%%
We can write down an expression for $\jmin$ by assuming $\Gamma>0$ and that maximum $e$ is achieved at $\omega=\pm\pi/2$. Then (see equation (52) of Paper III):
%%%%%%%%%%
\begin{align} 
j_{\rm min}&=\frac{\gamma j_-}{2}\left[ 1 + \sqrt{1+4\gamma^{-2}}\right]
= \frac{1}{2j_+^2\epsstr} \left[\epsGR + \sqrt{\epsGR^2+\epsweak^2}\right],
\label{eqn:gen_sol_1}
\end{align}
%%%%%%%%%%
where 
\begin{align}
    \epsweak \equiv 6(1+5\Gamma)j_+^2j_- \approx \left(720\Gamma\Sigma\right)^{1/2}\Theta^{1/2}.
    \label{eqn:epsweak}
\end{align}
%At this stage we have not said anything about $\Theta$, nor about the size of
%$\epsGR$ (i.e. we have not specified to the weak/moderate/strong GR regime).
%However, we know from Paper III that for a binary to
%start at $e\sim 0$ and to end up at $e \to 1$, it must initially reside in the
%weak-to-moderate GR regime and have $\Theta \ll 1$. Supposing this is true, for
%all orbits that maximise their eccentricity at $\omega = \pm\pi/2$ the solution
%$\jmin$ is given by equation \eqref{eqn:gen_sol_1}. 
It follows that for weak GR ($\epsGR \ll \epsweak$),
%%%%%%%%%%%%
\begin{align}
\label{eqn:scaling_weak}
\jmin \approx j_- \sim \Theta^{1/2} \gg \epsGR, \,\,\,\,\,\,\,\,\,\,\,\,\,\mathrm{(weak \,\,\, GR)}.
\end{align}
%%%%%%%%%%%%
Similarly, for moderate GR ($\epsweak \ll \epsGR \ll \epsstr$) and using the constancy of $\cos \imin$ (\S\ref{eqn:inclination_conservation}) we have 
%%%%%%%%%%%%
\begin{align}
\label{eqn:scaling_moderate}
\jmin \sim \epsGR \sim \Theta^{1/2}, \,\,\,\,\,\,\,\,\,\,\,\,\,\mathrm{(moderate \,\,\, GR)}.
\end{align}
%%%%%%%%%%%%
It follows from equations \eqref{eqn:jplusWTM},
\eqref{eqn:jminusWTM},
\eqref{eqn:scaling_weak} and\eqref{eqn:scaling_moderate}
that in the weak-to-moderate regime, provided $\Gamma \sim 1$, we always have
$j_+^2 \sim 1$ and $j_-^2 \lesssim \jmin^2 \sim \Theta \ll 1$. 

Note that in \S4 of Paper III we already arrived at the weak GR result \eqref{eqn:scaling_weak}. However we did not arrive at the same moderate GR result \eqref{eqn:scaling_moderate}; instead we found $\jmin \sim \epsGR \gg \Theta^{1/2}$.  The reason for the discrepancy is that in Paper III we implicitly 
assumed that $\Theta$ was kept fixed while $\epsGR$ was increased.
However, when the decay of $a$ is due to GW emission, one cannot change $\epsGR$
without also changing $\Theta$ --- see equation \eqref{eqn:Theta_of_a}. Accounting for this fact leads to \eqref{eqn:scaling_moderate}. 

%%%%%%%%%%%%%%%%%%%%%%%%%%%%%%%%%%%%%%%%%%%%%%%%%%%%%%%%%%%%
%%%%%%%%%%%%%%%%%%%%%%%%%%%%%%%%%%%%%%%%%%%%%%%%%%%%%%%%%%%%
%%%%%%%%%%%%%%%%%%%%%%%%%%%%%%%%%%%%%%%%%%%%%%%%%%%%%%%%%%%%

\subsubsection{Maximum angular momentum}

For all the phase space trajectories in which we are interested, $\jmin$ is given by the
same formula \eqref{eqn:gen_sol_1}. However it turns out (\S\S\ref{sec:secular_timescale}-\ref{sec:Numerical_Examples}) that to understand the behavior of slow mergers one must distinguish between qualitatively different
trajectories, and in particular to know their maximum angular momentum $\jmax$
(corresponding to minimum eccentricity $e_\mathrm{min}$), so we will devote some effort to 
this now.

The maximum $j$ can either be found at $\omega = \pm\pi/2$ (if the phase space
trajectory librates) or at $\omega=0$ (if it circulates). In the librating case
we find\footnote{To see this, recall that for librating trajectories $\jmax$ is a
solution to the quartic found by setting the first square bracket in
\eqref{eq:djdtGR} to zero. We can simplify this quartic by noting that,
since librating trajectories loop around fixed points at $\omega = \pm\pi/2$, they
necessarily have $\jmax^2 > j_{\mathrm{f},\pi/2}^2$.  We know from Figure
3 of Paper III that $j_{\mathrm{f},\pi/2}^2  \gg \Theta$, and from
equations
\eqref{eqn:scaling_weak}-\eqref{eqn:scaling_moderate} that
$\Theta \gtrsim j_-^2$, so we can ignore $j_-$ in the quartic and write
%%%%%%%%%%%
\begin{align}
 \jmax^3 - j_+^2 \jmax + \epsGR/\epsstr \approx 0 .
\end{align}
%%%%%%%%%%%%
Since in the weak-to-moderate GR regime we have $j_+^2 \sim 1$ and $\epsGR \ll
\epsstr$, the solution is obviously $\jmax \approx j_+$.}
%%%%%%%%%%%%
\begin{align}
\label{eqn:jmax_librating}
\jmax \approx j_+  \,\,\,\,\,\,\,\,\,\,\,\,\,\mathrm{(librating \,\,\, orbits).}
\end{align}
%%%%%%%%%%%%
Finding an approximate expression for $\jmax$ for circulating trajectories is more
complex, because there are two qualitatively different regimes of circulating
trajectory to consider.
The first type of circulating trajectory, which we call `high-$\jmax$', corresponds to
$\jmax \sim 1$ or $e_\mathrm{min} \sim 0$, i.e. the binary undergoes an
order-unity oscillation in eccentricity during each secular cycle. This is the
classic type of circulating trajectory undergone by, for instance, a binary with
$\Gamma > 1/5$ in the weak GR regime starting out with small eccentricity at
$\omega \approx 0$ --- see e.g. Figure
\ref{fig:Numerical_RX18_Fig3}f for illustration. The second
type of circulating solution, which we call `low-$\jmax$', corresponds to $\jmax \ll
1$ or $e_\mathrm{min} \sim 1$, so that the oscillation in eccentricity is
actually rather small despite $\emax$ being large --- see e.g. Figure
\ref{fig:Numerical_RX18_Fig3}h.  In this case we can say that the binary is trapped at high eccentricity.
Low-$\jmax$ circulating trajectorie are
important because every binary passes through this stage while in the moderate
GR regime during a slow merger, as a precursor to the strong GR
regime\footnote{For $0 < \Gamma \leq 1/5$, we know from Paper III
that high-$e$ circulating trajectories are immediately
formed once $\epsGR$ exceeds $6(1-5\Gamma)\Theta^{3/2} \ll
\epsweak$. Thus, one does not
necessarily need to be in the moderate GR regime to have low-$\jmax$ circulating
orbits. However most of our focus in this paper will be on low-$\jmax$ circulating
orbits that exist in the moderate GR regime, which occur for all $\Gamma$.}.

%Indeed, it will turn out that low-$\jmax$ circulating trajectories in the moderate GR
%regime have very similar $a(t)$ decay behavior to those in the strong GR
%regime, whereas high-$\jmax$ orbits show a qualitatively different behavior --- see
%\S\ref{sec:SMA_Evolution}.

\begin{figure}
\centering
\includegraphics[width=0.59\linewidth,clip]{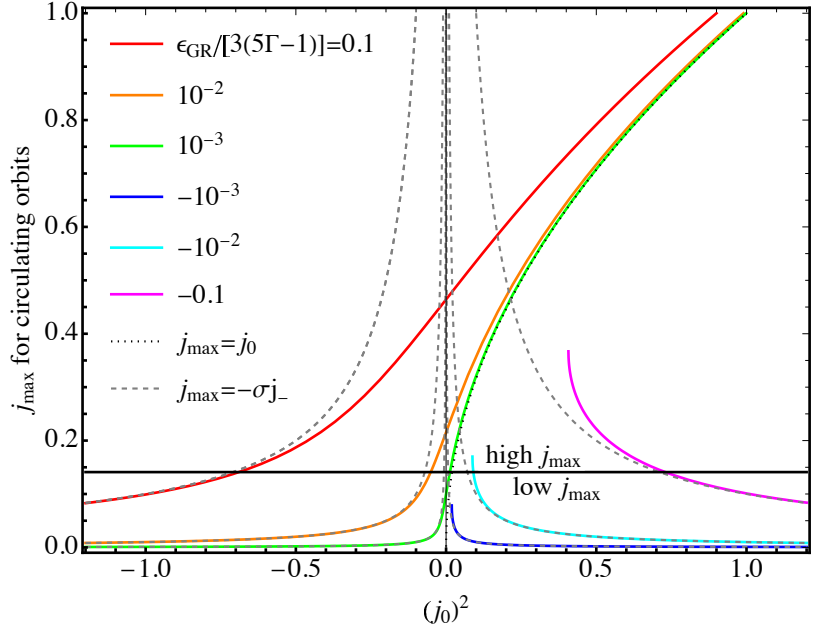}
\caption{Solid colored lines show $\jmax$ (the physical solutions to equation
\eqref{eqn:jmax_circulating}) for circulating trajectories as a function of
$j_0^2$, for different values of $\epsGR/[3(5\Gamma-1)]$, shown with different colors.
(We know from Papers II-III that
$j_0^2$ can be negative). A horizontal black line at $\jmax = 0.141$
($e_\mathrm{min} = 0.99$) separates `high-$\jmax$' and `low-$\jmax$' circulating solutions.
Grey dashed lines show the solution $\jmax = -\sigma j_-$, valid for low-$\jmax$
circulating trajectories (see equation \eqref{eqn:jmax_type_II}). The black
dotted line shows $\jmax = j_0$, valid for high-$\jmax$ circulating trajectories (equation
\eqref{eqn:jmax_type_I}).}
\label{fig:jmax_circulating}
\end{figure}

To make the distinction between high-$\jmax$ and low-$\jmax$ trajectories
 quantitative, recall from Appendix A3 of Paper III that for all circulating trajectories $\jmax$ is a solution to the
cubic equation
%%%%%%%%%%%%
\begin{align}
j_\mathrm{max}(j_\mathrm{max}^2-j_0^2) - \frac{\epsGR}{3(5\Gamma-1)} =0.
\label{eqn:jmax_circulating}
\end{align}
%%%%%%%%%%%%
One can solve this cubic analytically, but for simplicity here we will just plot the
solution.  Figure \ref{fig:jmax_circulating} shows $\jmax$ as a
function of $j_0^2$ (which can be positive or negative) for different values of $\epsGR/[3(5\Gamma-1)]$, shown with different colored solid lines.
In particular, red, orange and green lines correspond to $\Gamma > 1/5$ while blue, cyan and purple lines are for $\Gamma < 1/5$. We see from Figure \ref{fig:jmax_circulating} that for $\Gamma > 1/5$, circulating solutions exist for all values of $j_0^2$. However for $\Gamma < 1/5$ no solution exists below some (positive) value of $j_0^2$, consistent with what we found in Appendix B of Paper III.
%%%%%%%%%%%%%
%%%%%%%%%%%%%

We have chosen to split Figure \ref{fig:jmax_circulating} into two asymptotic regions, `high-$\jmax$'
($\jmax > 0.141$, i.e. $e_\mathrm{min} < 0.99$) and `low-$\jmax$' region ($\jmax < 0.141$,
i.e. $e_\mathrm{min} > 0.99$).
%All that matters is that circulating trajectories with $\jmax \sim 1$ (high-$\jmax$) behave qualitatively differently to those with $\jmax \ll 1$ (low-$\jmax$').
For high-$\jmax$ trajectories, provided $j_0^2$ is positive and $\mathcal{O}(1)$ we expect they are well approximated by ignoring the $\epsGR$ term in
\eqref{eqn:jmax_circulating}, so that
%%%%%%%%%%%%
\begin{align}
\jmax \approx j_0 \sim 1
 \,\,\,\,\,\,\,\, \mathrm{(high-}j_\mathrm{max}\,\,\,\mathrm{circulating \,\,\, orbits).}
\label{eqn:jmax_type_I}
\end{align}
%%%%%%%%%%%%
In Figure \ref{fig:jmax_circulating} we plot this solution with
a dotted black curve. For low-$\jmax$ trajectories, as long as $\jmax \ll \vert j_0 \vert$, we
find from \eqref{eqn:jmax_circulating} that
%%%%%%%%%%%%
\begin{align}
\jmax & \approx 
    \frac{\epsGR}{3(5\Gamma-1)(-j_0^2)} = -\sigma j_- \ll 1
\,\,\,\,\,\,\,\, \mathrm{(low-}j_\mathrm{max}\,\,\,\mathrm{circulating \,\,\, orbits).}
\label{eqn:jmax_type_II}
\end{align}
%%%%%%%%%%%%
We plot this solution with different grey dashed curves, using the same values of $\epsGR/[3(5\Gamma-1)]$ that we used for
the colored solid lines.   We see that for $\Gamma > 1/5$ (red, orange and green lines) the true solution interpolates between the two asymptotic solutions \eqref{eqn:jmax_type_I}, \eqref{eqn:jmax_type_II} as $j_0^2$ is varied.  For $\Gamma < 1/5$ (blue, cyan and purple lines), equation \eqref{eqn:jmax_type_II} provides a good approximation for sufficiently positive $j_0^2 > 0$.
Overall we see that different $\jmax$ curves touch the $-\sigma j_-$ solution approximately at $\jmax \approx 0.141$.  In other words, binaries on circulating trajectories transition from high-$\jmax$ to  low-$\jmax$ circulation around this point.  Expressing $\jmax$ in equation (\ref{eqn:jmax_type_II}) through $\emax$ and using equations \eqref{eqn:jminus_of_a} and \eqref{eqn:sigma_of_a} derived in the next section, we can calculate the semimajor axis at which this occurs, with the result
\begin{align}
    a \approx a_\mathrm{div}\left[ 1+ \frac{1-\emax}{10^{-2}}\right]^{2/7},
    \label{eqn:transition_adiv}
\end{align}
which is $\approx a_\mathrm{div}$ in most cases of interest since typically $1-\emax \ll 10^{-2}$. 
Thus a good rule of thumb is that circulating trajectories with $a>a_\mathrm{div}$ are high-$\jmax$, and circulating trajectories with $a<a_\mathrm{div}$ are low-$\jmax$.

%, where we
%saw that the existence of circulating solutions (which for $\epsGR=0$ just correspond
%to $\jmax = j_0$) requires the binary to reside in the very weak GR regime.
%Mathematically,
%a sufficiently positive $j_0^2$ is always required to render negative the left
%hand side of equation \eqref{eqn:jmax_circulating}.

In this discussion we have ignored one
possible regime, namely that of low-$\jmax$ circulating trajectories with small $\vert j_0
\vert$, i.e. $\vert j_0 \vert \lesssim \jmax \ll 1$. However as Figure
\ref{fig:jmax_circulating} shows, such solutions only exist for a
narrow range of $ j_0^2$ values centred around zero.  This regime is typically short-lived in the sense that a shrinking binary
 passes through it rather quickly on the way to merger (equivalently it is centered on a very narrow semimajor axis range around $a\approx a_\mathrm{div}$).  Throughout the rest of the paper we ignore
this intermediate case, i.e. we always assume that low-$\jmax$ circulating trajectories have $\vert j_0 \vert \gg
\jmax$. %(which obviously implies $\vert j_0 \vert \gg \jmin$).

%%%%%%%%%%%%%%%%%%%%%%%%%%%%%%%%%%%%%%%%%%%%%%%%
%\subsection{The strong GR regime}
%\label{sec:Strong_GR}
%%%%%%%%%%%%%%%%%%%%%%%%%%%%%%%%%%%%%%%%%%%%%%%%

%Finally, we recall from \S\ref{sec:very_strong_GR} that in the
%asymptotic limit of strong GR, cluster tides are negligible and the lowest order
%solution consists of pure precession at a rate 
%%%%%%%%%%%%
%\begin{align}
%\dot{\omega}_\mathrm{GR} 
%=\frac{3[G(m_1+m_2)]^{3/2}}{c^2 a^{5/2} j^2},
%\label{eqn:omega_dot_GR}
%\end{align}
%%%%%%%%%%%%
%In this limit there are no secular oscillations
%and hence there is no `secular timescale' to speak of.  

%%%%%%%%%%%%%%%%%%%%%%%%%%%%%%%%%%%%%%%%%%%%%%%%
%%%%%%%%%%%%%%%%%%%%%%%%%%%%%%%%%%%%%%%%%%%%%%%%
%%%%%%%%%%%%%%%%%%%%%%%%%%%%%%%%%%%%%%%%%%%%%%%%
%%%%%%%%%%%%%%%%%%%%%%%%%%%%%%%%%%%%%%%%%%%%%%%%
%%%%%%%%%%%%%%%%%%%%%%%%%%%%%%%%%%%%%%%%%%%%%%%%
\section{Phase space evolution and GR regimes for shrinking binaries}
\label{sec:phase_space_appendix}
%All the key quantities in \S\ref{sec:Framework} were written in terms
%f $\Gamma$, $\epsGR$, $\Theta$ and $\jmin$. For a given outer orbit, $\Gamma$
%is constant. And from equations \eqref{eq:epsGRformula} and
%\eqref{eqn:jmin_of_a} we know how $\epsGR$ and $\jmin$ scale with semimiajor
%axis $a$.  All that is left is to work out how $\Theta$ depends on $a$.

In \S\ref{sec:conservation_laws} we have seen how $\jmin$ and $\Theta$ depend on $a$
--- see equations \eqref{eqn:jmin_of_a} and
\eqref{eqn:Theta_of_a}. We can use these results to understand
how a binary moves through phase space as its semimajor axis shrinks.
To begin,
we substitute \eqref{eqn:jmin_of_a},
\eqref{eqn:Theta_of_a} and \eqref{eq:epsGRformula} into equations
\eqref{eqn:jplusWTM}-\eqref{eqn:j0WTM} to get $j_\pm^2$,
$j_0^2$ as explicit functions of semimajor axis:
%%%%%%%%%%%%
\begin{align}
j_+^2 &\approx \frac{10\Gamma}{1+5\Gamma}\left[ \cos^2\imin  + \left( \frac{d}{a}\right)^{7/2} \right]=\frac{10\Gamma}{1+5\Gamma}\cos^2\imin\left[ 1  + \left( \frac{\ell a_\mathrm{weak}}{a}\right)^{7/2} \right],
\label{eqn:jplus_of_a}
\\
j_-^2 &\approx \frac{2\pmin}{a} \left[ 1+ \frac{1}{\cos^2 \imin}\left( \frac{d}{a}\right)^{7/2}\right]^{-1},
\label{eqn:jminus_of_a}
\\
j_0^2 &\approx \frac{10\Gamma}{5\Gamma -1 }\left[\sin^2 \imin - \left( \frac{d}{a}\right)^{7/2} \right]=\frac{10\Gamma}{5\Gamma -1 }\sin^2 \imin\left[1 - \left( \frac{a_\mathrm{div}}{a}\right)^{7/2} \right],
\label{eqn:j0_of_a}
\end{align}
%%%%%%%%%%%%
where $d$ is defined in equation \eqref{eqn:d_def} and $\ell = [(\sqrt{2}-1)/2]^{-2/7} \approx 1.57$ --- see the definition (\ref{eqn:a_weak}).
Next
 we write down the important dimensionless quantities $\gamma$, $\sigma$ and
$\kappa$ (familiar from equations (49), (50) and (63) of Paper III respectively) as functions of $a$, as follows. First, by combining equations
\eqref{eqn:epsweak}, \eqref{eqn:jplus_of_a},
\eqref{eqn:jminus_of_a} and the definition \eqref{eqn:d_def} it is straightforward to show that 
%%%%%%%%%%%%%
\begin{align}
\label{eqn:gamma_of_a}
\gamma(a) \equiv \frac{2\epsGR}{\epsweak} \approx \frac{1}{\sqrt{\zeta(\zeta+1)}}, 
\,\,\,\,\,\,\,\,\,\,
\mathrm{where}
\,\,\,\,\,\,\,\,\,\,
\zeta \equiv (a/d)^{7/2}\cos^2 \imin.
\end{align}
%%%%%%%%%%%%%
Second, plugging \eqref{eq:epsGRformula}, \eqref{eqn:jminus_of_a} and
\eqref{eqn:j0_of_a} into
equation (50) of Paper III we get:
%%%%%%%%%%%%%
\begin{align}
\label{eqn:sigma_of_a}
\sigma(a) \approx \left[\left( \frac{d}{a} \right)^{7/2} \frac{1}{\cos^2 \imin} + 1\right]^{1/2} 
\left[ \left(\frac{a}{d} \right)^{7/2} \sin^2 \imin-1\right]^{-1}.
\end{align}
%%%%%%%%%%%%%
Third, we can take the ratio of \eqref{eqn:sigma_of_a} and
\eqref{eqn:gamma_of_a} to get $\kappa\equiv \sigma/\gamma$:
%%%%%%%%%%%%%
\begin{align}
\label{eqn:kappa_of_a}
\kappa(a) \approx \left[ \left( \frac{a}{d} \right)^{7/2} \cos^2 \imin + 1 \right] 
\left[ \left(\frac{a}{d} \right)^{7/2} \sin^2 \imin -1\right]^{-1}=
\left[ \left( \frac{a}{\ell a_\mathrm{weak}} \right)^{7/2} + 1 \right] 
\left[ \left(\frac{a}{a_\mathrm{div}} \right)^{7/2} - 1\right]^{-1}.
\end{align}
%%%%%%%%%%%%
These results lead naturally to the definitions of the critical semimajor axis values $a_\mathrm{sep}$ and $a_\mathrm{div}$ that we gave in \S\ref{sec:lengthscales}.
We now use these results, as well as the quantities $a_\mathrm{weak}$ (equation \eqref{eqn:a_weak}) and $a_\mathrm{strong}$ (equation \eqref{eqn:a_strong}), to understand more precisely how binaries move through
phase space and different GR regimes as $a$ decays.
% These considerations will help us
%develop approximate expressions for $\tsec$, $\Delta a$ and $a(t)$ in different
%regimes --- see
%\S\S\ref{sec:secular_timescale}-\ref{sec:SMA_Evolution}.
We begin with the regime $\Gamma > 1/5$, and then discuss $0<\Gamma\leq 1/5$. 

%%%%%%%%%%%%%%%
\begin{figure*}
\centering
\includegraphics[width=0.99\linewidth,clip]{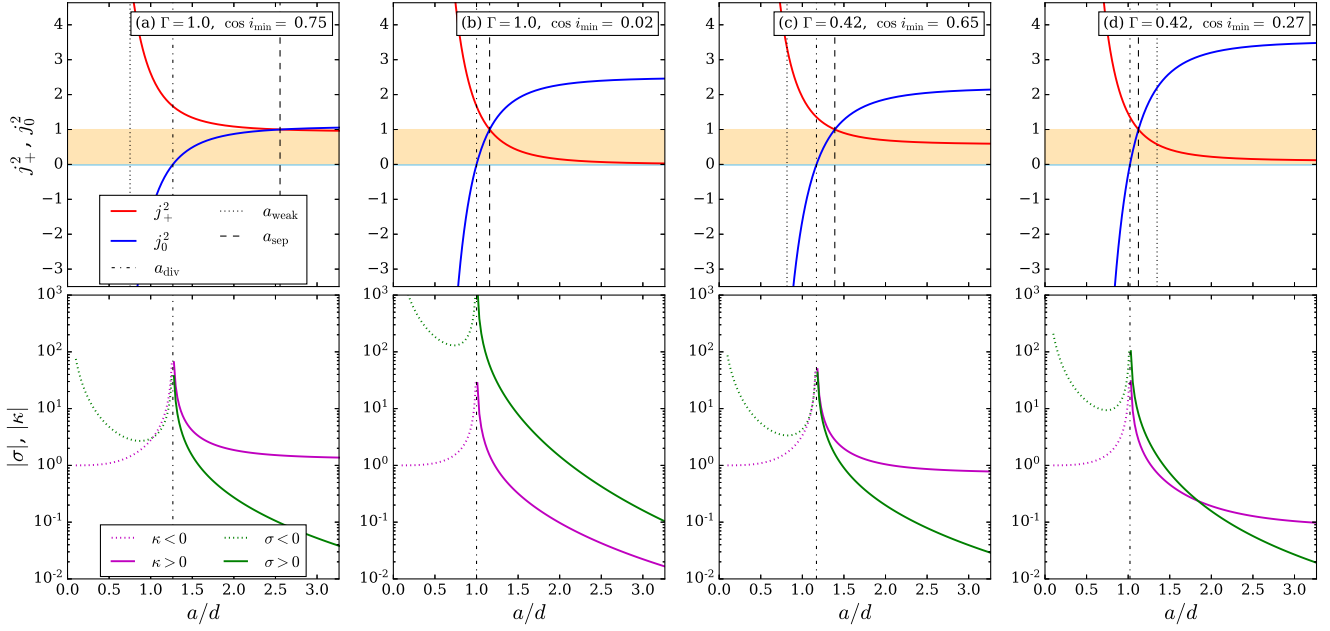}
\caption{Plots of the key quantities $j_+^2$, $j_0^2$, $\sigma$ and $\kappa$ as
functions of $a/d$, for different values of
$\Gamma>1/5$ and $\cos \imin$. In the lower panels, for which the vertical axis
is on a logarithmic scale, we show negative values of $\sigma, \kappa$ with
dotted curves and positive values with solid curves. In each panel we also show
$a_\mathrm{weak}$ (vertical dotted line), $a_\mathrm{sep}$ (vertical dashed
line) and $a_\mathrm{div}$ (vertical dot-dashed line). Finally in the upper row we show with
blue shading the (very thin) region $\vert j^2  \vert< (0.141)^2 \approx 0.02$, within which
the split into `high-$\jmax$' and `low-$\jmax$' circulating trajectories is invalid
(\S\ref{sec:high_ecc_no_GWs}), and with pale orange shading the region $(0.141)^2 <
j^2 <1$.
The values of $\Gamma$ and $\cos\imin$ in panels (a)-(d) are chosen to coincide with the examples shown in Figures \ref{fig:Numerical_LK}, \ref{fig:Numerical_RX18_Fig3}, \ref{fig:Numerical_Initially_WeakLib} and \ref{fig:Numerical_Initially_ModLib} respectively.}
\label{fig:jPlus_of_a_Gamma_Regime_I}
\end{figure*}
%%%%%%%%%%%%%%%
\begin{figure*}
\centering
\includegraphics[width=0.99\linewidth,clip]{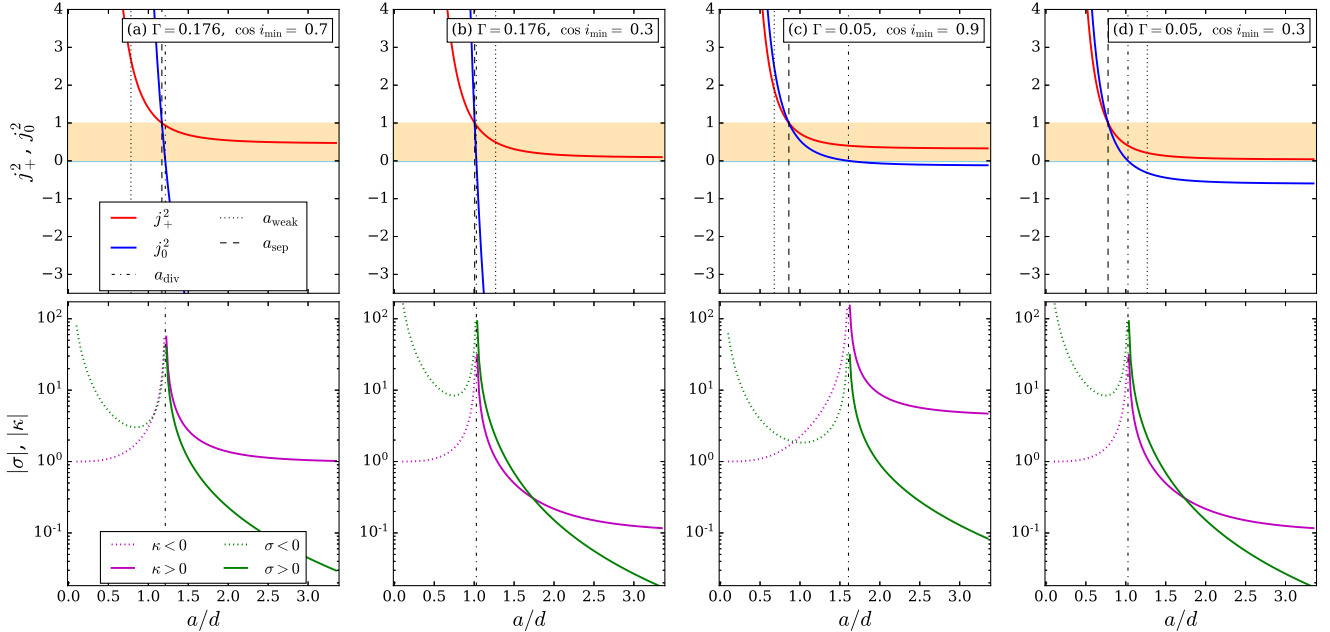}
\caption{As in Figure \ref{fig:jPlus_of_a_Gamma_Regime_I} but for
the regime $0 < \Gamma \leq 1/5$.  The choices of $\Gamma$ and $\cos\imin$ in panel (a) coincide with those from Figure \ref{fig:Numerical_Gampt176}.}
\label{fig:jPlus_of_a_Gamma_Regime_II}
\end{figure*}
%%%%%%%%%%%%%%%
%%%%%%%%%%%%%
\subsubsection{Phase space evolution for $\Gamma > 1/5$}
%%%%%%%%%%%%%

In Figure \ref{fig:jPlus_of_a_Gamma_Regime_I}  we plot $j_+^2$,
$j_0^2$, $\vert \sigma \vert$ and $\vert \kappa \vert$ as functions of $a/d$ for
various fixed values of $\Gamma$ and $\cos \imin$, according to equations
\eqref{eqn:jplus_of_a}, \eqref{eqn:j0_of_a},
\eqref{eqn:sigma_of_a} and \eqref{eqn:kappa_of_a}
respectively.
The choices of $\Gamma$ and $\cos\imin$ in panels (a)-(d) are chosen to coincide with the examples shown in Figures \ref{fig:Numerical_LK}, \ref{fig:Numerical_RX18_Fig3}, \ref{fig:Numerical_Initially_WeakLib} and \ref{fig:Numerical_Initially_ModLib} respectively.
We also show the critical values $a_\mathrm{weak}$  (dotted
vertical line), $a_\mathrm{sep}$ (dashed vertical line) and $a_\mathrm{div}$
(dot-dashed vertical line), defined in equations
\eqref{eqn:a_weak}-\eqref{eqn:a_div}. Additionally, in the
upper panels we show with blue shading the region $\vert j^2  \vert< (0.141)^2$, within which
the split into `high-$\jmax$' and `low-$\jmax$' circulating trajectories is invalid
(see the final paragraph of Appendix \ref{sec:high_ecc_no_GWs}).
We show with orange shading the region $(0.141)^2 < j^2 < 1$. In particular, by looking at
the runs of $j_+^2$ and $j_0^2$ and whether they lie in this orange region, we
will be able to infer the value of $\jmax$ and hence infer what type of phase
space trajectory the binary is on.
Without loss of generality, for each example (a)-(d) we can consider a binary
that starts at the extreme right of each panel, i.e. with $a\gg a_\mathrm{weak}$
(the weak GR regime), and follow it as $a$ decreases.

%\cmtrr{Let's redo panel (a) everywhere as it is not shown in Figures and is very different. This difference (i.e. absence of librating island) need to be very clearly spelled out (and violation of some conserved quantities need to be mentioned as well). I would not mind if panel (a) is dropped altogether, as it pertains to 'non-mergers' or  'very slow mergers'. } 

First we focus on panels (a) and (b), which are for $\Gamma=1$ (the LK limit). 
In panel (a) the binary `begins' at large $a$ with $0 <
j_+^2 < 1$ and $j_0^2 > 1$; this means that it is on a librating
trajectory in the weak GR regime, with $\jmax \approx j_+$ (equation
\eqref{eqn:jmax_librating}). Of course as $a$ is decreased $j_+^2$ is
always increased, while $j_0^2$ is decreased, and when $a=a_\mathrm{sep}$ the
two cross over, $j_0^2=j_+^2=1$.  At this point the binary switches to a high-$\jmax$
circulating trajectory with $\jmax \approx j_0$. In this case $a_\mathrm{sep}<
a_\mathrm{weak}$, so that the separatrix crossing occurs while the binary is
still in the weak GR regime.  
Once $a$ becomes smaller than $a_\mathrm{div}$ we quickly get $j_0^2$ values
that are strongly negative, and the binary transitions to a low-$\jmax$
circulating trajectory (Figure \ref{fig:jmax_circulating}) with $\jmax \approx
-\sigma j_- \ll 1$.  It will remain on such a trajectory until it gets trapped at
high eccentricity in the strong GR regime around $a\sim a_\mathrm{strong}$
(not shown here).

Example (b) shows very similar behavior to example (a), except that the
smaller value of $\cos\imin$ means that the three values $a_\mathrm{weak}$,
$a_\mathrm{sep}$ and $a_\mathrm{div}$ are now even more closely clustered
together around $a/d\approx 1$ (note also that $a_\mathrm{sep}$ is now very
slightly smaller than $a_\mathrm{weak}$). Because of this clustering, example
(b) is perhaps `cleaner' than (a): for $a$
significantly larger than $d$ the binary is clearly on a librating trajectory in the
weak GR regime, whereas for $a$ significantly smaller than $d$ it is clearly on
a low-$\jmax$ circulating trajectory in the moderate GR regime. In practice the
transitions between these two various regimes are not always so well
demarcated.

At this stage it is worth noting how different quantities scale with $a$ in each
regime. From examples (a)-(b) we see that in the weak GR regime ($a>
a_\mathrm{weak}$) we nearly always have $\vert j_+^2\vert, \vert j_0^2\vert \gg
0.1$, and both of these quantities scale very weakly with $a$.  
In the moderate GR regime ($a<a_\mathrm{weak}$) the scaling of $j_+^2$ and
$j_0^2$ with $a$ is much stronger, as we would expect from equations
\eqref{eqn:jplus_of_a}, \eqref{eqn:j0_of_a}. Moreover, in every case it is clear that
$\vert j_0 \vert^2$ lies in the blue shaded region only for a very narrow range
of semimajor axes surrounding $a_\mathrm{div}$ (equation \eqref{eqn:transition_adiv}), and so we were justified in
 ignoring the small $j_0$ regime when discussing low-$\jmax$ circulating trajectories in
Appendix \ref{sec:high_ecc_no_GWs}. Turning to the bottom panels, we see that
$\vert \sigma\vert $ and $\vert \kappa\vert $ both vary over several orders of
magnitude as $a$ is decreased. However, it is noteworthy that for $a$ far away
from $a_\mathrm{div}$, the value of $\vert \kappa\vert $ is usually
$\mathcal{O}(1)$ and scales weakly with $a$. 

Finally we turn to examples (c) and (d), which are for $\Gamma =0.42$.  The
physical interpretation of these examples is identical to those of (a) and (b),
demonstrating a broad uniformity of evolution for all binaries in the $\Gamma >
1/5$ regime.  In fact, this broad-brush picture can break down very close
to $\Gamma=1/5$, but we ignore this complication here.

%%%%%%%%%%%%%
\subsubsection{Phase space evolution for $0 < \Gamma \leq 1/5$}
%%%%%%%%%%%%%

In Figure \ref{fig:jPlus_of_a_Gamma_Regime_II} we plot the same
quantities as in Figure \ref{fig:jPlus_of_a_Gamma_Regime_I}, except
this time we focus on the regime $0<\Gamma\leq 1/5$.
In particular the choices of $\Gamma$ and $\cos\imin$ in panel (a) coincide with those from Figure \ref{fig:Numerical_Gampt176}.
We see that a rather different
phase space evolution emerges for $0<\Gamma\leq 1/5$ compared to $\Gamma > 1/5$.

First we consider panel (a), which is for $\Gamma =0.176$ and $\cos\imin = 0.7$.  
In this case, for large $a \gg d$ we have $j_+^2 \lesssim 1$ while $j_0^2$ is large and negative. This means that in the asymptotic weak GR regime the binary is on a
librating trajectory, with $\jmax \approx j_+$. 
%\sout{This is unsurprising since, as we saw
%in Paper II with GR switched off, for $0<\Gamma \leq
%1/5$, circulating trajectories very rarely reach high eccentricity.} \cmtch{Don't think that sentence tells us anything important.} 
However, once $a$
decreases below $a_\mathrm{div}$ in this plot, we see that $j_0^2$ becomes
positive (though still smaller than $j_+^2$). Soon $a$
reaches $a_\mathrm{sep}$, below which both $j_+^2$ and $j_0^2$ are greater than
unity: the binary has transitioned onto a low-$\jmax$ circulating
orbit (Figure \ref{fig:jmax_circulating}). We note that all of this
happens well before the binary reaches the moderate GR regime.  
This is not surprising because we know that a family of
high-eccentricity circulating trajectories (i.e. low-$\jmax$) naturally arises in the
$0<\Gamma\leq 1/5$ regime as soon as $\epsGR$ exceeds $6(1-5\Gamma)\Theta^{3/2}
\ll \epsweak$ (Paper III). The binary stays on its low-$\jmax$ circulating trajectory as $a$ shrinks into the moderate GR regime
$a<a_\mathrm{weak}$ and onward to the strong GR regime.

A very similar story holds in panels (b)-(d).  The only important difference is
that as we decrease $\Gamma$ or $\cos\imin$, or both, the value of $\jmax
\approx j_+$ for asymptotically weak GR ($a\gg d$) decreases. 
This means that
librating trajectories with high $e_\mathrm{max} \to 1$ in the very weak GR regime do
not reach low\footnote{This is essentially because the fixed points at $\omega =
\pm\pi/2, j=j_\mathrm{f} = (10\Gamma\Theta/(1+5\Gamma))^{1/4}$, sit at too high
an eccentricity.} $e_\mathrm{min}$ (recall that we have assumed $\jmin \ll 1$ in
deriving our expression for $j_+^2$). 
% Thus the
%%situation is like that in Figure , in which the only binaries that are
%capable of reaching very high eccentricity even with GR effectively switched off
%are those on librating trajectories that start out with $e\sim 1$ to begin with.
Said differently, for $0 < \Gamma < 1/5$, binaries that initially have $e\sim 0$ do not tend to reach $e\to 1$ --- that is, low minimum eccentricities are not typically associated with high maximum eccentricities,
so this is typically not the type of situation in which we are interested.

Finally we mention that in all examples shown in Figure
\ref{fig:jPlus_of_a_Gamma_Regime_II}, for $a$ sufficiently far from
$a_\mathrm{div}$ we again have $\vert \kappa \vert \sim \mathcal{O}(1)$ or smaller,
and $\kappa$ varies only weakly with $a$.

\section{Derivation of $\tsec$ formula in Regimes $\mathcal{C}$ and $\mathcal{D}$}
\label{sec:tsec_type2_appendix}

%%%%%%%%%%%%%%%%%%%%%%%%%%%%%%%%%%%%%%%%%
%%%%%%%%%%%%%%%%%%%%%%%%%%%%%%%%%%%%%%%%%
%%%%%%%%%%%%%%%%%%%%%%%%%%%%%%%%%%%%%%%%%
%%%%%%%%%%%%%%%%%%%%%%%%%%%%%%%%%%%%%%%%%

As $a$ crosses $a_\mathrm{div}$ and the binary enters regime $\mathcal{C}$, $j_0^2$ passes through zero and rapidly becomes strongly negative --- see equation (\ref{eqn:j0_of_a}) and Figures \ref{fig:jPlus_of_a_Gamma_Regime_I}-\ref{fig:jPlus_of_a_Gamma_Regime_II}. Those Figures also show that $j_+^2$ is typically large in amplitude for $a\lesssim a_\mathrm{div}$, certainly larger than $j^2$ which is limited by $j_\mathrm{max}^2$, and which is already in the low-$j_\mathrm{max}$ regime. Also, for most of the secular cycle we can neglect $j_-^2$ compared to $j^2$ since $\jmax = -\sigma j_-$ is well separated from $\jmin \approx \gamma j_-$. Taking the limit $j_-^2\ll j^2\ll j_+^2, j_0^2$ (which is most accurate in regime $\mathcal{D}$ with moderate GR), equation (\ref{eq:djdtGR}) reduces to
%%%%%%
\begin{align}  
\frac{\md j}{\md t} \approx
\pm \frac{6C}{Lj}
\sqrt{\vert 25\Gamma^2-1 \vert }\vert j_+ j_0 \vert \sqrt{(j-\jmin)(\jmax-j)}.
\label{eq:djdt_moderate_GR}
\end{align}
%%%%%%%
Plugging this into equation \eqref{eqn:tsec_exact}
and performing the integral we get 
%the secular timescale for Type
%2 circulating trajectories in the moderate GR regime:
%%%%%%
\begin{align}
    \tsec &\approx \frac{L \pi (\jmin + \jmax)}{6C \sqrt{\vert 25\Gamma^2-1 \vert }\vert j_+ j_0 \vert}
 \approx
    \frac{8 }{3A}\sqrt{\frac{G(m_1+m_2)}{\vert 25\Gamma^2-1\vert}} \times \frac{\pi}{2a^{3/2}}  
    \frac{(1-\kappa)\jmin}{\vert j_+ j_0 \vert },
\label{eqn:tsecTypeII}
\end{align}
%%%%%%
where in the second line we used $\jmin + \jmax \approx \gamma j_-
-\sigma j_- = (1-\kappa)\jmin$ (see equation
\eqref{eqn:jmax_type_II}). Using equations (\ref{eqn:jmin_of_a}), (\ref{eqn:jplus_of_a}), (\ref{eqn:j0_of_a}), (\ref{eqn:kappa_of_a}) for $j_\mathrm{min}$, $j_+^2$, $j_0^2$, $\kappa$, correspondingly, we find
\begin{align}
    \tsec &\approx 
    \frac{2\pi}{15 \Gamma A}\frac{\sqrt{2G(m_1+m_2)p_\mathrm{min}}}{a^2\cos\imin \sin\imin}f(a),~~~~{\rm with} ~~~f(a)=\left(\frac{a}{a_\mathrm{div}}\right)^{7/4}\frac{2+\left( \frac{a}{\ell a_\mathrm{weak}} \right)^{7/2}-\left(\frac{a}{a_\mathrm{div}} \right)^{7/2}}{\sqrt{1+\left( \frac{\ell a_\mathrm{weak}}{a} \right)^{7/2}}\left|1-\left(\frac{a}{a_\mathrm{div}} \right)^{7/2}\right|^{3/2}}.
\label{eqn:tsecCD}
\end{align}
Different limits of this expression in regimes $\mathcal{C}$ and $\mathcal{D}$ are explored in \S \ref{sec:RegC} and \S \ref{sec:RegD}.

\section{Derivation of an approximate formula for semimajor axis decay}
\label{sec:Delta_a_appendix}

Assuming the binary reaches very high maximum eccentricity $e_\mathrm{max} \to
1$ we can approximate equation \eqref{eqn:Delta_a_sec_cycle} as
%%%%%%%%%%%%%
\begin{align}
\Delta a \approx -\frac{2\lambda_1}{a^3}\int_{\jmin}^{\jmax} 
\frac{\md j}{j^7}
\left(\frac{\md j}{\md t} \right)^{-1},
\label{eqn:Delta_a_formal}
\end{align}
%%%%%%%%%%%%%%%
where $\lambda_1 \equiv (1+73/24+37/96) \lambda_0 = (170/3)G^3 c^{-5}
m_1m_2(m_1+m_2)$. In general $\md j/\md t$ --- given in equation
\eqref{eq:djdtGR} --- is so complicated that even this approximate
integral is intractable.  However, noting the very strong $j^{-7}$ dependence in
\eqref{eqn:Delta_a_formal} we expect the integral to be dominated by
the contributions from very high eccentricity, i.e. $j \ll j_+, \vert j_0
\vert$. In this limit we can approximate $\md j/\md t$ using equation
(48) of Paper III. Moreover, since we know that the
minimum $\jmin$ is a zero of the first square bracket in
that equation, we can write it as
%%%%%%%%%%%%%%%
\begin{align}
\frac{\md j}{\md t} &\approx
\pm \frac{3Aa^{3/2}}{4\sqrt{G(m_1+m_2)}j^{3/2}}
\sqrt{(25\Gamma^2-1)j_+^2 (-j_0^2)
(j-\jmin)(j+\vert j_\alpha \vert)(j_\sigma - j)},
\label{eqn:djdt_for_Delta_a}
\end{align}
%%%%%%%%%%%%%%%
where $j_\alpha \equiv \gamma j_- [1-\sqrt{1+4\gamma^{-2}}]/2 < 0$ is the other
root of the first square bracket in
equation
(48) of Paper III, and $j_\sigma \equiv -\sigma j_-$.
\footnote{Using the results of Appendix \ref{sec:high_ecc_no_GWs} one can 
check that the sign of the quantity inside the square root is positive.  For
instance, for $\Gamma > 1/5$ we recall that low-$\jmax$ circulating trajectories have
$j_0^2 < 0$ and $\jmax = -\sigma j_- = j_\sigma$, while Type
1 circulating trajectories have $j_0^2 > 0$, $j_\sigma < 0$ and $\jmax \sim 1$.}

We
now take \eqref{eqn:djdt_for_Delta_a} and plug it into
\eqref{eqn:Delta_a_formal}. Defining 
%%%%%%%%%%%%%%%
\begin{align}
x_\mathrm{max} \equiv j_\mathrm{max}/\jmin, \,\,\,\,\,\,\,\,\,\,\,
x_\alpha \equiv j_\alpha/\jmin, \,\,\,\,\,\,\,\,\,\,\,
x_\sigma \equiv j_\sigma/\jmin,
\end{align}
%%%%%%%%%%%%%%%
and using \eqref{eqn:jmin_of_a}, the result is
%%%%%%%%%%%%%%%
\begin{align}
\Delta a \approx &-\lambda_2
\times \frac{\xi(x_\mathrm{max},  x_\alpha, x_\sigma)}{a^{3/2}\vert j_+ j_0\vert},
\label{eqn:Delta_a_of_xmax}
\end{align}
%%%%%%%%%%%%%%%
where $\lambda_2 \equiv
1360G^{7/2}m_1m_2(m_1+m_2)^{3/2}/[9c^5A(2\pmin)^3\sqrt{\vert 25\Gamma^2-1
\vert}]$
is independent of $a$, and
%%%%%%%%%%%%%%%
\begin{align}
\label{eqn:dimensionless_integral}
\xi(x_\mathrm{max},  x_\alpha, x_\sigma)
\equiv 
\int_{1}^{x_\mathrm{max}} \frac{\md x}{x^{11/2}\sqrt{(x-1)(x+\vert x_\alpha \vert)\vert x_\sigma - x\vert }}.
\end{align}
%%%%%%%%%%%%%%%

We can simplify this result in the limit of weak GR. 
In this limit we have $\jmin  \ll \jmax$ so that $x_\mathrm{max} \gg 1$. We also
have $j_\sigma < 0$, so that $\vert x_\sigma - x \vert = x + \vert x_\sigma
\vert$. In this case the integral in
\eqref{eqn:dimensionless_integral} is completely dominated by the
contribution from $x\approx 1$, and so we may take the upper limit of the
integral to $x_\mathrm{max} \to \infty$ with impunity. Since $\jmin
\approx j_-$ and $\gamma \ll 1$ in this limit (see equation \eqref{eqn:gamma_of_a}), we can simply replace $x_\sigma$
with $\sigma$ and $x_\alpha \to 1$. An excellent approximation to the
resulting integral (accurate to within a few percent
over several decades of $\vert \sigma \vert$) is then given by equation \eqref{eqn:dimensionless_integral_not_II}.

%%%%%%%%%%%%%%%%%%%%%%%%%%%%%%%%%%%%%%%%%
\section{Relation to Randall \& Xianyu (2018)}
\label{sec:RX18}
%%%%%%%%%%%%%%%%%%%%%%%%%%%%%%%%%%%%%%%%%
Throughout the main text we referred to the paper by
\citet{Randall2018-uq} --- hereafter RX18 --- 
several times.  
The RX18 paper largely inspired the present work, since those authors are among the few who have attempted to gain
an analytical understanding of LK-driven slow mergers (indeed it is from their
paper that we have taken the terminology `slow merger').  
In particular, to our knowledge RX18 were first to
(i) calculate $\Delta a$ explicitly, and (ii) comment upon the decrease in
$\tsec$ as the binary shrinks and offer an explanation thereof. On the other
hand, we feel that both (i) and (ii) as presented in RX18 can
be improved.  In this Appendix we explain how our calculations differ from those of
RX18 regarding points (i) and (ii) (\S\S\ref{sec:Delta_a_RX18} and \ref{sec:tsec_RX18} respectively). 

\begin{figure*}
\centering
   \includegraphics[width=0.99\linewidth]{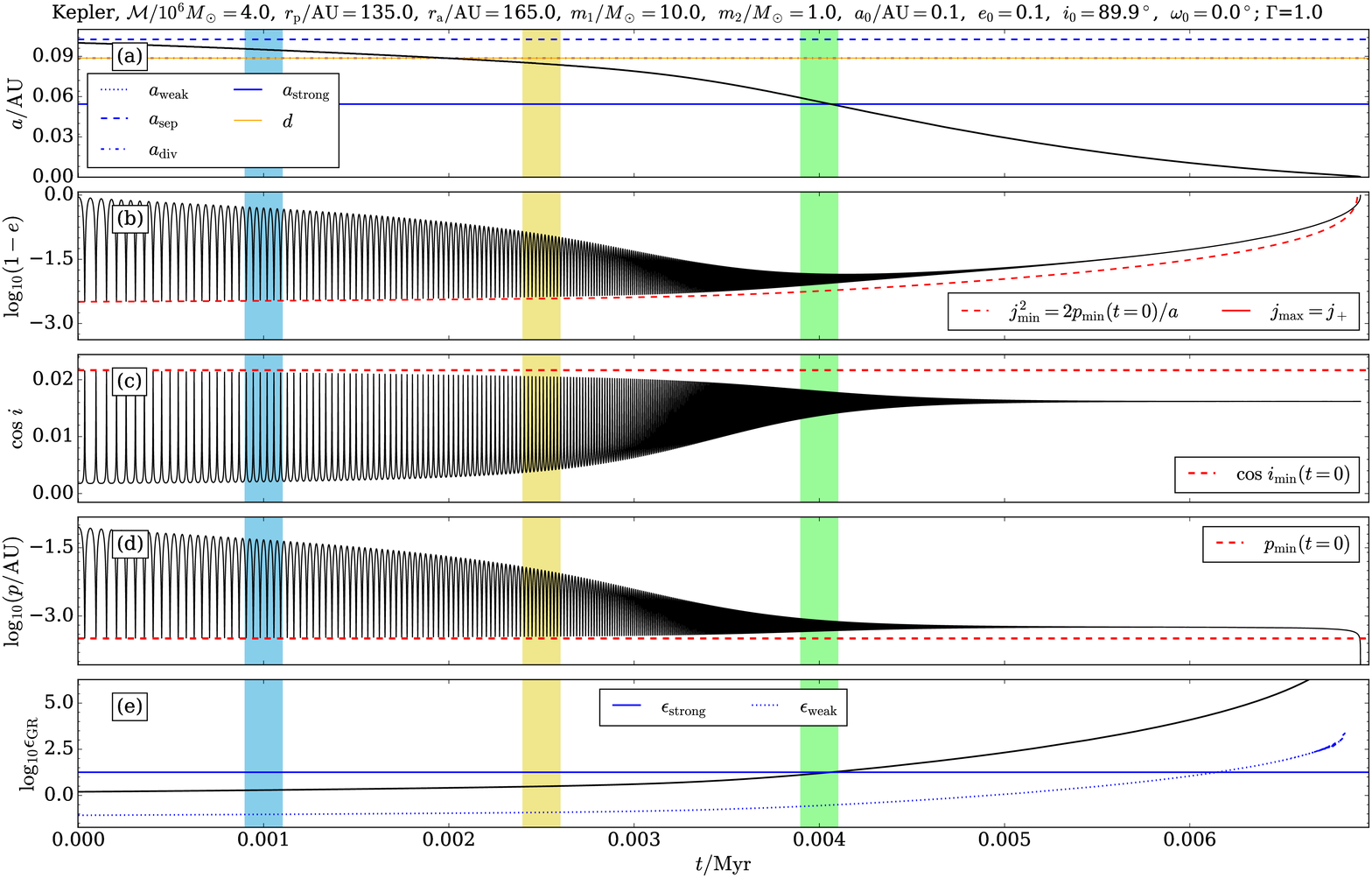}
    \includegraphics[width=0.99\linewidth]{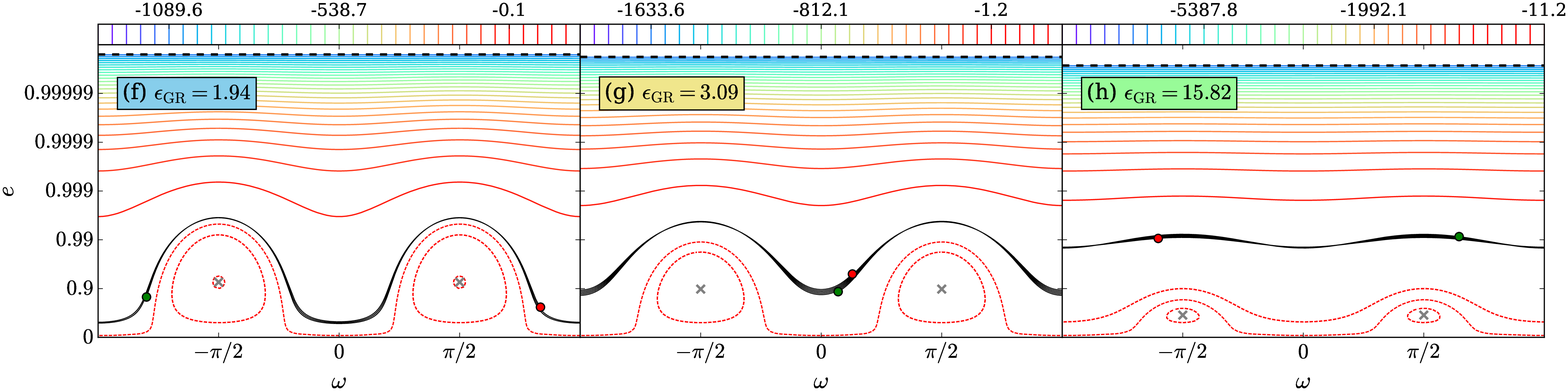}
    \includegraphics[width=0.99\linewidth]{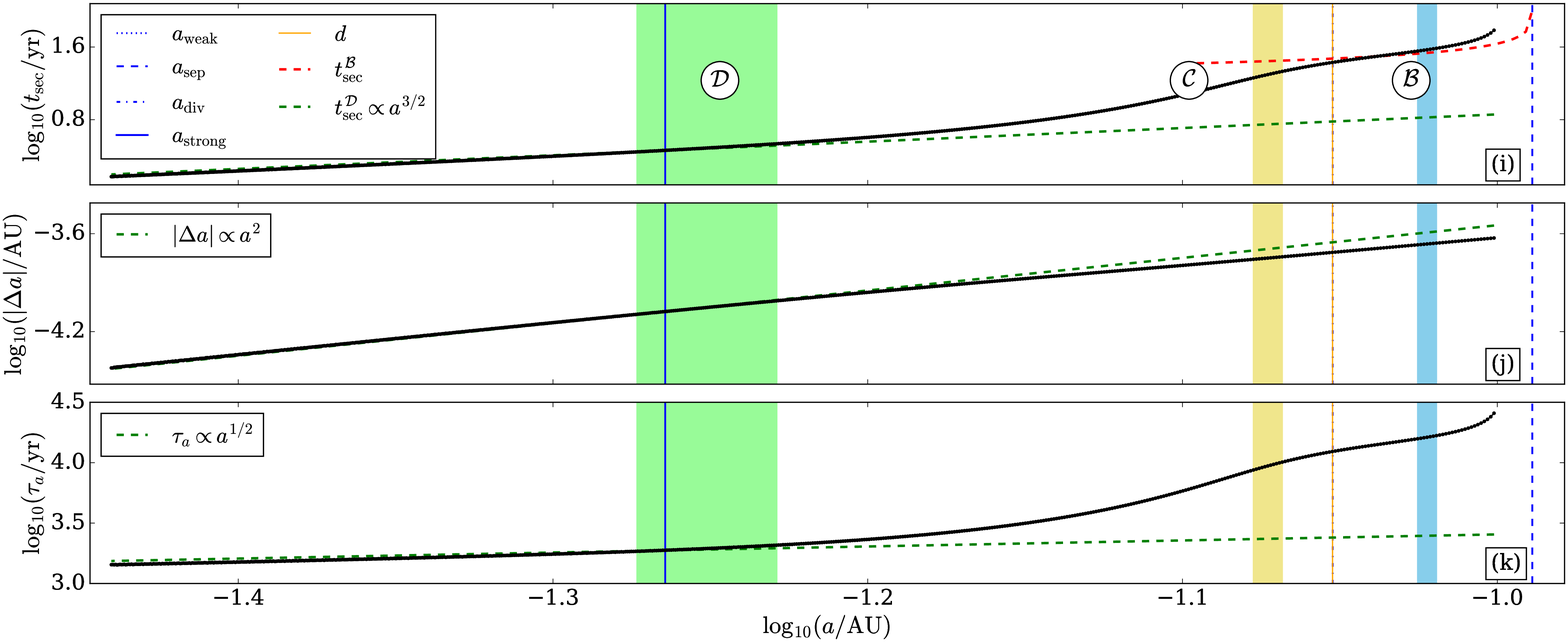}
\caption{Reproducing Figure 3 of \citet{Randall2018-uq}. In this case a binary with
$m_1=m_2=10M_\odot$ orbits a supermassive black hole (i.e. Kepler
potential, $\Gamma=1$) of mass $\mathcal{M} = 4\times 10^6M_\odot$. The outer orbit has
semimajor axis $a_\mathrm{g} = (\ra+\rp)/2 = 150$ AU and eccentricity
$e_\mathrm{g} = (\ra-\rp)/(\ra+\rp)=0.1$. Note that $a_\mathrm{div}$ and $d$ overlap almost exactly, which follows from the fact that in this example $\cos \imin \ll 1$ (see equation \eqref{eqn:a_div}).}
    \label{fig:Numerical_RX18_Fig3}
\end{figure*}

To begin we present Figure \ref{fig:Numerical_RX18_Fig3}.
This Figure reproduces exactly the numerical example shown in RX18's Figure 3, from which those authors drew several of their conclusions. Specifically, it follows the evolution of a binary of $m_1=m_2=10M_\odot$ and $a_0 = 0.1$AU as it orbits a SMBH of mass $4\times 10^6M_\odot$. We see that in this example the binary
 sits from the start in the moderate (rather than weak) GR regime on a circulating phase space trajectory, and that the
secular timescale does indeed decrease as the binary shrinks.
The merger occurs after around $t=7000$ yr.
We will refer to this Figure frequently throughout the remainder of this section.

%%%%%%%%%%%%%%%%%%%%%%%%%%%%%%%%%%%%%%%%%%%%%%%%
%%%%%%%%%%%%%%%%%%%%%%%%%%%%%%%%%%%%%%%%%%%%%%%%
\subsection{Calculation of $\Delta a$}
\label{sec:Delta_a_RX18}
%%%%%%%%%%%%%%%%%%%%%%%%%%%%%%%%%%%%%%%%%%%%%%%%
%%%%%%%%%%%%%%%%%%%%%%%%%%%%%%%%%%%%%%%%%%%%%%%%

RX18 begin their calculation of $\Delta a$ by writing down their
equation (55), the first two lines of which are identical to our
equation \eqref{eqn:Delta_a_sec_cycle} if we evaluate the final
bracket at $e=e_\mathrm{max}$. One is then faced with the computation of an
integral, $\Delta a \propto \int \md t (1-e^2(t))^{-7/2}$, over one
secular cycle. To perform this integral in \S\ref{sec:SMA_Evolution} we
changed variables from $t \to j \in (\jmin,\jmax)$ and hence wrote down equation
\eqref{eqn:Delta_a_formal}. On the other hand, RX18
choose to compute the integral by first approximating $e(t)$ as a quadratic in time
(see their equation (53)). In particular, using our notation
and letting the maximum eccentricity occur at $t=0$ without loss of generality,
their equation (52) reads
%%%%%%%%%%%%%%
\begin{align}
    e(t) = \emax + \frac{1}{2}\left( \frac{\md^2 e}{\md t^2}\right)_{t=0} t^2.
\end{align}
%%%%%%%%%%%%%%
RX18 then plug this into 
$\int \md t (1-e^2(t))^{-7/2}$ and 
integrate over $t\in (-\infty, \infty)$ to get $\Delta a$. The result is their second equation (55),
which in our notation and 
evaluating at $e_\mathrm{max} \approx 1$ reads
%%%%%%%%%%%%%%
\begin{align}
    \Delta a_{\mathrm{RX18}} \approx - \frac{544 G^3 m_1m_2(m_1+m_2)}{9c^5 a^3 \jmin^6} 
    \times \left\vert \frac{\md^2 e}{\md t^2}\right\vert^{-1/2}_{t=0}.
    \label{eqn:RX18_Delta_a}
\end{align}
%%%%%%%%%%%%%%
Finally, RX18 evaluate $\ddot{e}\vert_{t=0}$ using their equation (53).

However, RX18's method for computing $\Delta a$ implicitly makes two assumptions which are not true in general, as we now explain.

\begin{enumerate}
    \item The assumption
that $e(t)$ is quadratic for small $t$ is equivalent to the assumption that
$j(t)$ is quadratic for small $t$. We know from Paper III
that this quadratic approximation is only good if the binary is in the weak GR regime ($\epsGR \ll \epsweak$) and it has
$\sigma \ll 1$ (equation \eqref{eqn:sigma_of_a}).  While these conditions do hold for many binaries of interest (i.e. see the early stages of Figures \ref{fig:Numerical_LK} and \ref{fig:Numerical_Initially_WeakLib}, for which $\sigma \approx 0.02$ and $0.08$ respectively), they are not true for the RX18 calculation shown in
Figure \ref{fig:Numerical_RX18_Fig3} --- this example begins in the moderate GR regime (panel (e)) and has $\sigma \approx 14.8$.

\item The equation that
RX18 quote for $\ddot{e}\vert_{t=0}$ --- namely their equation
(53) --- is a poor approximation in general. To see this,  we compute the `exact' value of $\ddot{e}\vert_{t=0}$ directly by
differentiating $e = (1-j^2)^{1/2}$ twice, using the DA equations of motion (see equations (12)-(13) of Paper III),
and demanding that at $t=0$, $j =
\jmin$, $\md j /\md t = 0$ and $\omega = \pm\pi/2$.  Without any approximations
we find
%%%%%%%%%%%%%%
\begin{align}
    \left( \frac{\md^2 e}{\md t^2}\right)_{t=0} = 
    -\frac{60\Gamma C}{L} \frac{(\jmin^2 - \Theta)\emax}{\jmin}
    \left( \frac{\md\omega}{\md t}\right)_{t=0}.
    \label{eqn:d2edt2}
\end{align}
%%%%%%%%%%%%%%
For this to coincide with equation (53) of RX18 in the LK ($\Gamma=1$) limit, one must have $(\jmin^2 - \Theta) \approx \jmin^2 = (1-e_\mathrm{max}^2)$,
which is only true if\footnote{It is easy to show that the condition \eqref{eqn:RX18_assumption} is also required
to make equation (54) of  
RX18 agree with equation (12) of Paper III at maximum
eccentricity.}
%%%%%%%%%%%%%%
\begin{align}
    \label{eqn:RX18_assumption}
    \jmin^2 \gg \Theta, \,\,\,\,\,\,\,\,\, \,\,\,\,\, \mathrm{i.e.}\,\,\,\,\, \,\,\,\,\, \cos^2 \imin \ll 1.
\end{align}
%%%%%%%%%%%%%%
The condition
\eqref{eqn:RX18_assumption} \textit{does} happen to be true in the
specific numerical example shown in Figure \ref{fig:Numerical_RX18_Fig3}, but it is certainly
not true in general, as we have seen in several numerical examples (Figures \ref{fig:Numerical_LK}, \ref{fig:Numerical_Initially_WeakLib}, \ref{fig:Numerical_Initially_ModLib} and \ref{fig:Numerical_Gampt176}). In fact, we know from Appendix \ref{sec:high_ecc_no_GWs} that
if a slow-merging binary is initially in the weak GR regime then it has $\jmin^2 \sim \Theta$ all the way into the moderate GR regime and beyond, so in general one should use the formula (\ref{eqn:d2edt2}). 
\end{enumerate}

We can make a direct comparison between our method of computing $\Delta a$ and that of RX18 as follows. Let us follow the RX18 method and use equations
\eqref{eqn:RX18_Delta_a} and \eqref{eqn:d2edt2},
evaluating $\md \omega/\md t$ at maximum eccentricity using
equation (12) of Paper III --- we call the result $\Delta a_\mathrm{RX18}$.  We then compare the result to our equation for $\Delta a$, namely
\eqref{eqn:Delta_a_of_xmax}.
Using $\jmin^2 = 2\pmin/a$ and after some algebra we arrive at
%%%%%%%%%%%%%%%%%%%%%%%%%%%%%%%%%%%%%%%%%%%%%%%%
\begin{align}
    \frac{\Delta a_\mathrm{RX18}}{\Delta a} = \frac{8}{15} 
    \frac{\sqrt{\vert 25\Gamma^2-1\vert }}{10\Gamma} \frac{\vert j_+ j_0\vert }{\xi} \frac{\jmin^2}{\sqrt{\jmin^2-\Theta}} 
    \left( 10\Gamma \Theta -(1+5\Gamma)\jmin^4 + \frac{\epsGR\jmin}{6}\right)^{-1/2}.
\label{eqn:Delta_a_ratio}
\end{align}
%%%%%%%%%%%%%%%%%%%%%%%%%%%%%%%%%%%%%%%%%%%%%%%%
We can make sense of \eqref{eqn:Delta_a_ratio} by evaluating the
right hand side in the weak and moderate GR regimes.

In the weak GR regime we
have $\epsGR \jmin \ll \Theta$ (equation \eqref{eqn:scaling_weak}).  If we also assume $\jmin^4 \ll \Theta$ (see
Appendix C of Paper III for justification) and ignore the $a$-dependent terms in
\eqref{eqn:jplus_of_a}, \eqref{eqn:j0_of_a}, we get
%%%%%%%%%%%%%%%%%%%%%%%%%%%%%%%%%%%%%%%%%%%%%%%%
\begin{align}
    \frac{\Delta a_\mathrm{RX18}}{\Delta a} \approx \frac{8}{15\xi} \approx \sqrt{1+\vert \sigma\vert }.
\end{align}
%%%%%%%%%%%%%%%%%%%%%%%%%%%%%%%%%%%%%%%%%%%%%%%%
where to get the second equality we used \eqref{eqn:dimensionless_integral_not_II}. Note
that for $\sigma \ll 1$ we recover $\Delta
a_\mathrm{RX18} = \Delta a$, i.e. our calculation coincides precisely with that of
RX18 when we make the approximations that they
(implicitly) did, namely weak GR and $\sigma \ll 1$. However, we emphasize that neither of these approximations is actually valid for the example shown in Figure \ref{fig:Numerical_RX18_Fig3}.

In the moderate GR regime we assume that the $\epsGR$ term dominates the final
bracket in \eqref{eqn:Delta_a_ratio}, and that the $a$-dependent
terms dominate equations \eqref{eqn:jplus_of_a},
\eqref{eqn:j0_of_a}.  With these assumptions we get
%%%%%%%%%%%%%%%%%%%%%%%%%%%%%%%%%%%%%%%%%%%%%%%%
\begin{align}
    \frac{\Delta a_\mathrm{RX18}}{\Delta a} &= \frac{8\sqrt{60\Gamma}}{15} \left( \frac{d}{a}\right)^{7/2}
    \frac{1}{\xi}\frac{\jmin^2}{\sqrt{\jmin^2-\Theta}} \sqrt{\frac{6}{\epsGR\jmin}}
     \nn
    \\
    &\sim \frac{\sqrt{10\Gamma}}{\xi} \left( \frac{d}{a}\right)^{7/2}  \frac{\jmin}{\epsGR},
\label{eqn:Delta_a_ratio_mod}
\end{align}
%%%%%%%%%%%%%%%%%%%%%%%%%%%%%%%%%%%%%%%%%%%%%%%%
with $\xi$ given in equation
\eqref{eqn:dimensionless_integral_type_II} (and plotted in Figure
\ref{fig:dimensionless_integral_type_II}). All three fractions in
\eqref{eqn:Delta_a_ratio_mod}
are $\mathcal{O}(1)$ or larger.  Thus we typically
have $\Delta a_\mathrm{RX18}/\Delta a  \gg 1$, meaning that the method of
RX18 can seriously overestimate the value of $\Delta a$ in the moderate GR
regime.

%%%%%%%%%%%%%%%%%%%%%%%%%%%%%%%%%%%%%%%%%%%%%%%%
%%%%%%%%%%%%%%%%%%%%%%%%%%%%%%%%%%%%%%%%%%%%%%%%
\subsection{Decrease in $\tsec$ with time}
%%%%%%%%%%%%%%%%%%%%%%%%%%%%%%%%%%%%%%%%%%%%%%%%
%%%%%%%%%%%%%%%%%%%%%%%%%%%%%%%%%%%%%%%%%%%%%%%%
\label{sec:tsec_RX18}

As we mentioned in \S\ref{sec:Introduction}, the decrease in $\tsec$
with time during a slow merger was first pointed out by RX18 in their \S3.1, when discussing the example shown in Figure \ref{fig:Numerical_RX18_Fig3}.
When interpreting this counter-intuitive scaling of $\tsec(a)$ physically, RX18
noted that smaller $a$ (larger $\epsGR$) promotes faster apsidal precession,
which is obviously true.  They then claimed that this faster precession
directly leads to a shorter secular period.  They also claimed that it was directly responsible for the corresponding
increase in maximum eccentricity with time and decrease in minimum eccentricity
with time as the binary shrinks (Figure \ref{fig:Numerical_RX18_Fig3}b). 

This interpretation is not quite right, and also does not explain
why in the librating regime $\tsec$ \textit{increases} with shrinking $a$. In reality, in the weak-to-moderate regime, GR precession is unimportant except during an
extremely high eccentricity episode. Typically these extreme eccentricity
episodes last a very short time compared to the secular period. In other words, for most phase space trajectories the
second (GR) term in equation (56) of RX18 is completely
negligible during the majority of the evolution, so barely affects $\tsec$. What
GR precession \textit{does} do, when coupled with GW emission, is to alter the
phase space morphology, and to periodically nudge the binary onto a new phase
space trajectory every time it reaches high eccentricity (note how closely the
contours of $H^*$ are bunched at these high eccentricities in Figure \ref{fig:Numerical_RX18_Fig3}f-h). As $a$ is decreased
and $\epsGR$ is increased, after passing from libration to circulation the binary gets pushed ever further away from the
separatrix, towards the low-$\jmax$
circulating region where \eqref{eqn:t_sec_D} applies. As
long as this process continues the binary gets pushed to higher minimum
eccentricity (smaller and smaller $\jmax$), even though its $\emax$ is getting
smaller. On average the binary spends more and more time at `high' (say
$e\gtrsim 0.9$) eccentricities where cluster tide-driven secular evolution is
fast (since the binary angular momentum is small). We emphasize that this last statement is true regardless of GR precession:
indeed, the binary typically does not care about GR precession directly when,
say, $e=0.9$. Thus, whereas RX18 attributed the evolution of
$\tsec$ and $e_\mathrm{min/max}$ to fast GR-aided $\omega$ precession during the whole secular cycle, both of
these phenomena are present even in the weak GR regime where apsidal precession
is nearly always negligible --- see Figure
\ref{fig:Numerical_Initially_WeakLib}.
%\footnote{Mathematically, the tidal
%contribution to $\dot{\omega}$ is proportional to $j^{-1}$, while
%$\dot{\omega}_\mathrm{GR} \propto j^{-2}$, so although GR dominates at extremely
%high eccentricities (extremely small $j$), the cluster tides dominate
%otherwise.}.

%%%%%%%%%%%%%%%%%%%%%%%%%%%%%%%%%%%%%%%%%%%%%%%%%%



\bibliographystyle{apj}
\bibliography{Bibliography} % if your bibtex file is called example.bib

%%%%%%%%%%%%%%%%% APPENDICES %%%%%%%%%%%%%%%%%%%%%

\end{document}